\RequirePackage{fix-cm}
\documentclass{svjour3}                     
\smartqed  
\usepackage{graphicx}
\usepackage{epsfig}
\usepackage{multirow}
\usepackage{makecell}

\begin{document}

\title{Simulation of force-insensitive optical cavities in cubic spacers
}


\author{Eugen Wiens         \and
        Stephan Schiller 
}


\institute{
              Institut f\"ur Experimentalphysik, Heinrich-Heine-Universit\"at D\"usseldorf,
          D\"usseldorf, Germany \\ \\
              Tel.: +49-211-81-12317\\
              \email{step.schiller@hhu.de}           
}

\date{Received: date / Accepted: date}

\maketitle

\begin{abstract}
We analyze the properties of optical cavities contained in spacers
with approximate octahedral symmetry and made of different materials,
following the design of Webster and Gill (S.~Webster, P.~Gill, Optics Letters \textbf{36}(18), 3572 (2011)). We show
that for isotropic materials with Young's modulus less than $200$~GPa,
the Poisson's ratio $\nu$ must lie in a ``magic'' range $0.13<\nu<0.23$
in order to null the influence of the forces supporting the spacer.
This restriction can be overcome with the use of anisotropic materials
such as silicon. A detailed study aiming at identification of all
suitable crystal orientations of silicon with respect to the resonator
body is performed and the relation to the Poisson's ratio and the
Young's modulus along these orientations is discussed.  We also perform
an analysis of the sensitivity of the cavity performance to errors
in spacer manufacturing. We find that the orientation of the {[}110{]} or {[}100{]}
crystallographic directions oriented along one of the three optical axes of the resonator provides low sensitivities to imprecise manufacturing and interesting options for fundamental physics experiments.
\end{abstract}


\section{Introduction}
\label{sec:Intro}
Optical Fabry-P\'erot resonators are widely used in different fields
of optics and metrology. As passive optical resonators they can provide
the frequency reference for obtaining laser waves with ultra-stable
frequencies for interrogation of transitions in atomic clocks \cite{Huntemann2016,Ludlow2015,Nicholson2015},
for gravitational wave detectors, or for fundamental tests of space-time
structure \cite{Antonini2005,Eisele2009,Wiens2016}. Transfer of the
frequency stability of laser waves into the microwave region potentially
enables their application in radars and in navigation of deep space
probes. High demands on frequency stability of laser light set forth
by these applications require an optical resonator with a low sensitivity
to vibrations.

Optical resonators are usually made of a spacer and two mirrors optically
contacted to it. The frequency stability of such a resonator is determined,
once evacuated, by the length stability of the resonator's spacer.
Vibrations transferred from the surroundings to the resonator change
the distance between the mirrors and tilt them, degrading the frequency
stability that it can provide. To fulfill the requirement of low vibration
sensitivity a careful design of the shape of the resonator and of
the supporting frame is needed.

The design with the lowest sensitivity to vibrations so far was presented
by Webster and Gill \cite{Webster2011}. The cavity structure (Fig.~\ref{fig:Cavity-Dimensions},
top row) consists of a cube-shaped body made of ultra-low expansion
glass ULE material with a side length of 50 mm. It is held inside
a frame (not shown in the figure) by four supports acting at four
tetrahedrically oriented cube vertices. Three cavities are contained
in the body. The cubic (more precisely: octahedral) symmetry of the
cube-like spacer causes, upon action of a body force density (gravity or acceleration)
oriented in arbitrary direction, an equal displacement of the centers
of opposing faces, and therefore zero differential displacement. This
makes the three cavities (completely) insensitive to accelerations
along any axis. Experimentally, finite acceleration sensitivities
are observed: of the three sensitivity coefficients, the smallest
was $k_{y}=1\times10^{-13}/{\rm g}$, the largest $k_{x}=2.5\times10^{-11}/{\rm g}$.
Values of this order can be explained by imperfections in fabrication
or mounting.

Furthermore, the cube vertices are truncated to a depth of 6.7~mm.
This value was determined by Finite Element Analysis (FEA) simulations
and ensures that the external forces acting at the support points,
if equal, do not shift the position of the centers of the faces of
the cube. This means that the three cavity lengths are insensitive
to the support forces.

The material ULE has the advantageous property of a zero Coefficient
of Thermal Expansion (CTE) at or near room temperature and therefore
makes the resonator insensitive to thermal fluctuations. The drawback
of ULE is its slow dimensional change due to its amorphous nature
and the moderate Young's modulus $E=$67.6 GPa, a value that is relevant
if one considers deviations of the spacer from ideal symmetry (see
Sec.~\ref{sec:Effect-of-imperfections}). Another fundamental limitation
is the thermal Brownian noise of the ULE spacer \cite{Numata2004a,Notcutt2006}.
The mirror substrates, usually made of the same or from a similar
material (fused silica), also contribute to the thermal noise \cite{Davila2017}.

Reduction of the operating temperature of resonators down to cryogenic
temperatures is an approach that can reduce thermal noise \cite{Numata2004a}.
This has motivated the development of optical resonators cooled to
cryogenic temperature \cite{Seel1997}. Cryogenic resonators operated
at particular temperatures or close to zero absolute temperature also
exhibit an ultra-low CTE, which relaxes the requirements on temperature
stability \cite{Matei2017,Zhang2017,Wiens2014,Wiens2015,Seel1997}.
High-performance cryogenic optical resonators have so far been crystals,
which also enjoy the advantage of a long-term drift orders of magnitude
smaller than ULE \cite{Wiens2016,Hagemann2014}.

In our work we present an analysis of the extension of the
design of Webster and Gill \cite{Webster2011} to other materials
beside ULE, in particular to materials that may be used advantageously
at cryogenic temperatures. Because of the vibrational noise present
in closed-cycle cryostats, it is particularly important to develop
resonators with low acceleration sensitivity. In addition, the analysis
seeks to answer the question whether it is possible to achieve an
even lower acceleration sensitivity than possible with ULE when considering
the influence of manufacturing errors.
%
\section{Spacer geometry and modeling method}
\label{sec:Geometry_And_Modelling}
The main goal of the modeling is to find shapes and materials that
lead to an insensitivity of the cavities contained in the cubic spacer
to the strength of the forces acting on four vertices of the spacer.
The sensitivity to acceleration arises automatically from the assumed
octahedral symmetry of the spacer, and requires, at a first glance,
no particular simulation. However, when there are deviations from
symmetry, simulations allow to determine the acceleration sensitivity.
We performed simulations using a commercial FEA software (Ansys).
The FEA computation yields the fractional length change of each optical
cavity upon application of a set of forces.

\begin{figure}[tb]
\begin{minipage}[t]{0.48\columnwidth}%
\begin{center}
\includegraphics[width=0.9\textwidth]{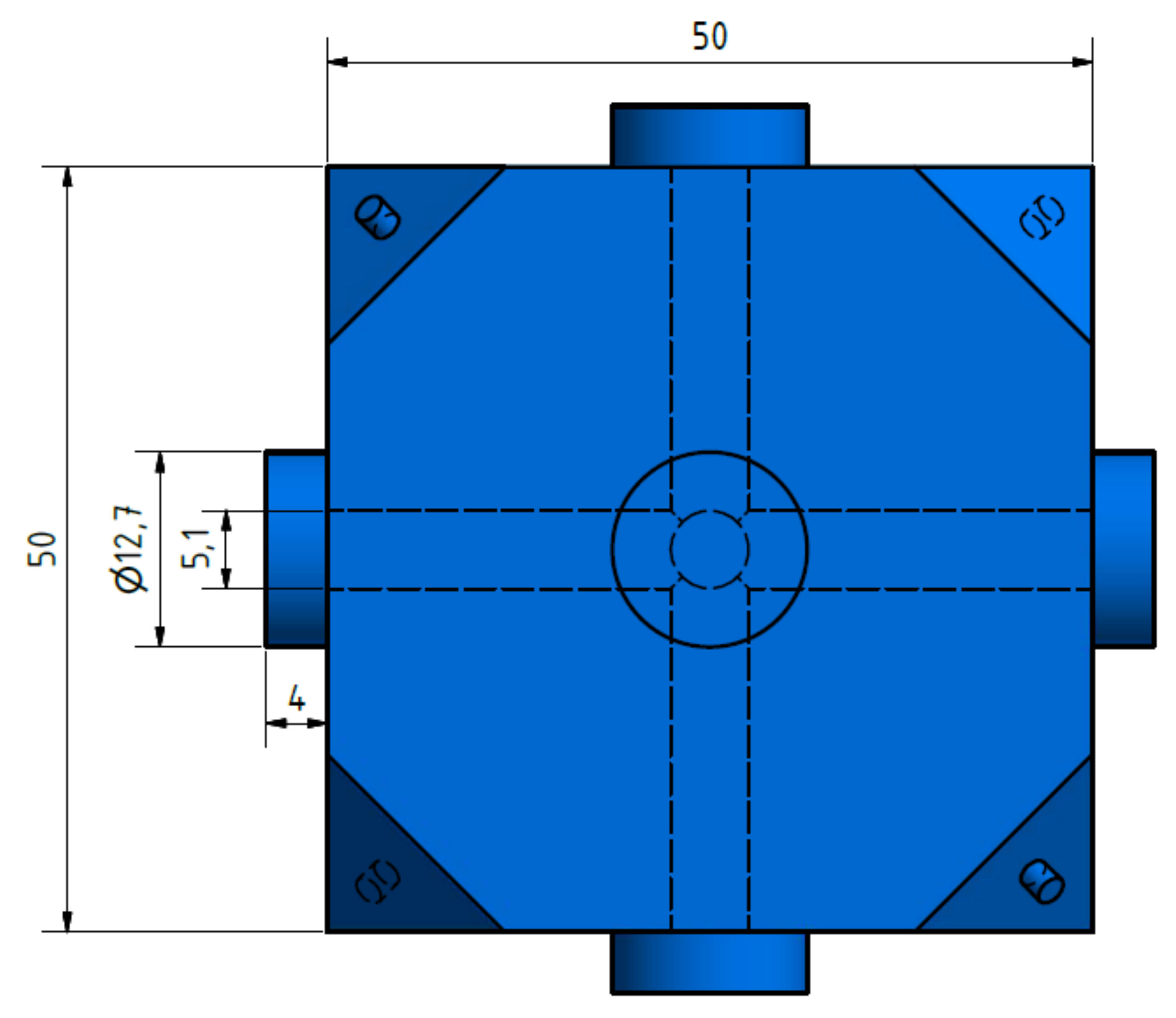}
\par\end{center}%
\end{minipage}\quad{}%
\begin{minipage}[t]{0.48\columnwidth}%
\begin{center}
\includegraphics[width=0.8\textwidth]{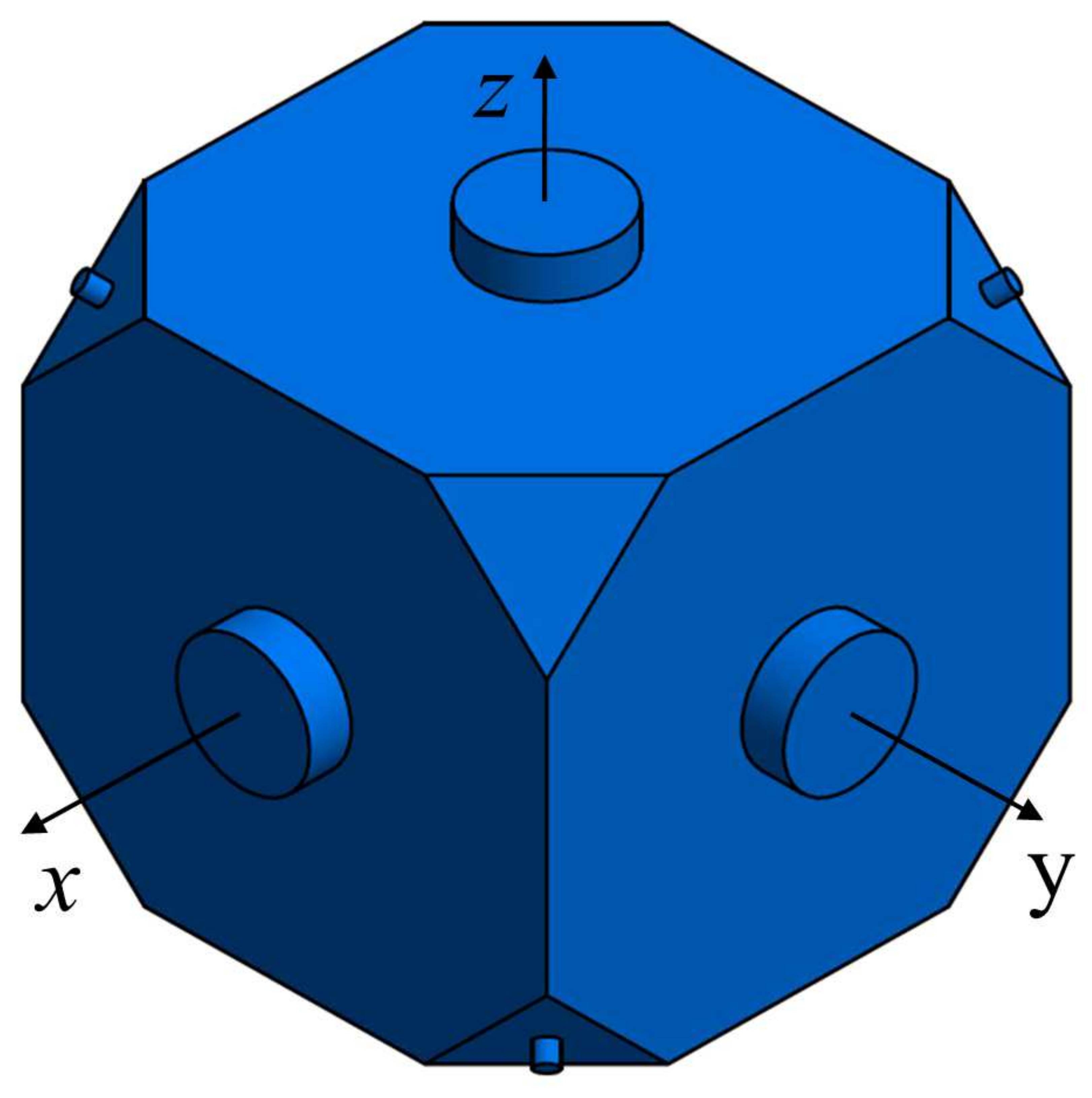}
\par\end{center}%
\end{minipage}\medskip{}
\begin{minipage}[t]{0.48\columnwidth}%
\begin{center}
\includegraphics[width=0.75\textwidth]{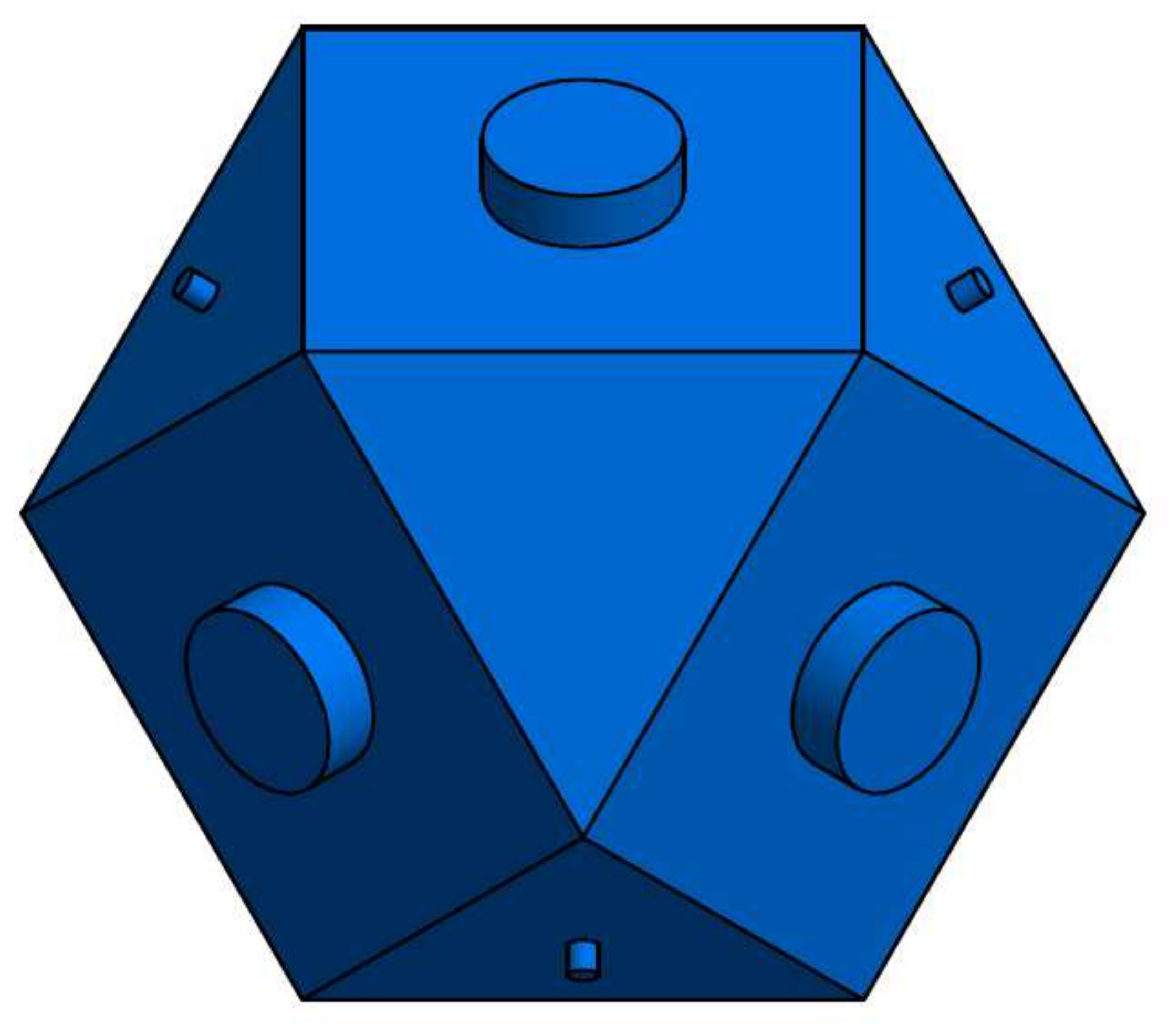}
\par\end{center}%
\end{minipage}\quad{}%
\begin{minipage}[t]{0.48\columnwidth}%
\begin{center}
\includegraphics[width=0.6\textwidth]{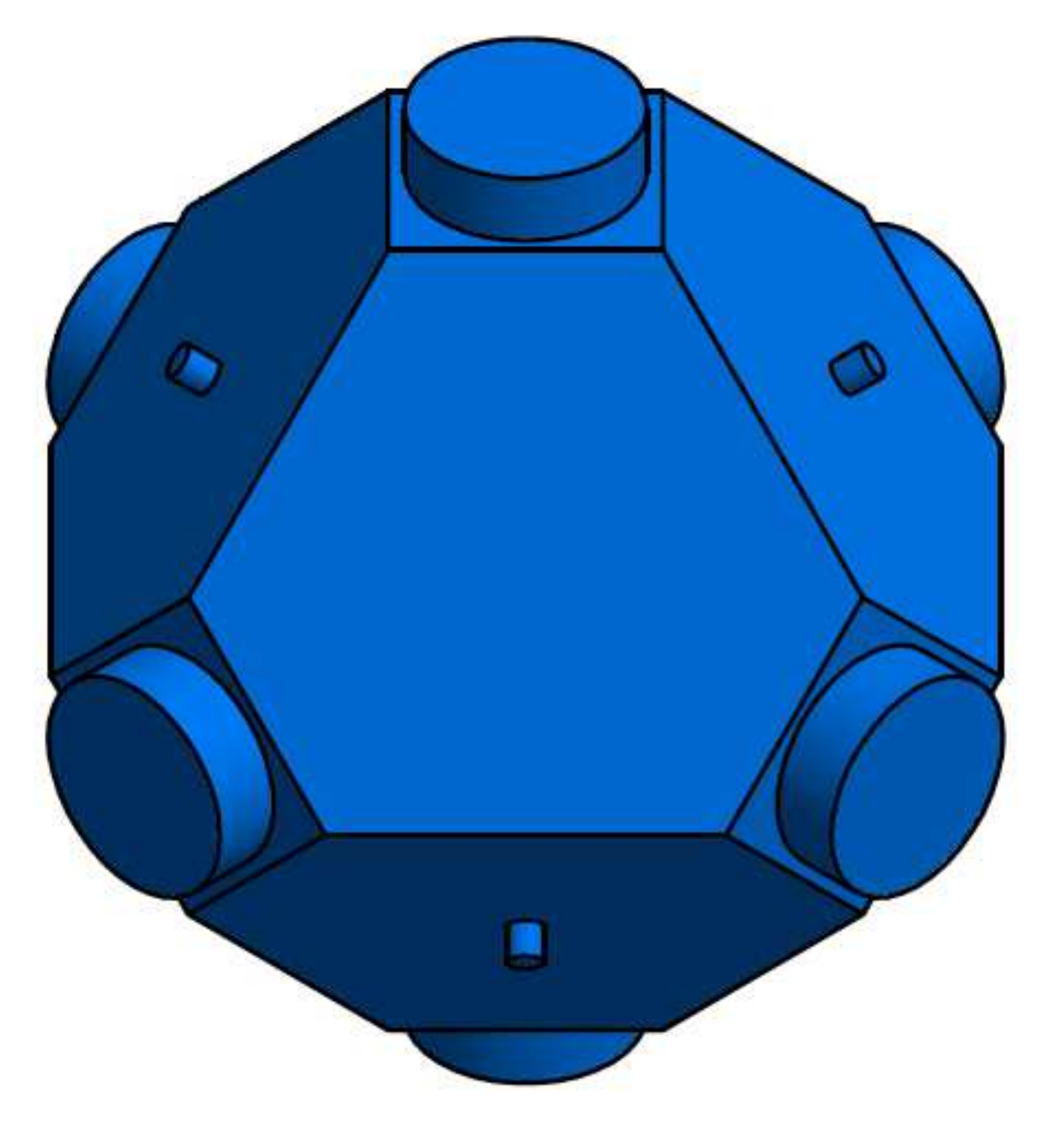}
\par\end{center}%
\end{minipage}\caption{\label{fig:Cavity-Dimensions}Shapes and dimensions of cubic spacers.
Top, left: cavity dimensions used in the FEA simulations. Top, right:
orientation of the resonator with respect to the laboratory reference
frame defined by $(x,\,y,\,z).$ The shape is that derived by Webster
and Gill. Bottom, left: transformation of the cube to its dual Platonic
solid, the octahedron, when the cut depth of the vertices is equal
to 14.47~mm. Bottom, right: cavity with a large cut depth, 23~mm. }
\end{figure}

For concreteness, we chose the same dimensions for the spacer block
as in \cite{Webster2011}: a cube with side length $L=50$~mm. Its
density and mass are denoted by $\rho$ and $m$, respectively. The
edges of the cube lie along the space-fixed coordinate system axes
$(x,\,y,\,z).$ Although only a single cavity, here the $x$-cavity,
is usually of interest, three mutually orthogonal cavities are formed
by three through holes along the directions $x,\,y,\,z$, so as to
preserve octahedral symmetry. For concreteness, they have $R_{b}=2.55$~mm
radius.  A total of six mirror substrates of the same material as
the block, having a diameter of $12.7$~mm and a thickness of $4$~mm, are attached to the end faces of the spacer. The substrates and the block are assumed to form
a single unit (see Fig.~\ref{fig:Cavity-Dimensions}).

 No pumping holes were included in the simulation. Since in the actual
manufacturing the hole diameter could be chosen small, we expect that
its effect on the mechanical properties would be minor. The eight
corners of the cavity are truncated to a depth $d$ which is a free
parameter.

The cavity is always simulated with four ``holding'' forces applied normal to four of the truncated corners. They each have an arbitrarily chosen but realistic magnitude of $F_{c}=1$~N, and are applied via four cylindrically shaped supports
(here, having $r=2\,$mm diameter) rigidly attached to the resonator
at the four corners with a tetrahedral symmetry. We consider two cases:

(1) the application of only the four holding forces $F_{c}$, i.e.
gravity is ignored. The sensitivity to support force strength, $\Delta L_{i}(F_{c})/L=(L_{i}(F_{c})-L)/L$,
is calculated.

(2) In presence of $F_{c}$, an additional  acceleration $a_j$ acting along $j=x,\,y,\,z$ is applied. This simulates acceleration of the cavity support (and thus of the cavity body) or the gravitational acceleration. 
For this case we define the acceleration sensitivity $k_{ij}=\Delta L_{i}(a_{j})/(a_{j}\,L)$,
where $\Delta L_{i}(a_{j})$ is the additional cavity length change
when $a_{j}$ is added.
%
%
\section{Spacer made of ULE}
\label{sec:ULE_Spacer}
\begin{figure}
\begin{centering}
\begin{center}
\includegraphics[width=0.48\textwidth]{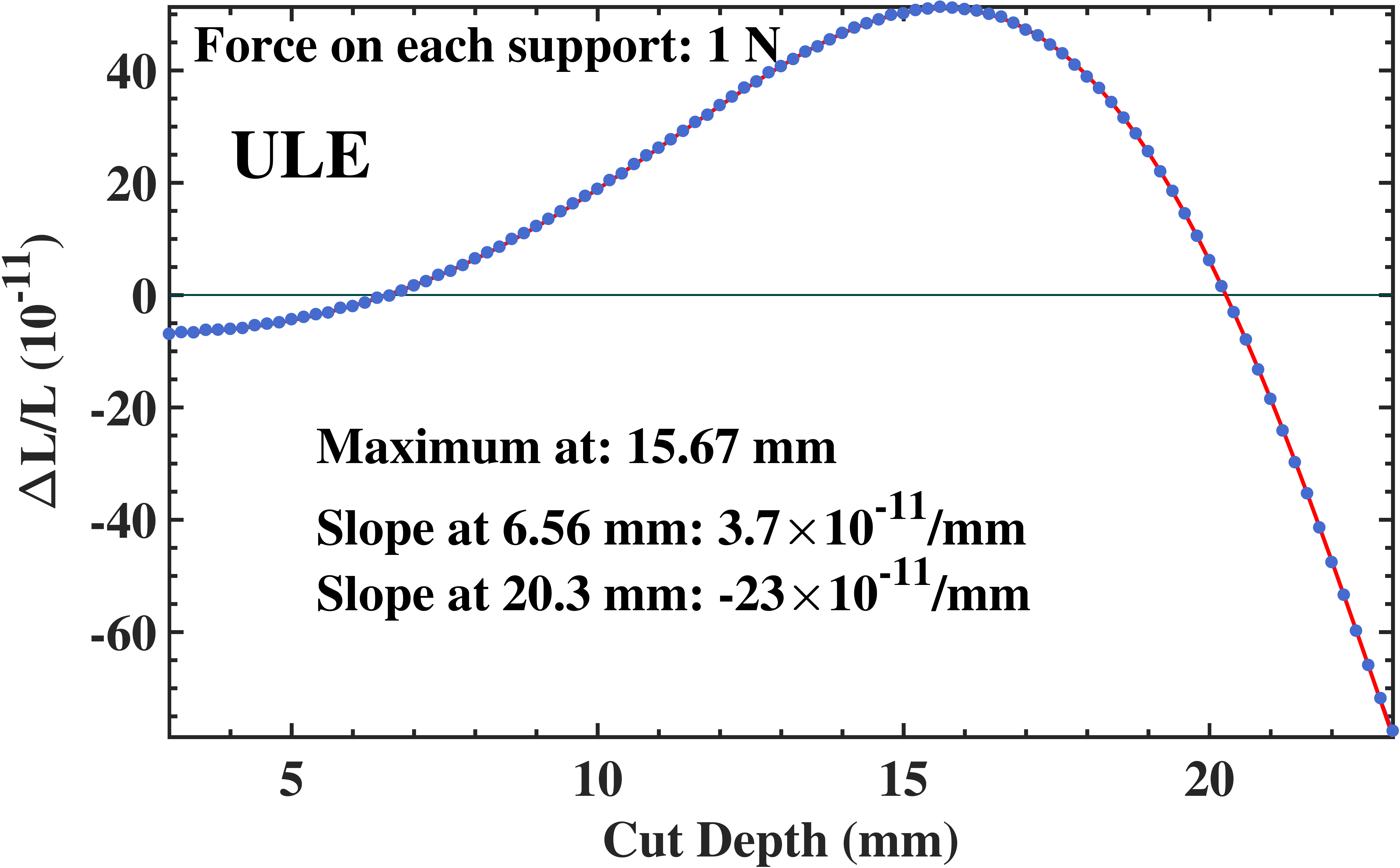}
\quad{}
\includegraphics[width=0.46\textwidth]{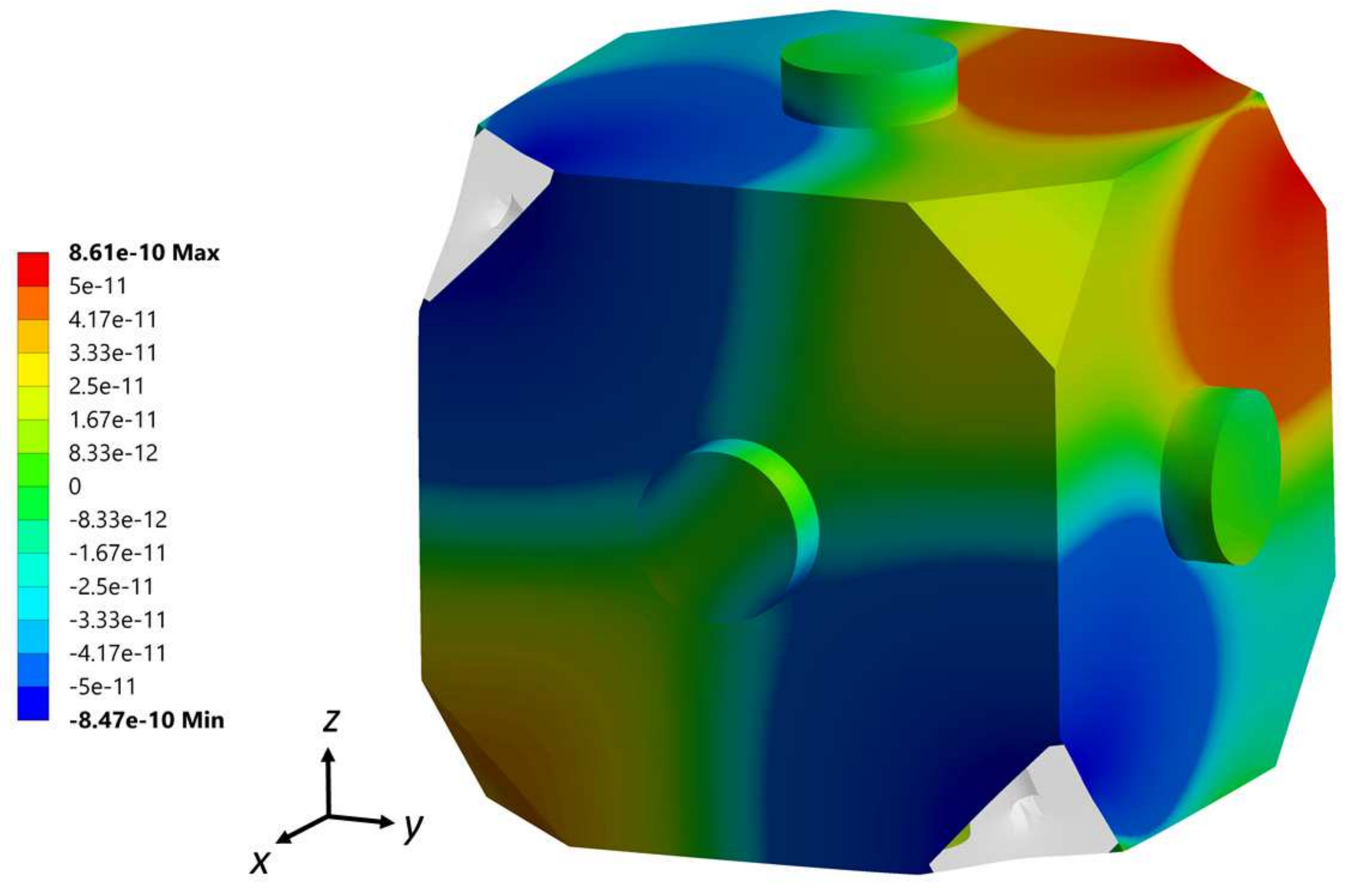}
\end{center}
\bigskip{}
\begin{center}
	\includegraphics[scale = 0.15]{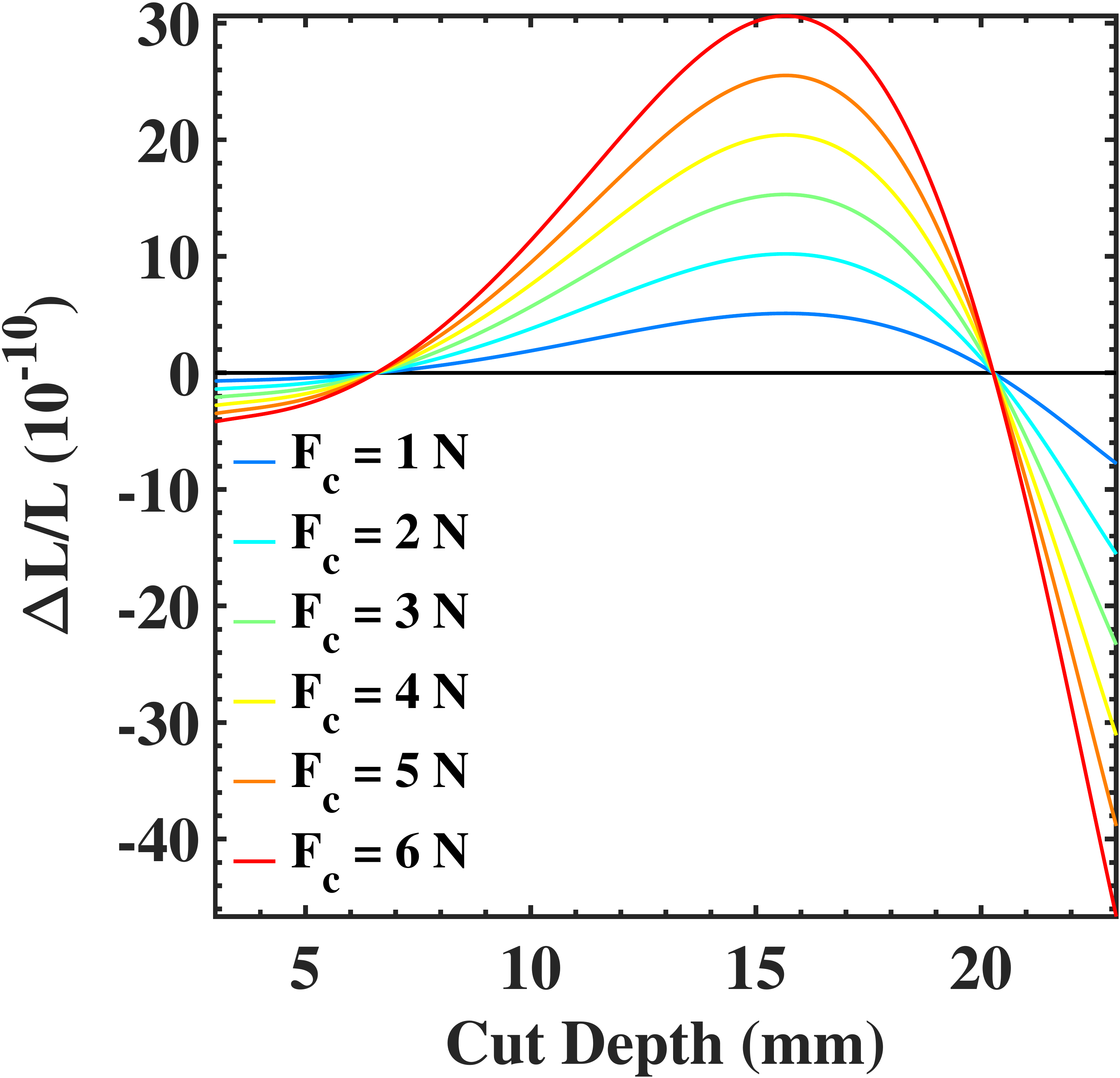}
	\quad{}
	\includegraphics[scale = 0.15]{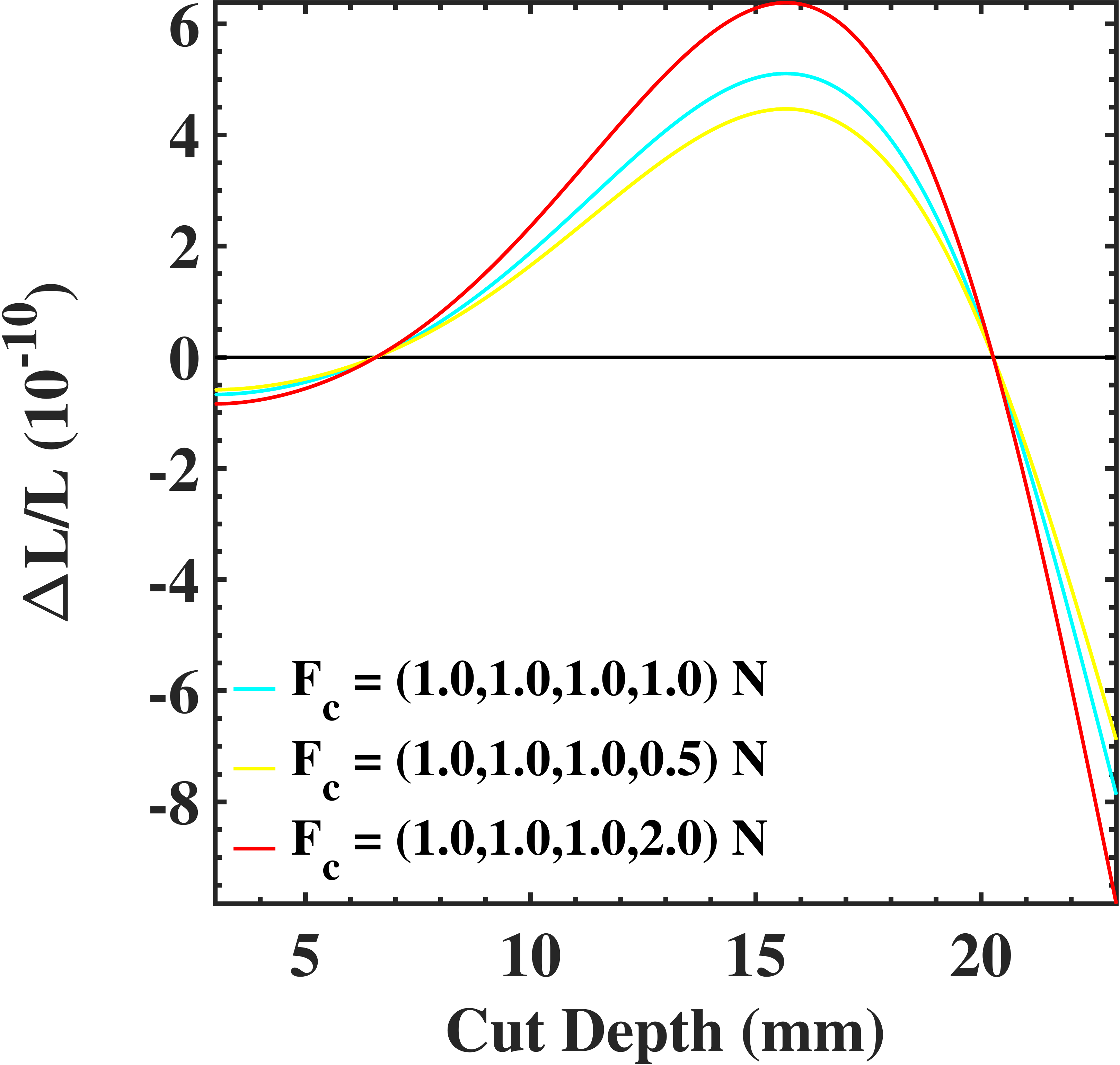}
\end{center}
\bigskip{}
\begin{center}
\includegraphics[scale = 0.15]{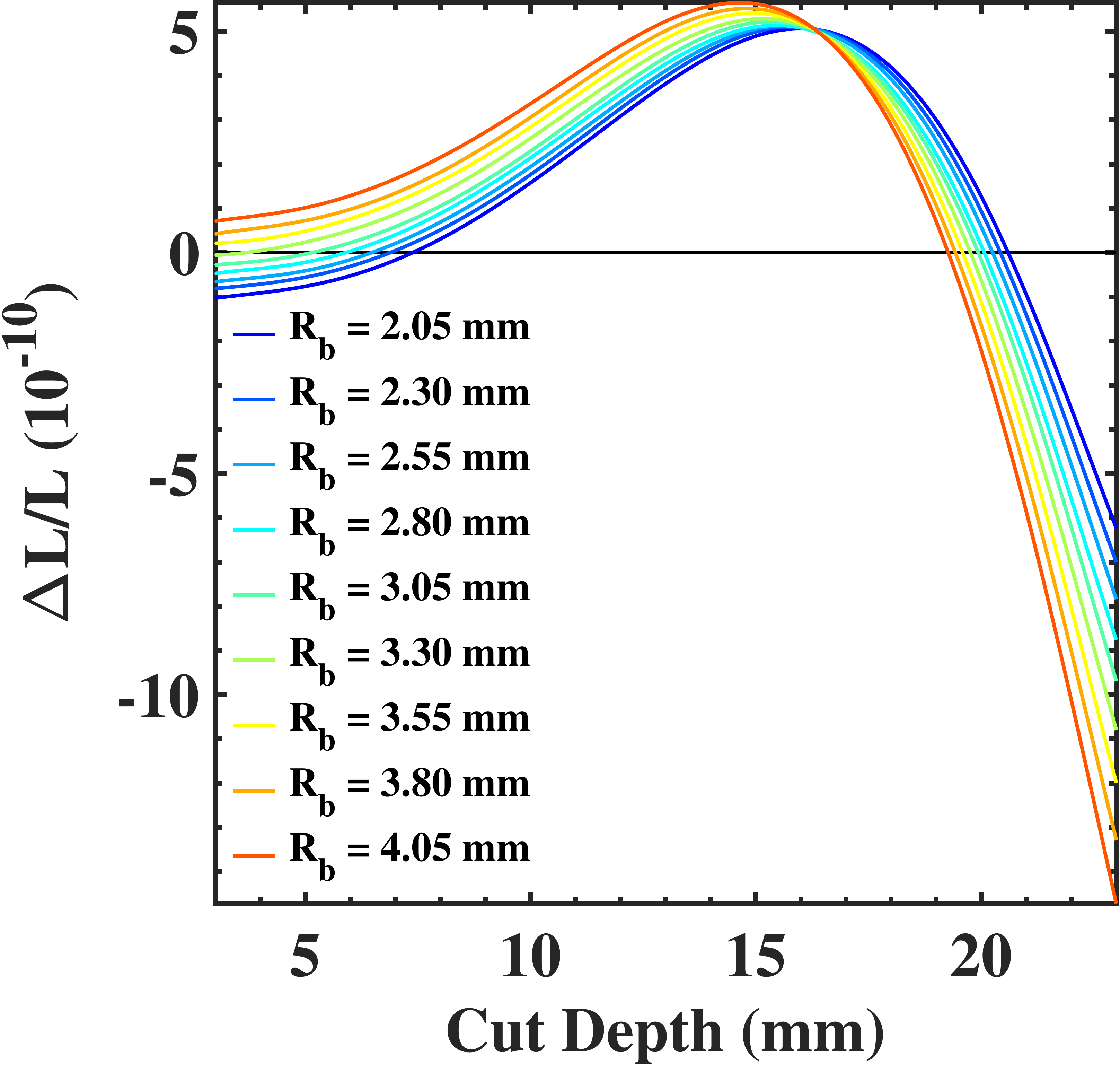}
\quad{}
\includegraphics[scale = 0.15]{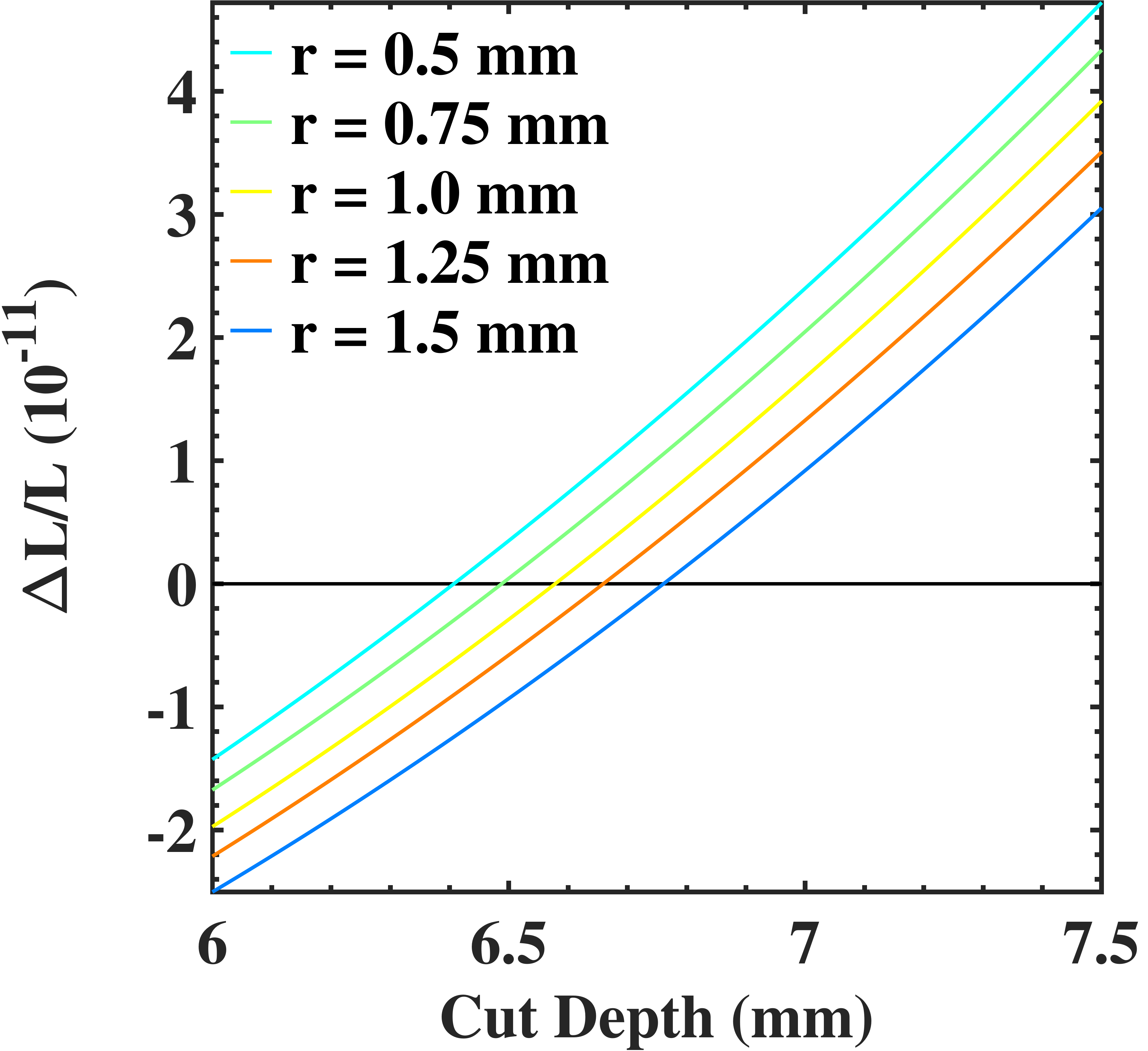}
\end{center}
\end{centering}
\caption{\label{fig:ULE Cavity Simulation Results}  Top left: fractional
length change $\Delta L(F_{c})/L$ of the cavity in an ULE block,
vs. the cut depths of the vertices. Top right: axial ($x$-axis) displacement
at the optimum cut depth of $6.6$~mm. The scale is in meter. Middle left: fractional length change $\Delta L(F_{c})/L$ as function of force $F_{c}$ applied at four support points. Middle right: fractional length change $\Delta L(F_{c})/L$ as function of force variation at one of the four supports. Magnitude of force applied at each of the four supports is depicted in the brackets.  Bottom
left: fractional length change $\Delta L(F_{c})/L$ as a function
of cut depth, for various bore diameters $R_{b}$. Bottom right: fractional
length change for different supports radii. }
\end{figure}

The computation was tested on a ULE cube with six mirror substrates
made of ULE. The substrates considered in \cite{Webster2011} were
from fused silica; this difference is minor. The cut depth was varied
in the interval between $3$~mm and $23$~mm. The fractional length
change $\Delta L_{x}(F_{c})/L$ of the $x$-cavity occuring when the
holding forces are applied, is depicted in Fig.~\ref{fig:ULE Cavity Simulation Results},
top left. Initially, for small cut depths, is is negative. This means
that the distance between the mirrors is reduced upon application
of the forces. At a cut depth of $6.6$~mm it crosses zero for the
first time with a slope of $3.7\times10^{-11}/$mm. The cube deformation
for this case is seen in Fig.~\ref{fig:ULE Cavity Simulation Results}
top right. Clearly, the central part of the mirror on the $+x$-face
of the cube does not have any $x$-displacement. From symmetry, also
the $-x$-face remains unaffected, and this results in $\Delta L_{x}(F_{c})/L=0$.
After passing the cut depth of $14.5$~mm, the shape of the spacer
becomes octahedral. Soon after, at a cut depth of $15.7$~mm, $\Delta L_{x}(F_{c})/L$
is maximum and then starts to decrease with increasing cut depth.
The second zero crossing is reached at a cut depth of $20.3$~mm
with a slope of $-23\times10^{-11}/$mm.  Among the two zero-sensitivity
cut depths, the smaller one, $6.6$~mm, is clearly more preferable
since the slope is six times smaller and so the shape is more forgiving
in case of fabrication errors.

To justify our arbitrary choice of force magnitude $F_{c}=1$~N acting at each of the four supports, we analyzed the influence of force magnitude on optimum cut depth. We found no dependence of the position of zero-sensitivity cut depth on $F_{c}$, when we varied the latter in the range $1$~N$ < F_{c} < 1$~kN (see Fig.~\ref{fig:ULE Cavity Simulation Results}, middle left panel. Due to the large difference in scale, only simulation results for forces $F_{c}\leq6$~N are presented there). The sensitivity to force $F_{c}$ at zero crossing cut depth is $18\times10^{-12}$/N$\cdot$mm.

We studied the influence of unequal forces at the four supports by holding the force $F_{c}$ at a constant magnitude of $F_{c} = 1$~N at three support points and varying the force acting at the fourth support  by a factor 2. Note that this is possible without causing an overall resonator displacement since in the simulations 
each support surface is allowed to move only along the direction perpendicular to it. The overall effect of force variation at one support is equivalent to the application of additional longitudinal forces and generation of transversal forces at three remaining support surfaces.  The results of this simulation are presented in Fig.~\ref{fig:ULE Cavity Simulation Results}, middle right panel. We found no dependence of optimal cut depth on variation of force at one support.

The optimum cut depth is found to increase with the size of the supports,
as displayed in Fig.~\ref{fig:ULE Cavity Simulation Results}, bottom
right. This could explain the small difference of $0.1$~mm in optimum
cut depth between the result presented here and in \cite{Webster2011}.
Another crucial geometry parameter that has an influence on the position
of the zero crossing is the size of the cavity bores. Fig.~\ref{fig:ULE Cavity Simulation Results},
bottom left, shows that the zero sensitivity cut depth near 6.6~mm
only exists if the bore radius is below $3.5$~mm. The second zero
crossing at near $20.3$~mm exists for all studied bore sizes, but
has a slope that increases with increasing bore size.
%
%
%
\subsection{Influence of shape on sensitivity}
\label{sec:Influence_Shape_On_Sensitivity}
\begin{figure}
\begin{centering}
\includegraphics[width=0.3\textwidth]{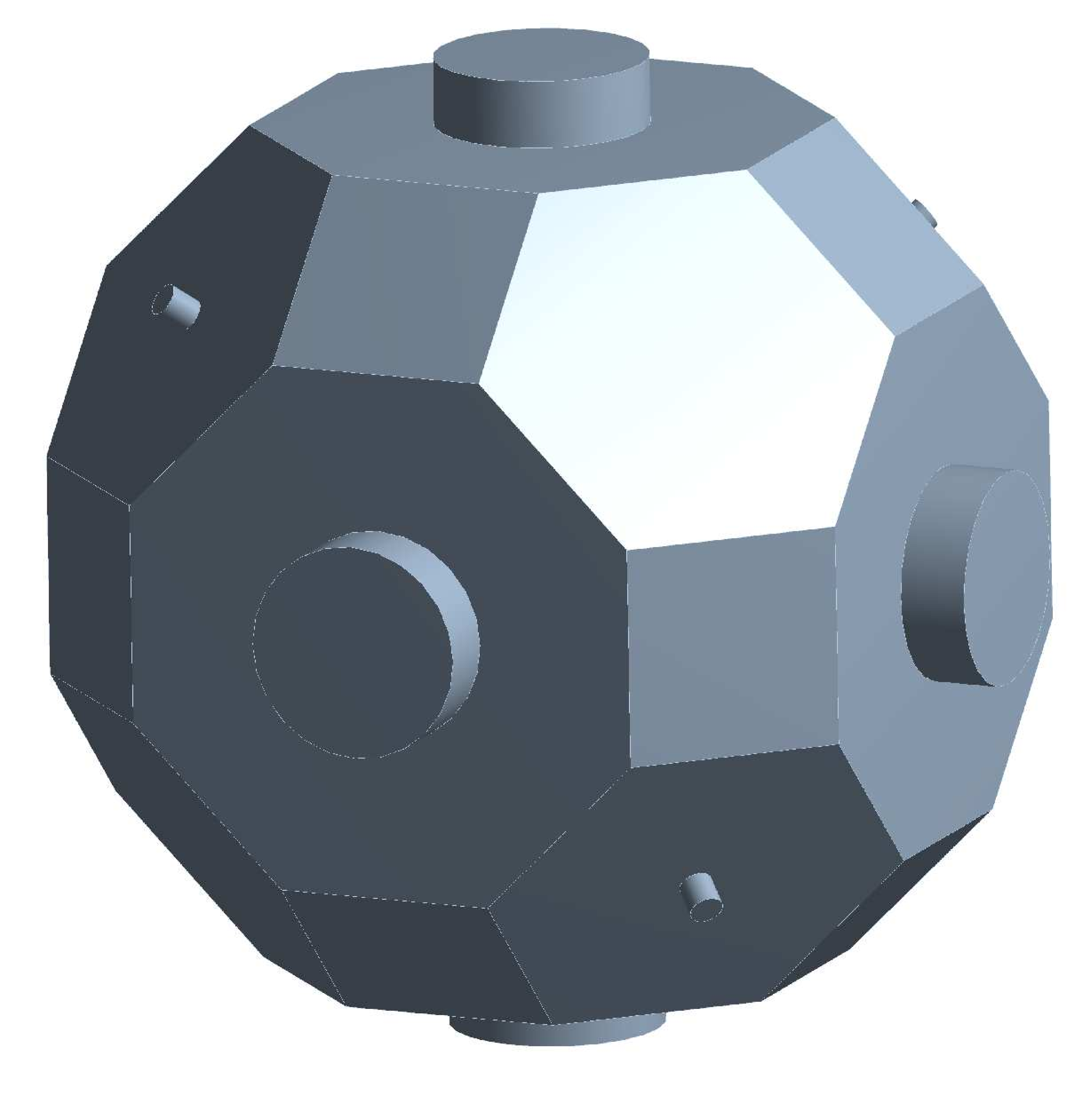}%
\quad{}%
\includegraphics[width=0.3\textwidth]{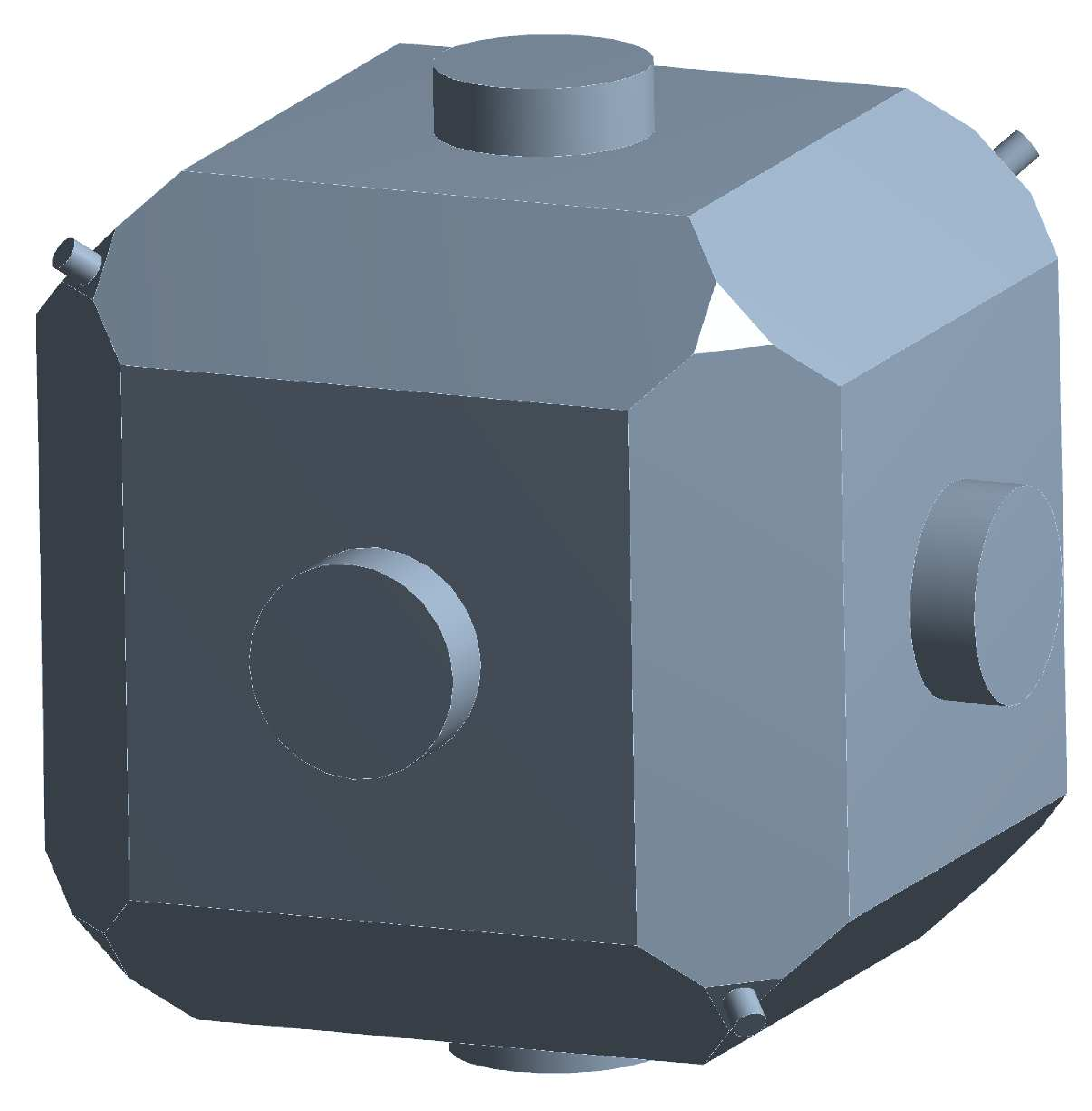}%
\quad{}%
\includegraphics[width=0.3\textwidth]{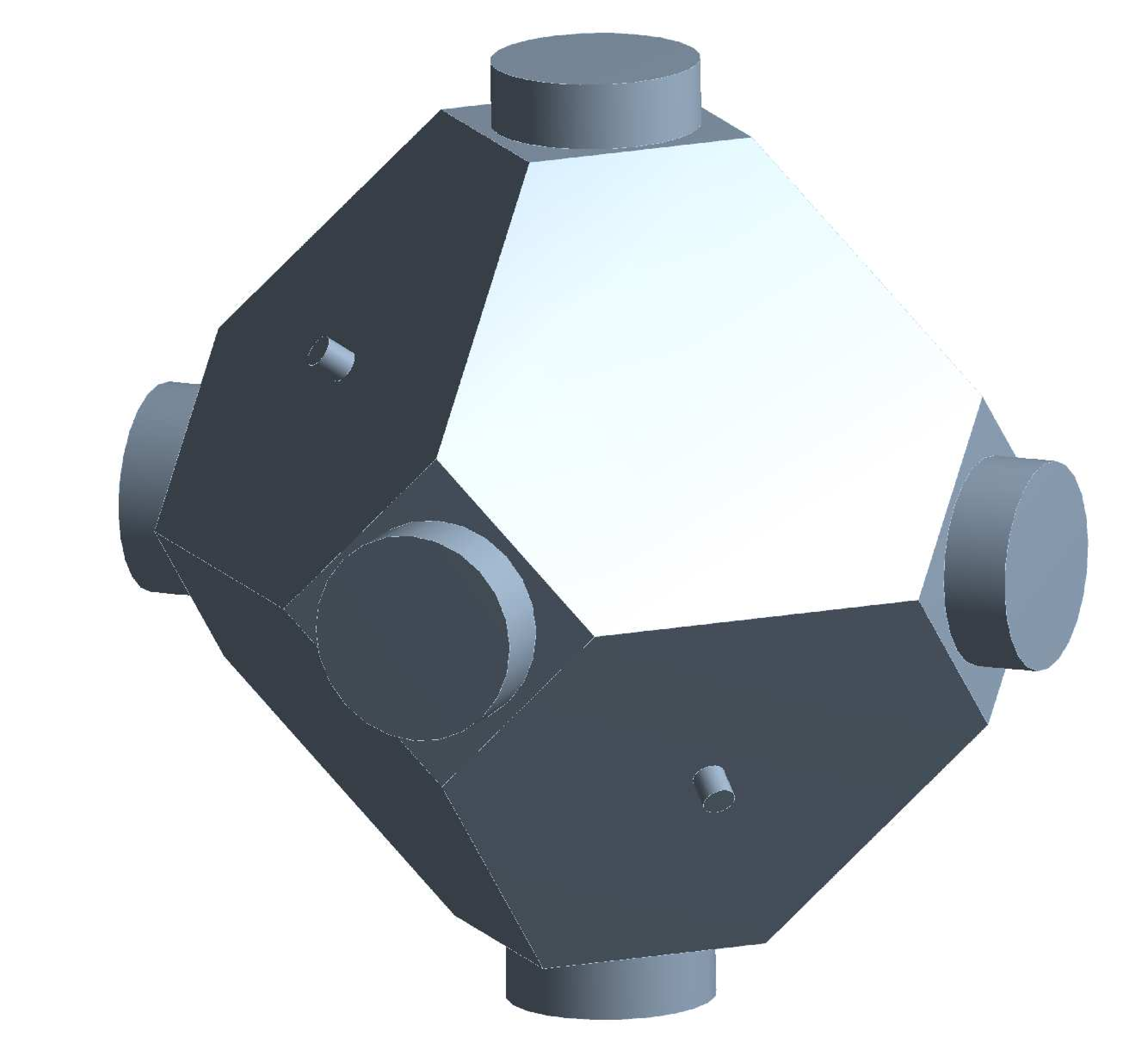}%
\par\end{centering}
\begin{centering}
\bigskip{}
\par\end{centering}
\begin{centering}
\includegraphics[width=0.3\textwidth]{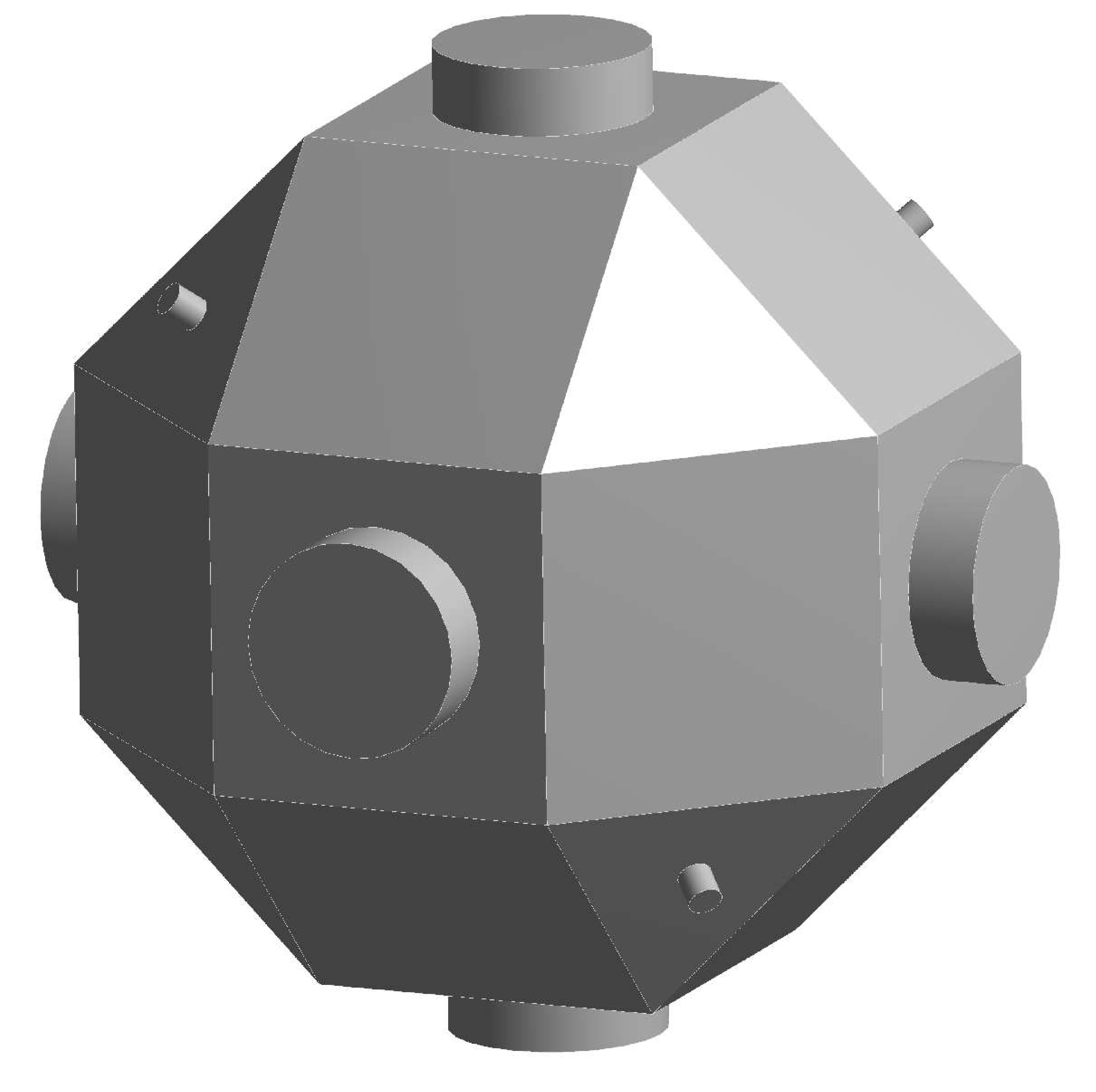}%
\quad{}%
\includegraphics[width=0.3\textwidth]{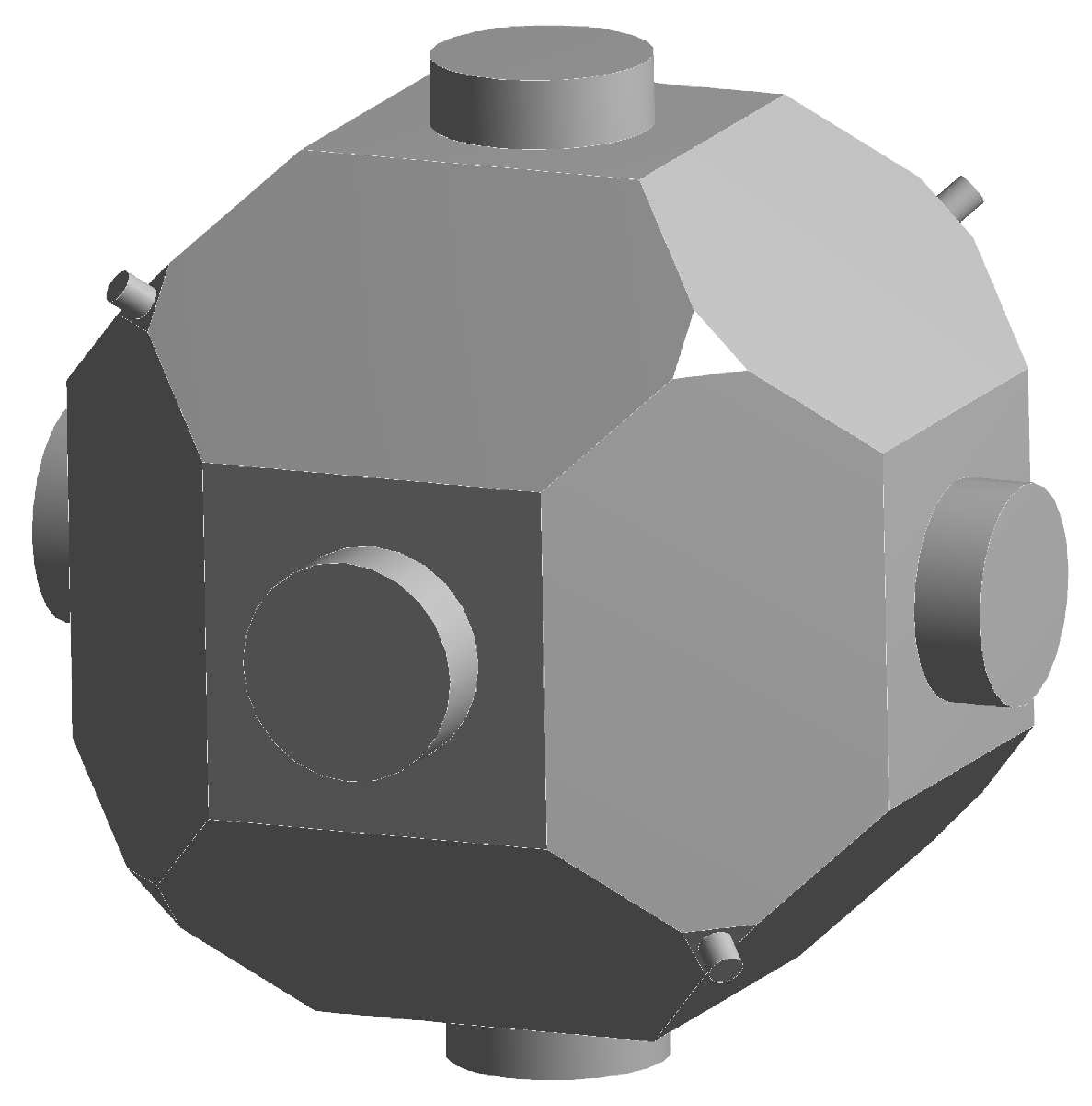}%
\quad{}%
\includegraphics[width=0.3\textwidth]{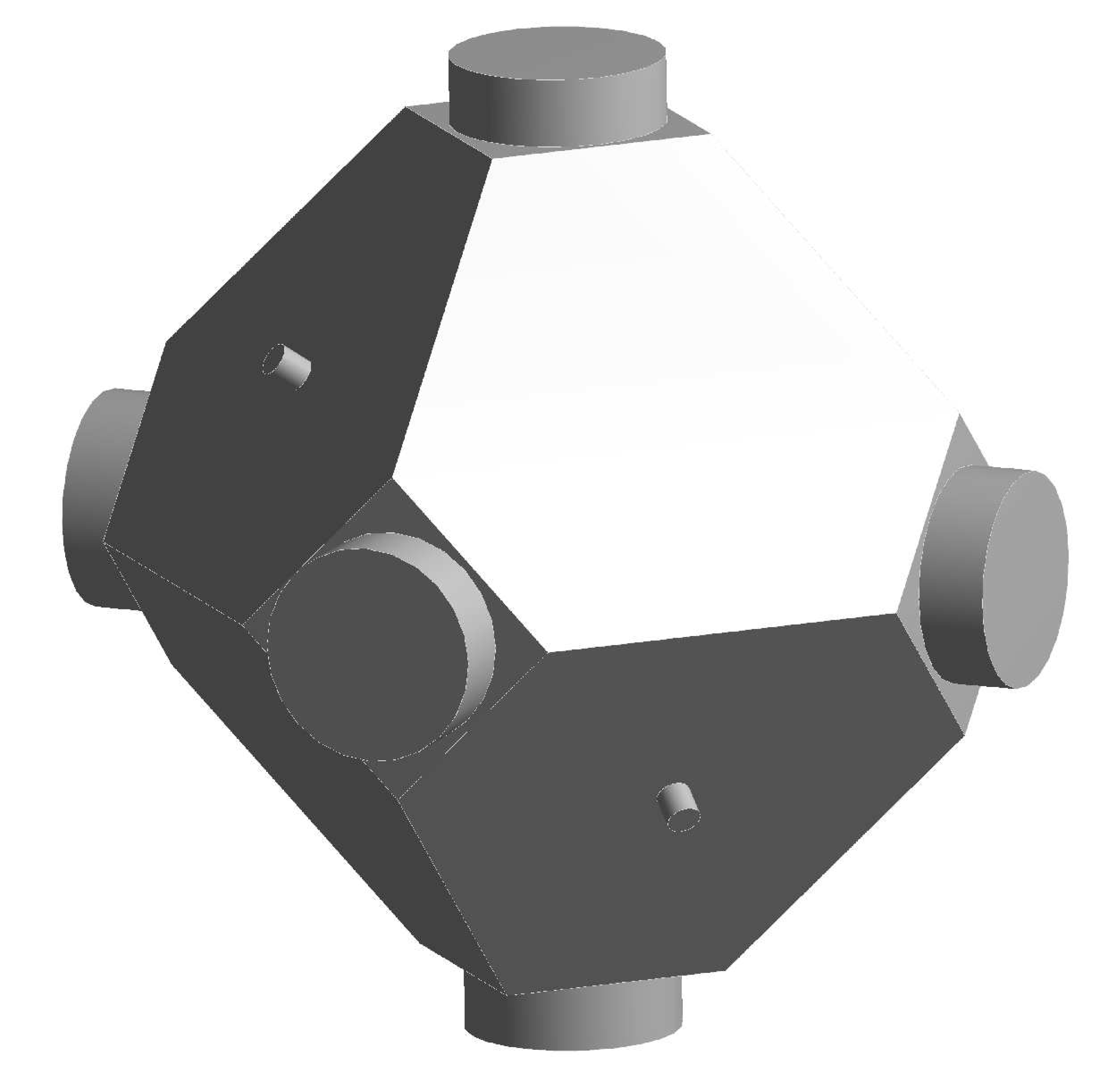}%
\bigskip{}
\par\end{centering}
\centering{}%
\includegraphics[width=0.3\textwidth]{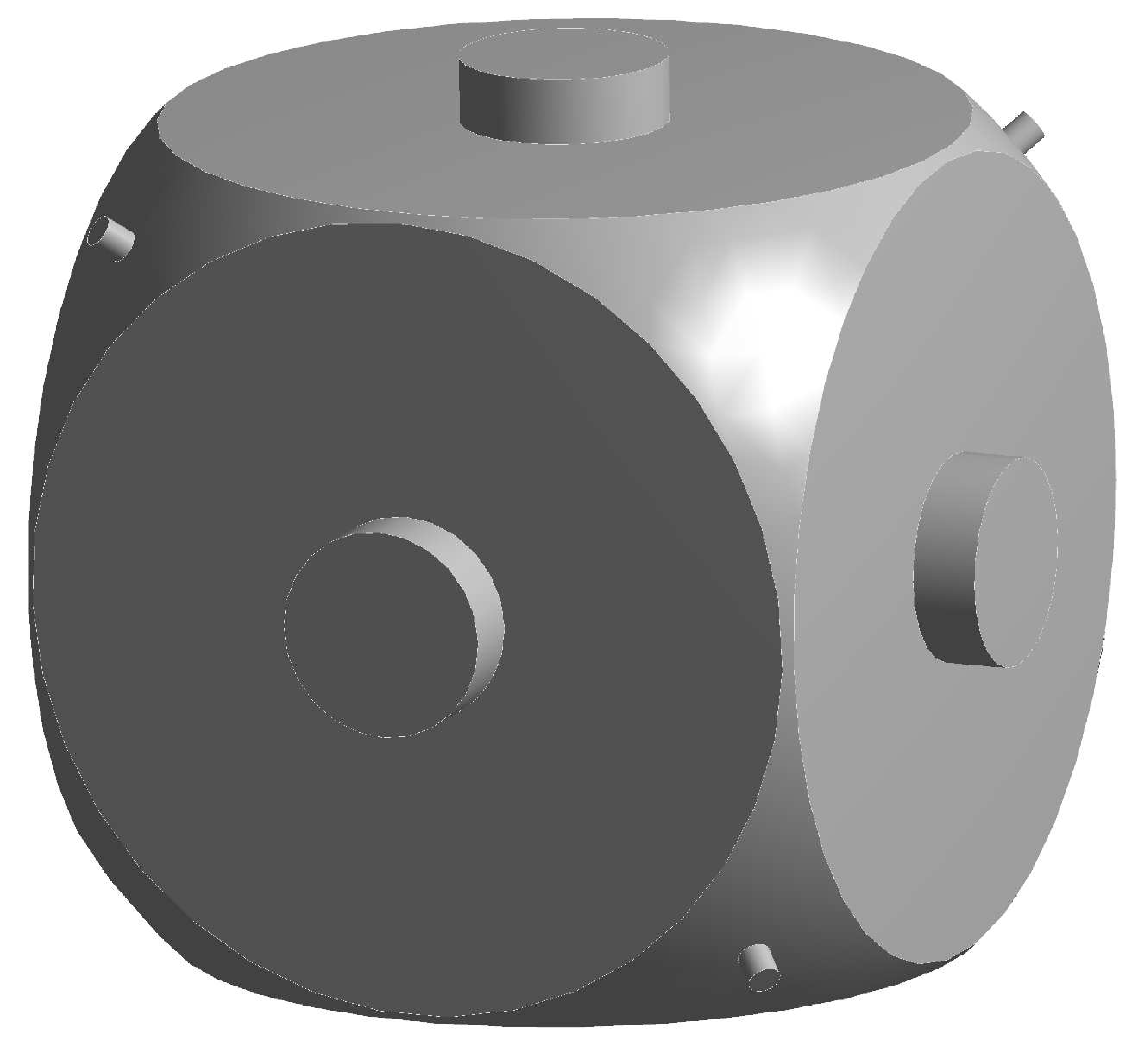}%
\quad{}%
\includegraphics[width=0.3\textwidth]{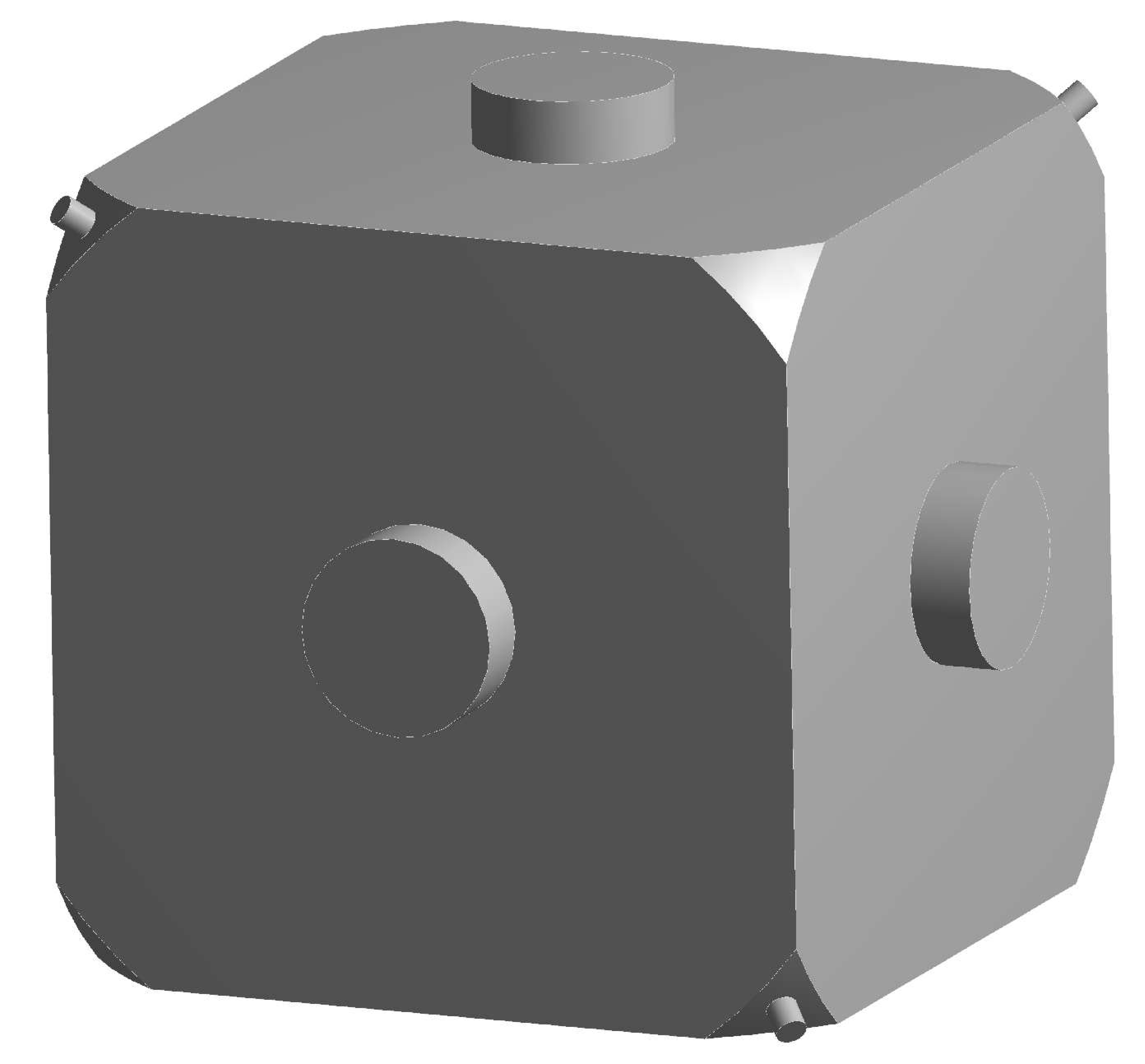}%
\quad{}%
\includegraphics[width=0.3\textwidth]{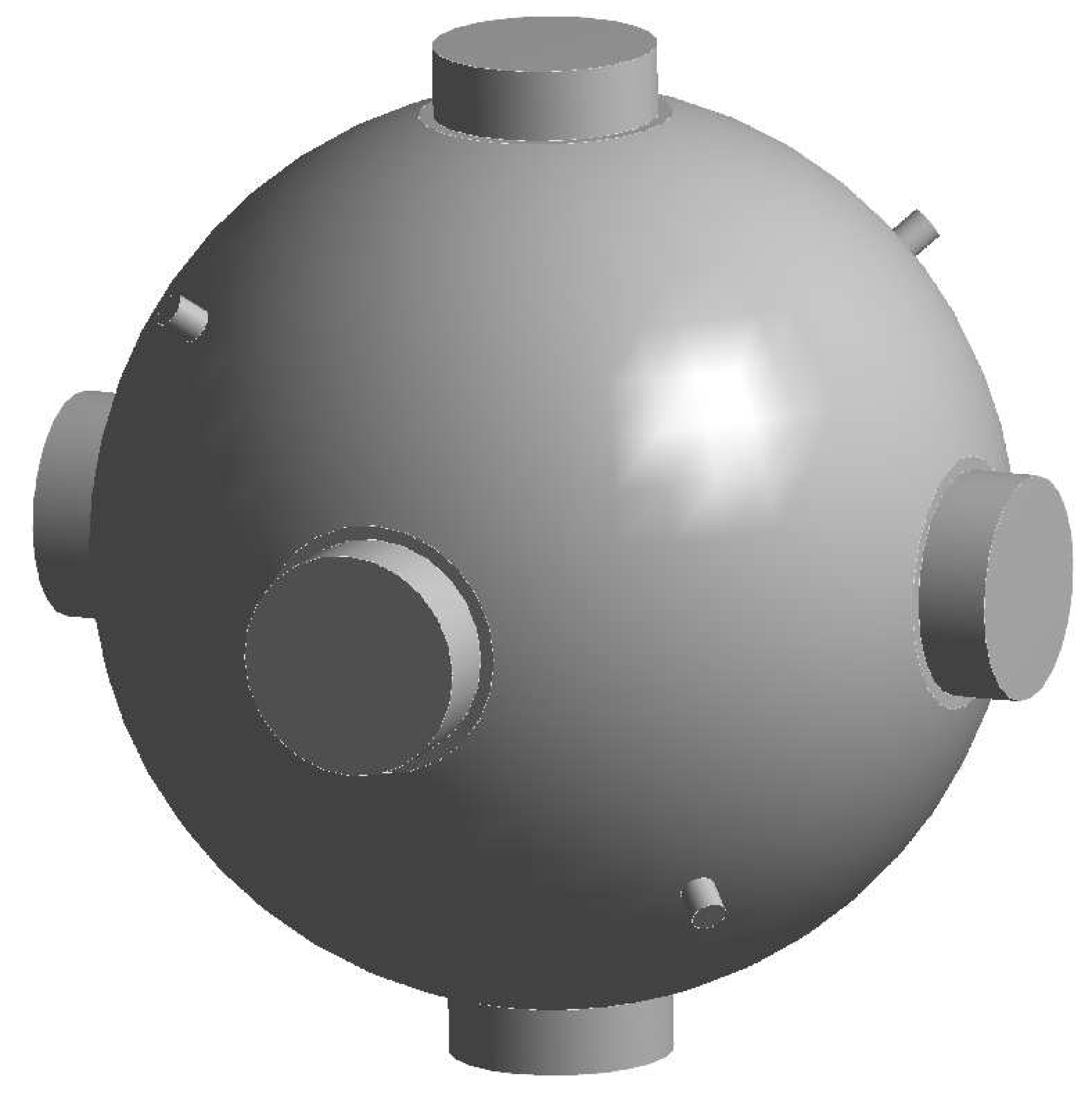}%
\caption{\label{fig:Different-Resonator-Shapes} Different simulated ULE resonator
shapes, with the distance between the mirrors for all three cavities
fixed at $50$~mm. Top row: Left, great rhomb-cube-octahedron with
the edge length of polygons being $13.1$~mm and the two extremes
with hexagons pulled out (middle) or in (right) by $7$~mm. Middle
row: Left, rhomb-cube-octahedron with the edge length of polygons
being $20.7$~mm and the two extremes with triangles pulled out (middle)
by $3.3$~mm or in (right) by $6.4$~mm. Note that the geometries
of the top row, right and middle row, right are identical. Bottom
row: Left, spherical cube resonator with circular faces of $49$~mm
diameter, and the two extremes with a cut depth of $3.3$~mm (middle)
and $17.3$~mm (right).}
\end{figure}

\begin{figure}
\begin{centering}
\begin{center}
\includegraphics[width=0.9\textwidth]{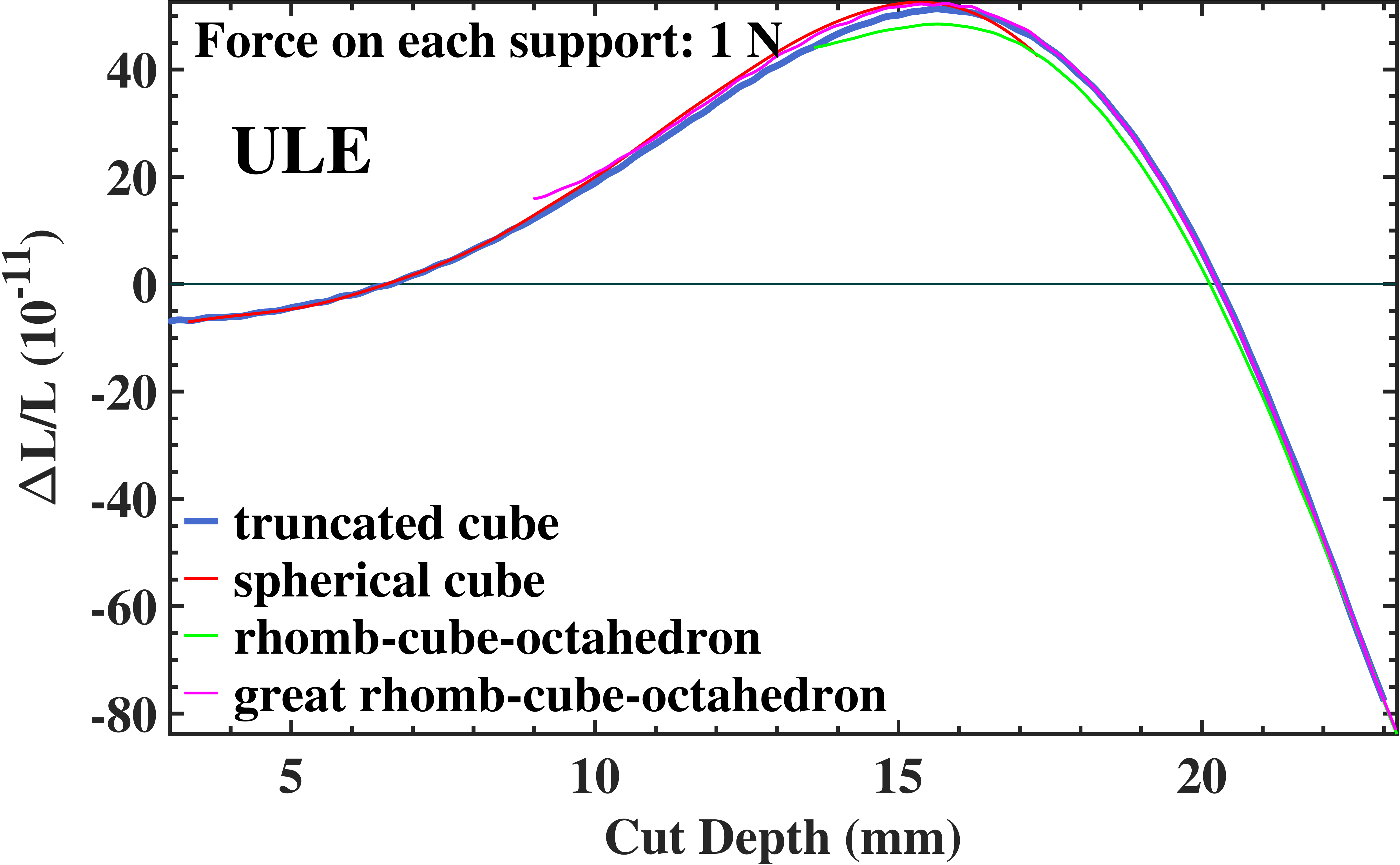}
\par\end{center}%
\par\end{centering}
\caption{\label{fig:Different-Resonator-Shapes-Results} Sensitivity $\Delta L_{x}(F_{c})/L$
as function of cut depth for the truncated cube, for the great rhomb-cube-octohedron,
for the rhomb-cube-octahedron, and for the spherically shaped cube. }
\end{figure}

To study the influence of the resonator's shape on the sensitivity
to the support forces we studied other cavity block shapes with 
octahedral symmetry, such as the great rhomb-cube-octahedron \cite{Cromwell2008},
the rhomb-cube-octahedron \cite{Cromwell2008} and the spherically
shaped cube (see Fig.~\ref{fig:Different-Resonator-Shapes}). All
these bodies can be produced from the cube-shaped resonator by cutting
out parts of the block in a symmetric way. The distance between the
mirrors was kept at $50$~mm for all shapes. Analogous to the truncated
cube geometry already discussed, the supports were set in tetrahedral
configuration and a force of $1$~N applied on each. The cut depth
was varied equally for all of them, within the limitations of the
respective geometry. The results are presented in Fig.~\ref{fig:Different-Resonator-Shapes-Results}.
The zero-sensitivity cut depths are the same for all geometries, with
the only difference being the corresponding limitations in cut depth.
These results suggest that the dominant features depend only on the
bulk properties of the material. 

%
\section{Cubic cavities made from conventional optical materials
\label{sec:Cubic-cavities-conventional-optical-materials}}
Nexcera, SiC and Zerodur are well-known materials used for manufacturing
optical components, in particular mirror substrates. Near room temperature
Nexcera and Zerodur exhibit a zero thermal expansion coefficient $\alpha$. In contrast, for the material SiC it is finite, but comparatively small \cite{RohmAndHaas2008}.
 Tab.~\ref{tab:Isotropic Material Properties} summarizes relevant
physical properties of the materials. It is well-known that a high
specific stiffness $E/\rho$ leads to low cavity acceleration sensitivities.
This is the reason for including SiC in the present analysis. Also
listed in the table is the Poisson ratio $\nu$, defined as the negative ratio of transverse strain to longitudinal strain. Thus, Poisson's ratio is responsible
for the redistribution of strain in the directions normal to the direction
of the applied force.
%
%
\subsection{Support force sensitivity}
\label{sec:Support-Force-Sensitivity}

\begin{figure}
\begin{centering}
\begin{center}
\includegraphics[width=0.45\textwidth]{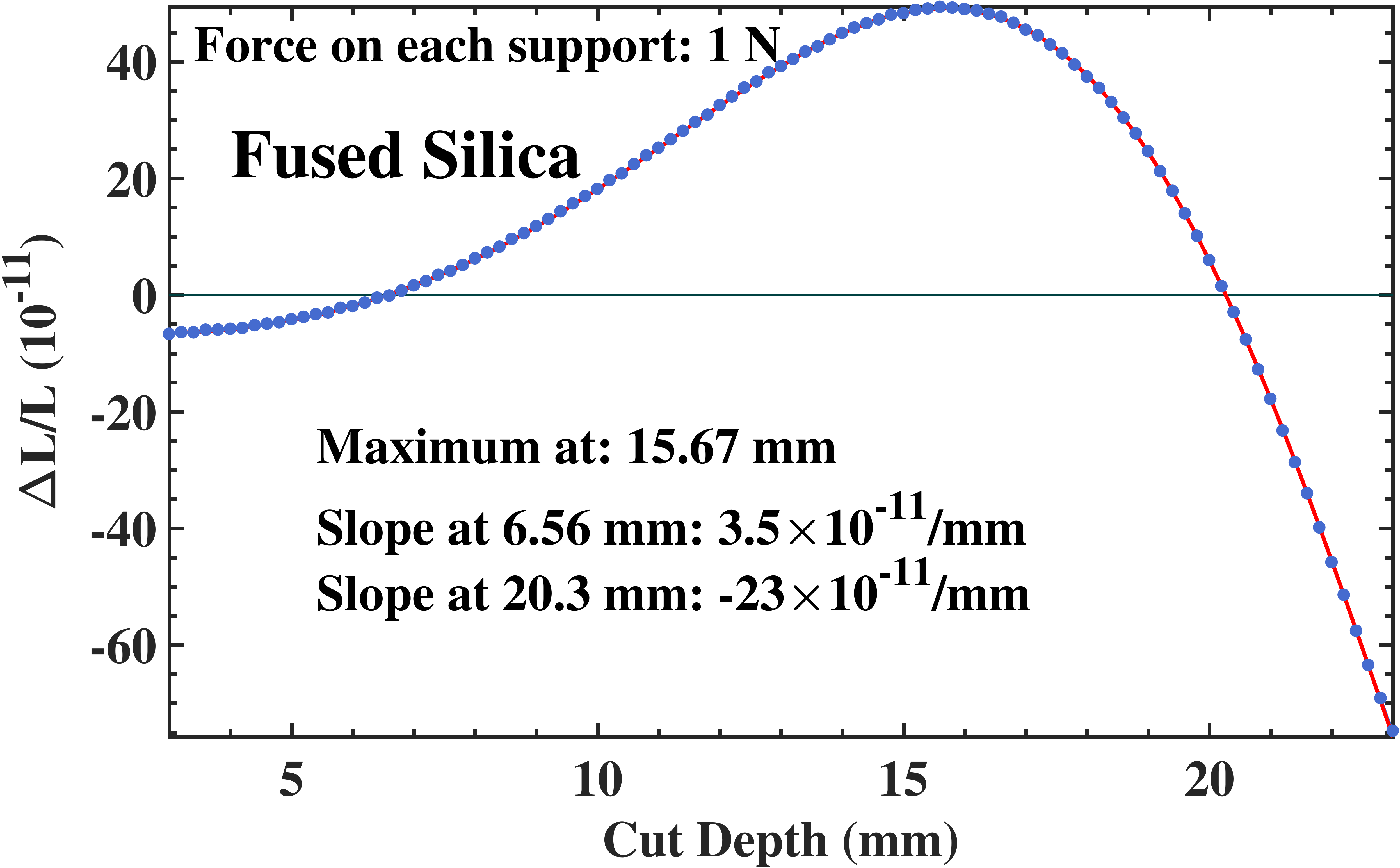}
\quad{}%
\includegraphics[width=0.45\textwidth]{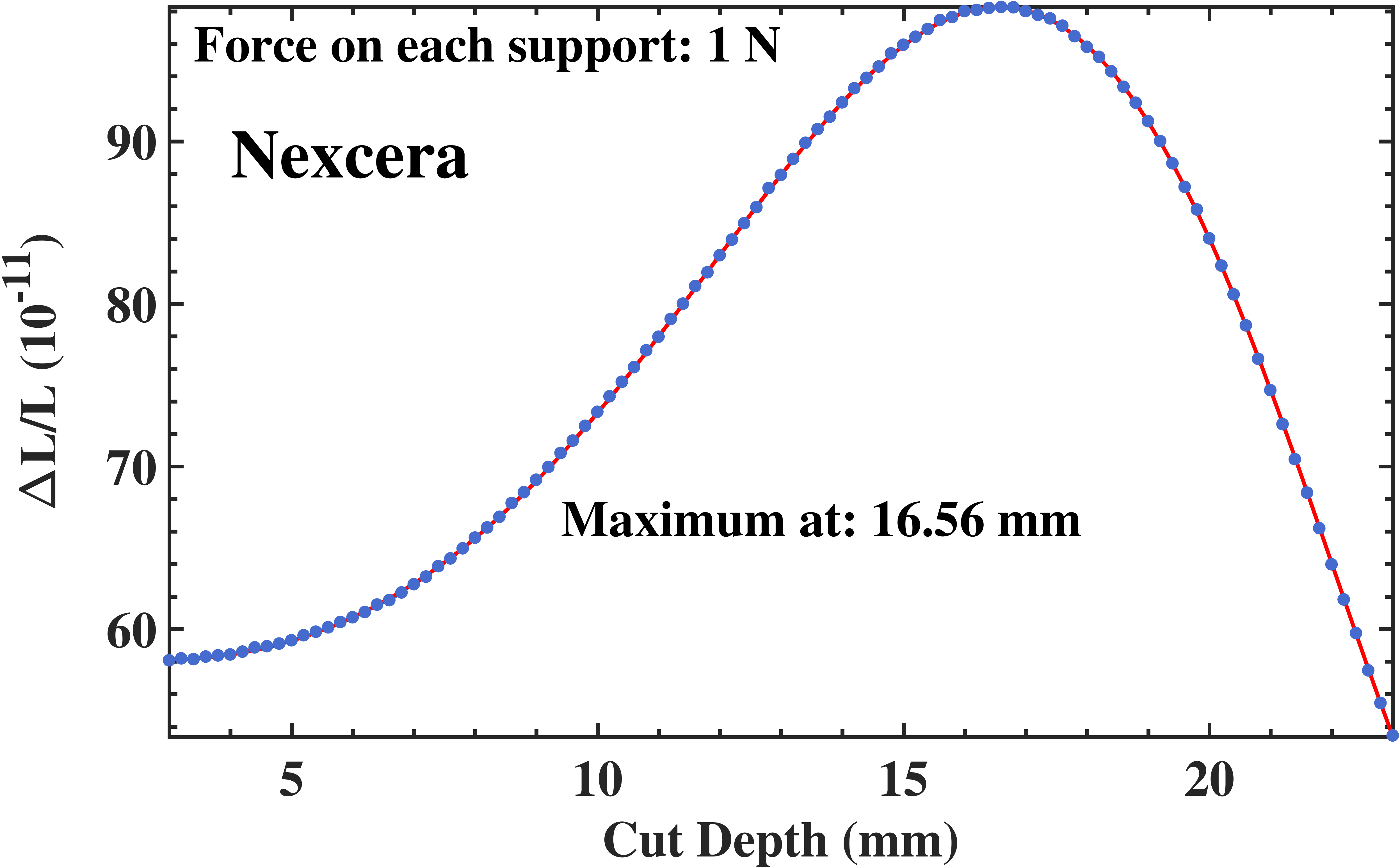}%
\par\end{center}%
\end{centering}
\medskip{}
\begin{centering}
\begin{center}
\includegraphics[width=0.45\textwidth]{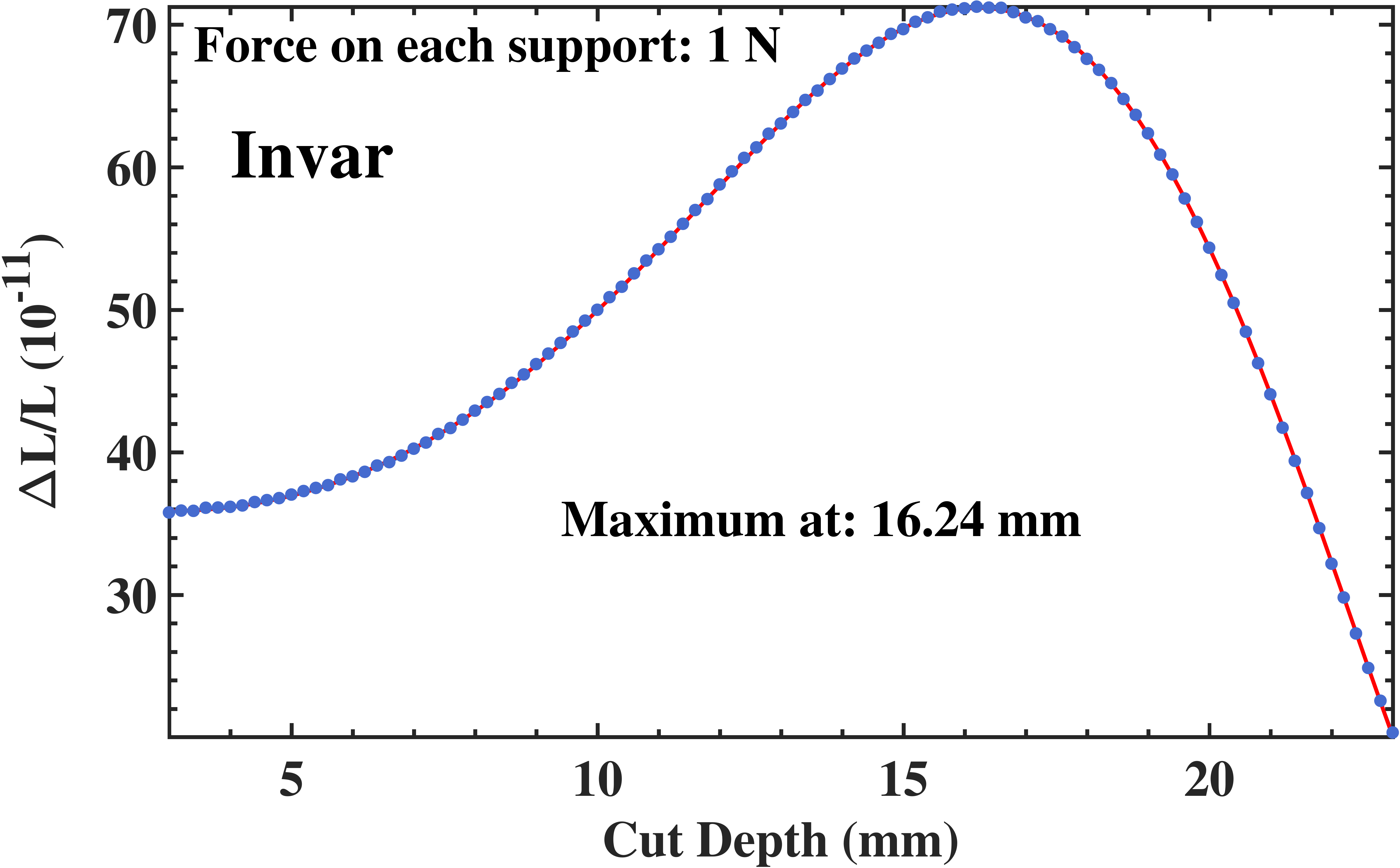}
\quad{}%
\includegraphics[width=0.45\textwidth]{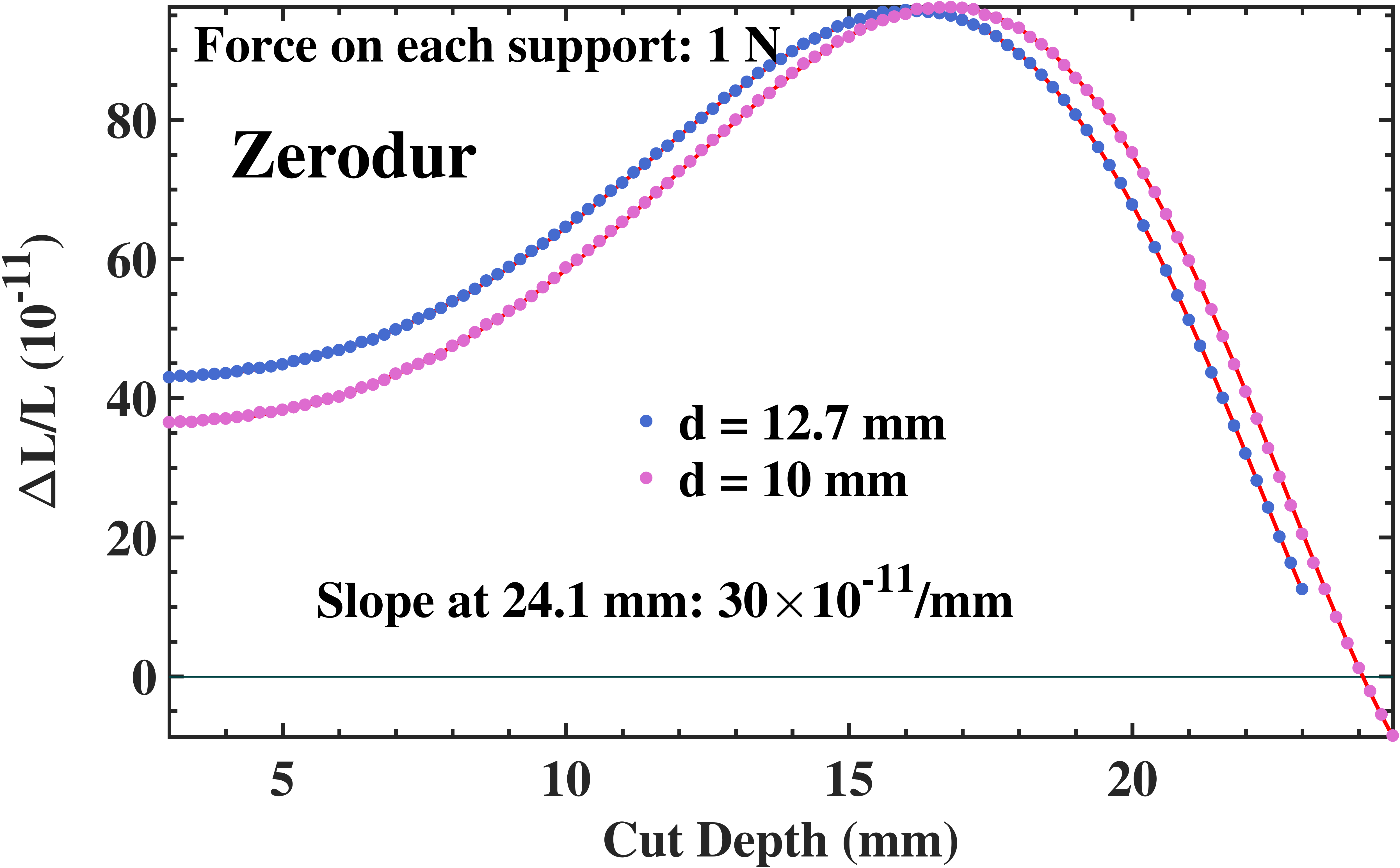}%
\par\end{center}%
\end{centering}
\medskip{}
\begin{centering}
\begin{center}
\includegraphics[width=0.45\textwidth]{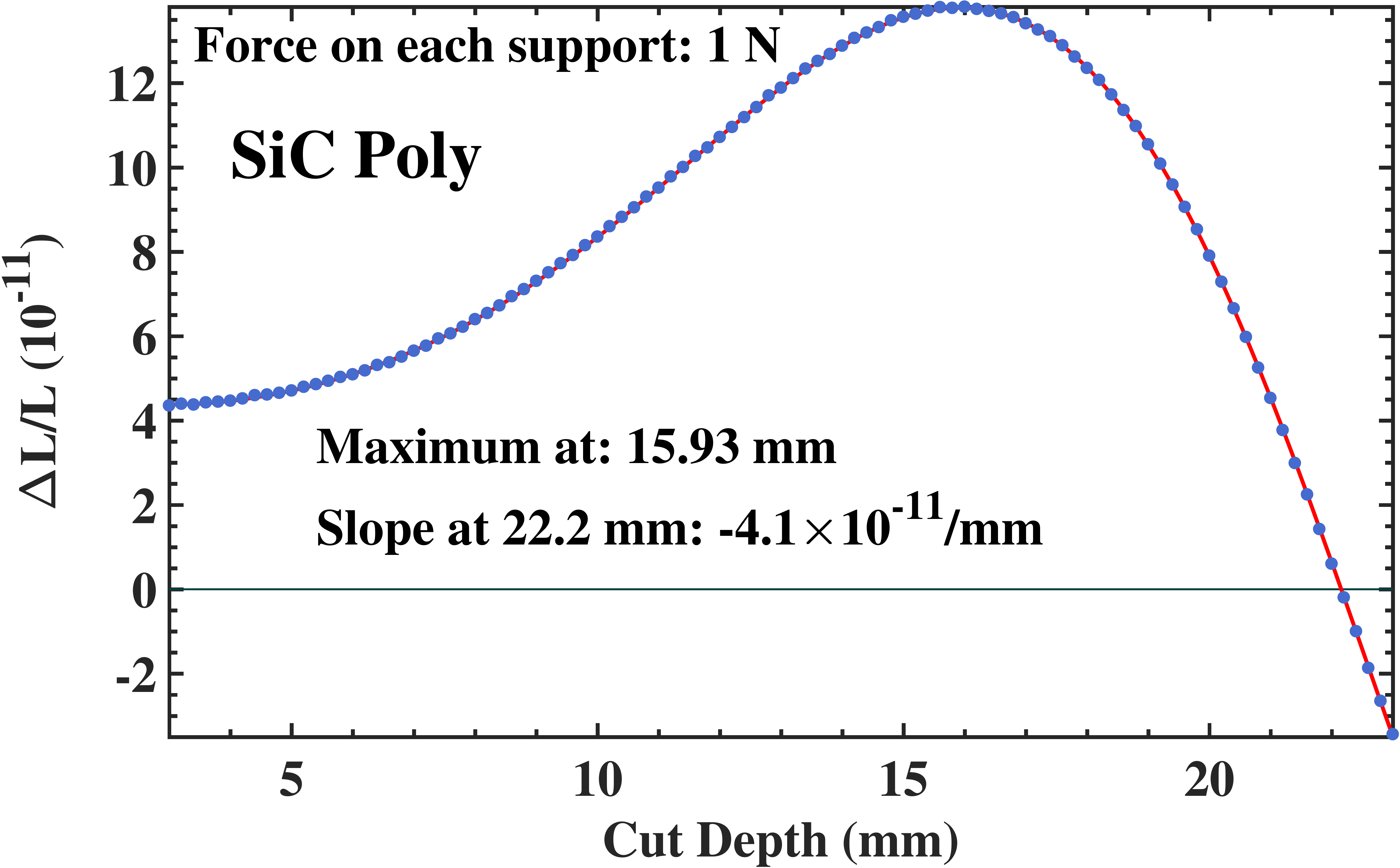}
\quad{}%
\includegraphics[width=0.45\textwidth]{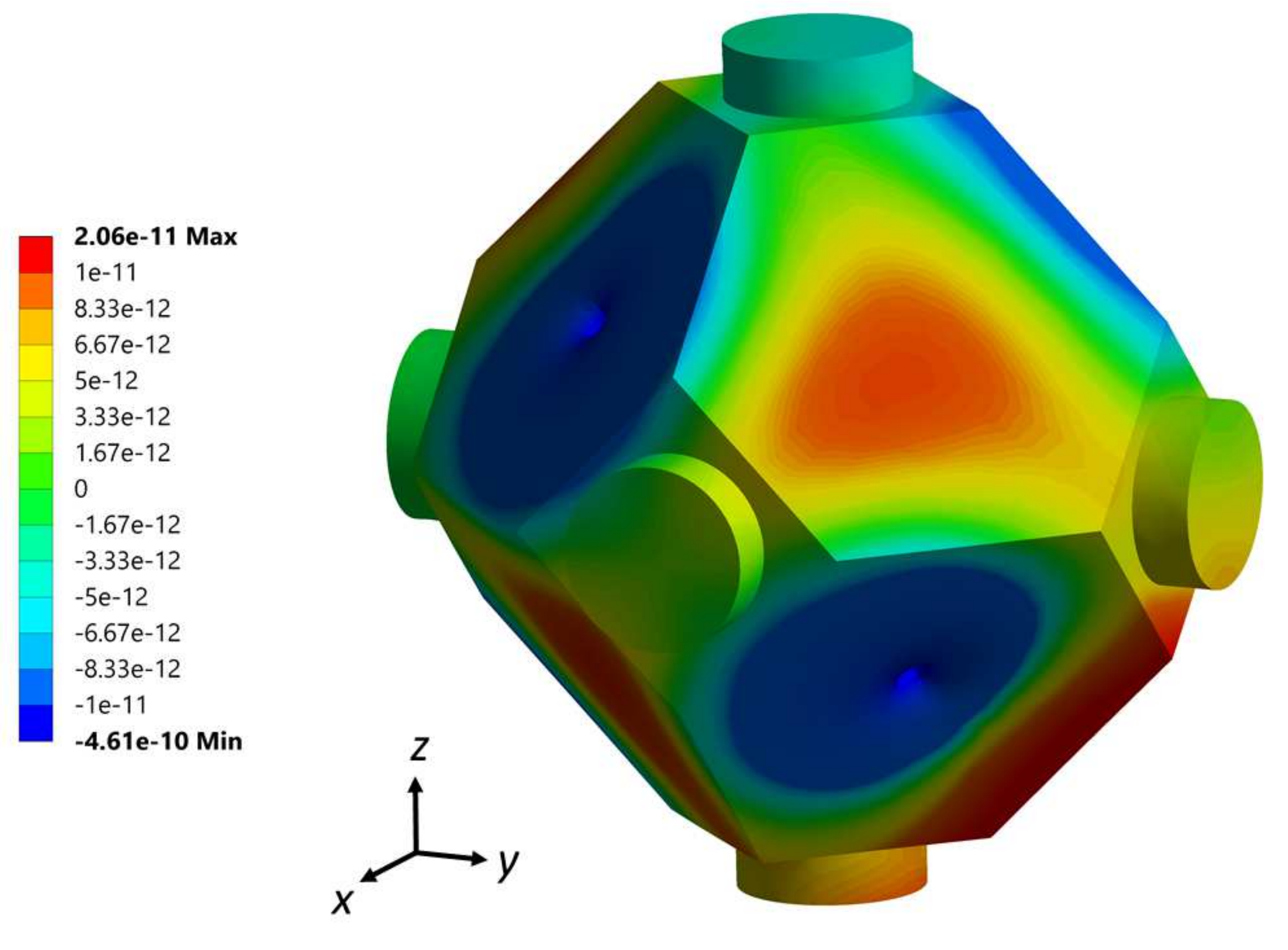}%
\par\end{center}%
\end{centering}
\caption{\label{fig:ULE vs Nexcera vs SiC} Fractional length change of a
resonator made of different materials and having different vertex
cut depth. A force of 1~N acts at each support. Bottom right: axial
$x$-deformation of the resonator made of $\beta$-SiC, for a cut
depth of $22.2$~mm. The scale is in m.}
\end{figure}

Our simulations show that for Nexcera, Invar and Zerodur, there does
not exist an optimal cut depth (see Fig.~\ref{fig:ULE vs Nexcera vs SiC}).
However, the sensitivity of Zerodur is low at high cut depth. To increase
the latter we reduced the diameter of the mirrors from half-inch to
$10$~mm. This results in a zero crossing of the fractional length
change at a cut depth of $24.1$~mm, with a slope of $30\times10^{-11}$/mm. 
  ULE is a modification of fused silica. Therefore, the fused silica
resonator has zero sensitivity to $F_{c}$ at essentially the same
cut depth as the ULE resonator, but with a reduced slope because of
its slightly higher Young's modulus.

Polycrystalline $\beta$-SiC has zero sensitivity at a cut depth of
$22.2$~mm and a slope which is comparable with ULE at 6.6~mm, due
to the much larger Young's modulus. Fig.~\ref{fig:ULE vs Nexcera vs SiC},
bottom right shows the deformation of the $\beta$-SiC block having
the zero-sensitivity geometry. 

The overall ("peak-peak") variation of support force sensitivity over the complete range of cut depths is inversely proportional to the Young's modulus. It is largest for
ULE ($130\times10^{-11}$) and lowest for $\beta$-SiC ($17\times10^{-11}$).
Fractional length change at a $6.6$~mm cut depth is the highest
for Nexcera, followed in decreasing order by Zerodur, Invar, polycrystalline $\beta$-SiC and the ULE (for which it is zero). With exception of Zerodur and
Invar, that switch places, this sequence corresponds with the Poisson's
ratio value. 
\begin{table}
\begin{centering}
\begin{tabular}{|c|c|c|c|c|c|}
\hline 
Material & $\rho$ {[}g/$\mathrm{cm^{3}}${]} & $E$ {[}GPa{]} & $\nu$ & $E/\rho$ {[}${\rm MJ/kg}${]} & $\alpha$  {[}$10^{-6} \cdot \rm{K}^{-1}${]}\tabularnewline
\hline 
\hline 
ULE \cite{Corning2008} & 2.21 & 67.6 & 0.17 & 30.59 & 0$\pm$0.03\tabularnewline
\hline 
Nexcera \cite{KrosakiHarima2017} & 2.58 & 140 & 0.31 & 54.26 & $<$0.05\tabularnewline
\hline 
Zerodur \cite{Schott2013} & 2.53 & 90.3 & 0.24 & 35.69 & 0$\pm$0.1\tabularnewline
\hline 
$\beta$-SiC, polycrystalline \cite{RohmAndHaas2008} & 3.21 & 466 & 0.21 & 145.2 & 2.2\tabularnewline
\hline 
Fused Silica \cite{Schott2013} & 2.2 & 70.2 & 0.17 & 31.9 & 0.5\tabularnewline
\hline 
Invar \cite{IVPV2017} & 8.05 & 141 & 0.259 & 17.52 & 1.0\tabularnewline
\hline 
\end{tabular}
\par\end{centering}
\caption{\label{tab:Isotropic Material Properties}Comparison of some mechanical
and thermomechanical properties of the considered isotropic materials. Invar, an alloy
with low thermal expansion coefficient at room temperature, is included
for reference. }
\end{table}

\begin{figure}
\begin{centering}
\begin{center}
\includegraphics[width=0.45\textwidth]{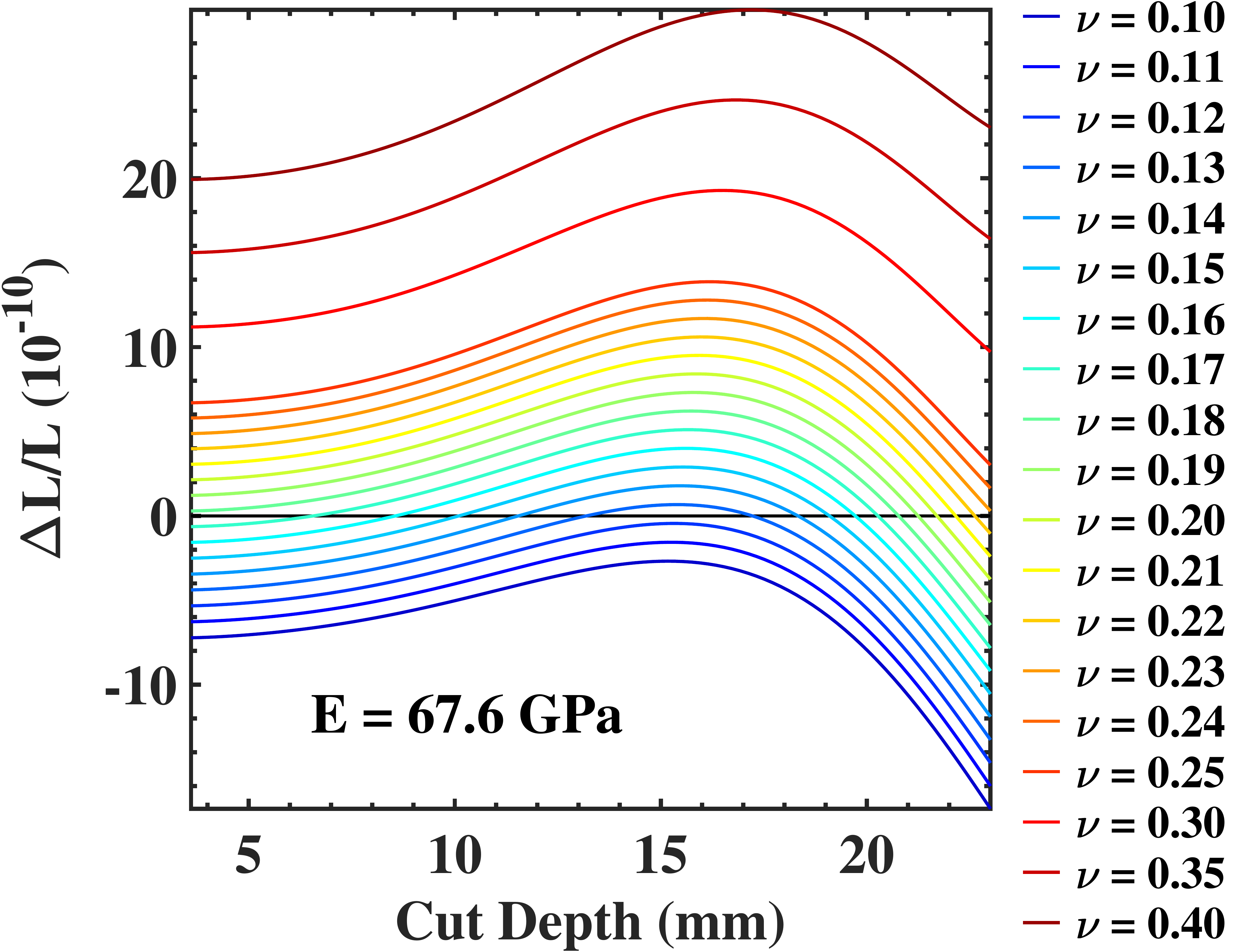}%
\quad{}%
\includegraphics[width=0.45\textwidth]{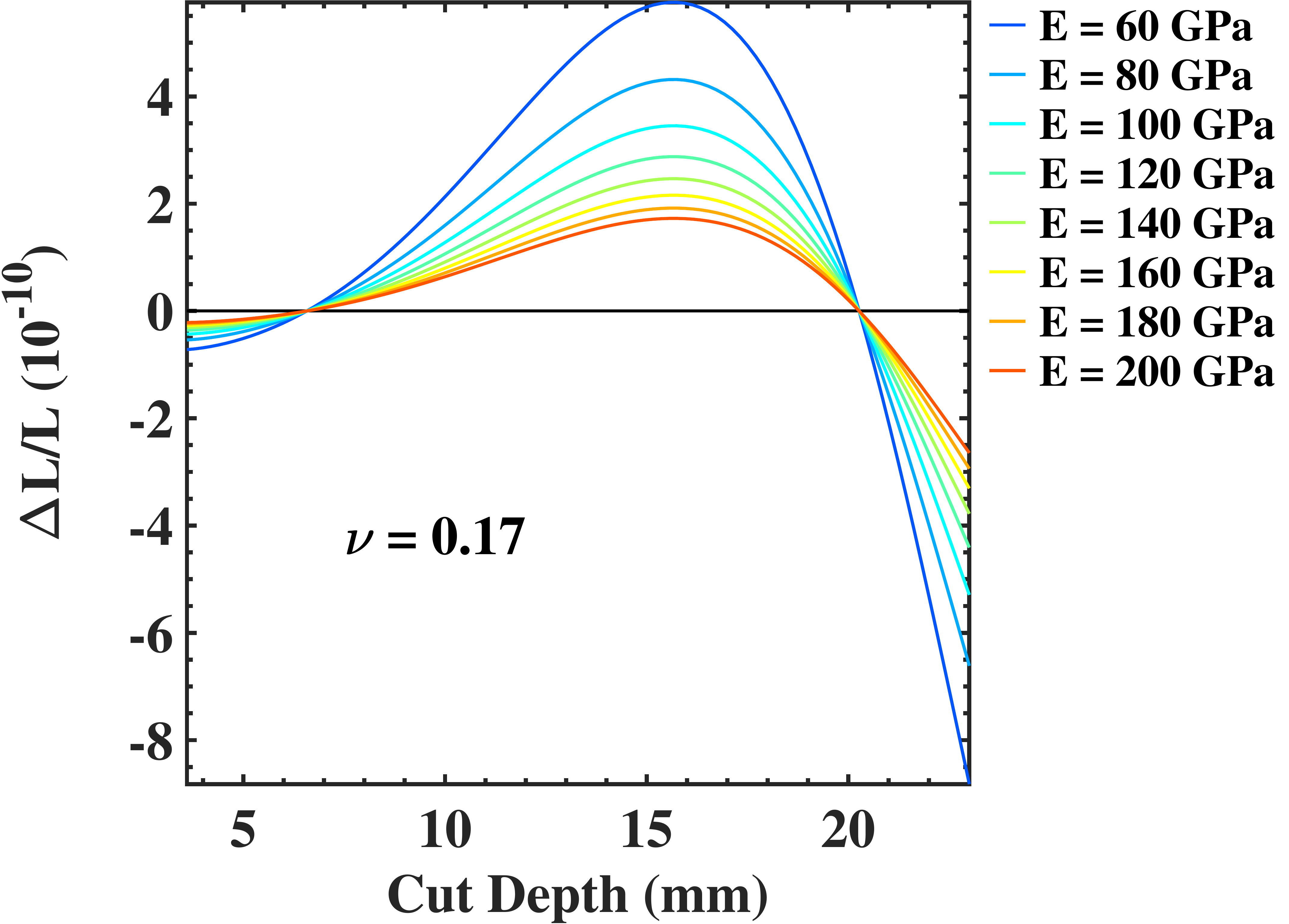}%
\par\end{center}%
\end{centering}
\medskip{}
\begin{centering}
\begin{center}
\includegraphics[width=0.45\textwidth]{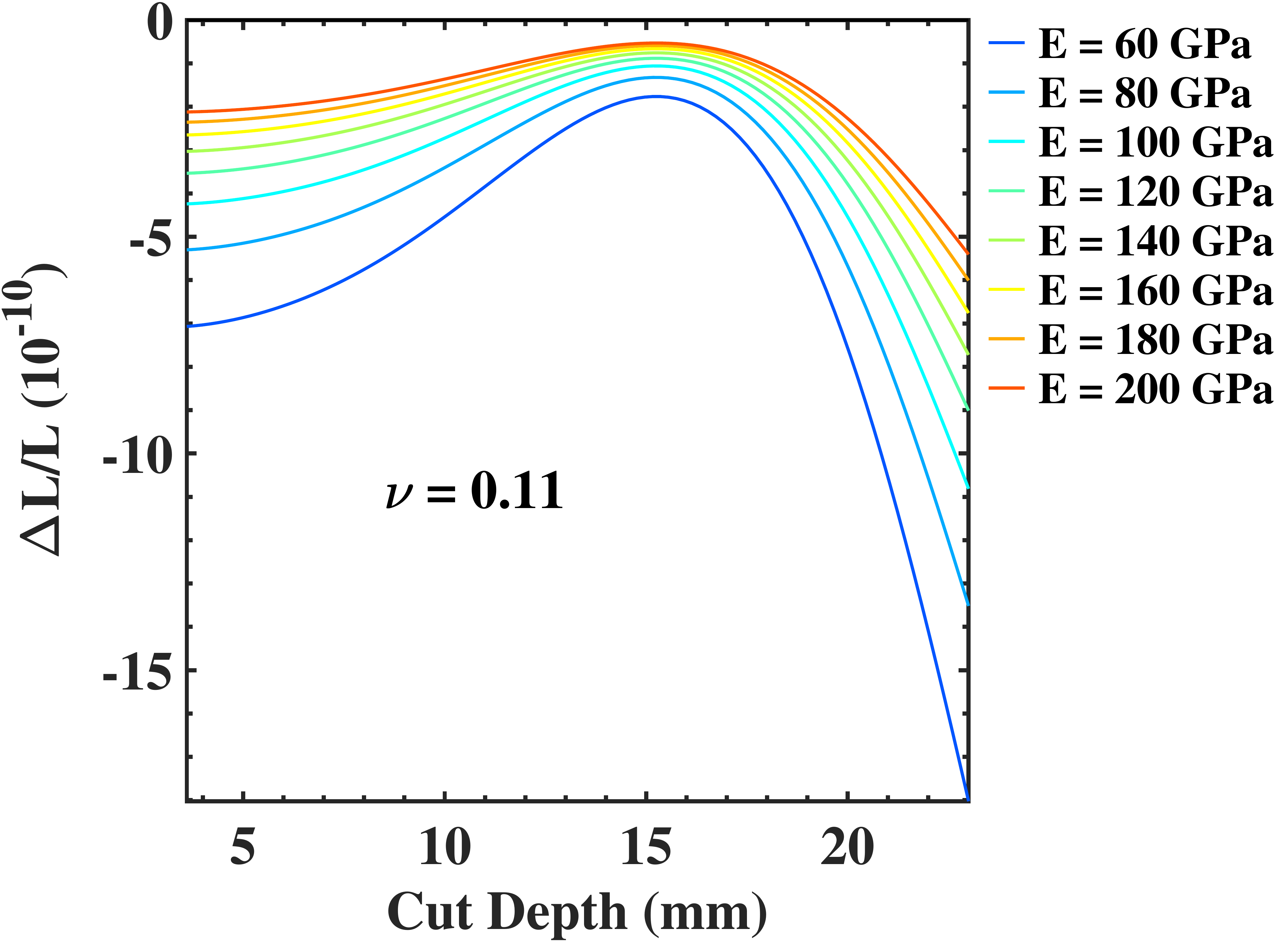}%
\quad{}%
\includegraphics[width=0.45\textwidth]{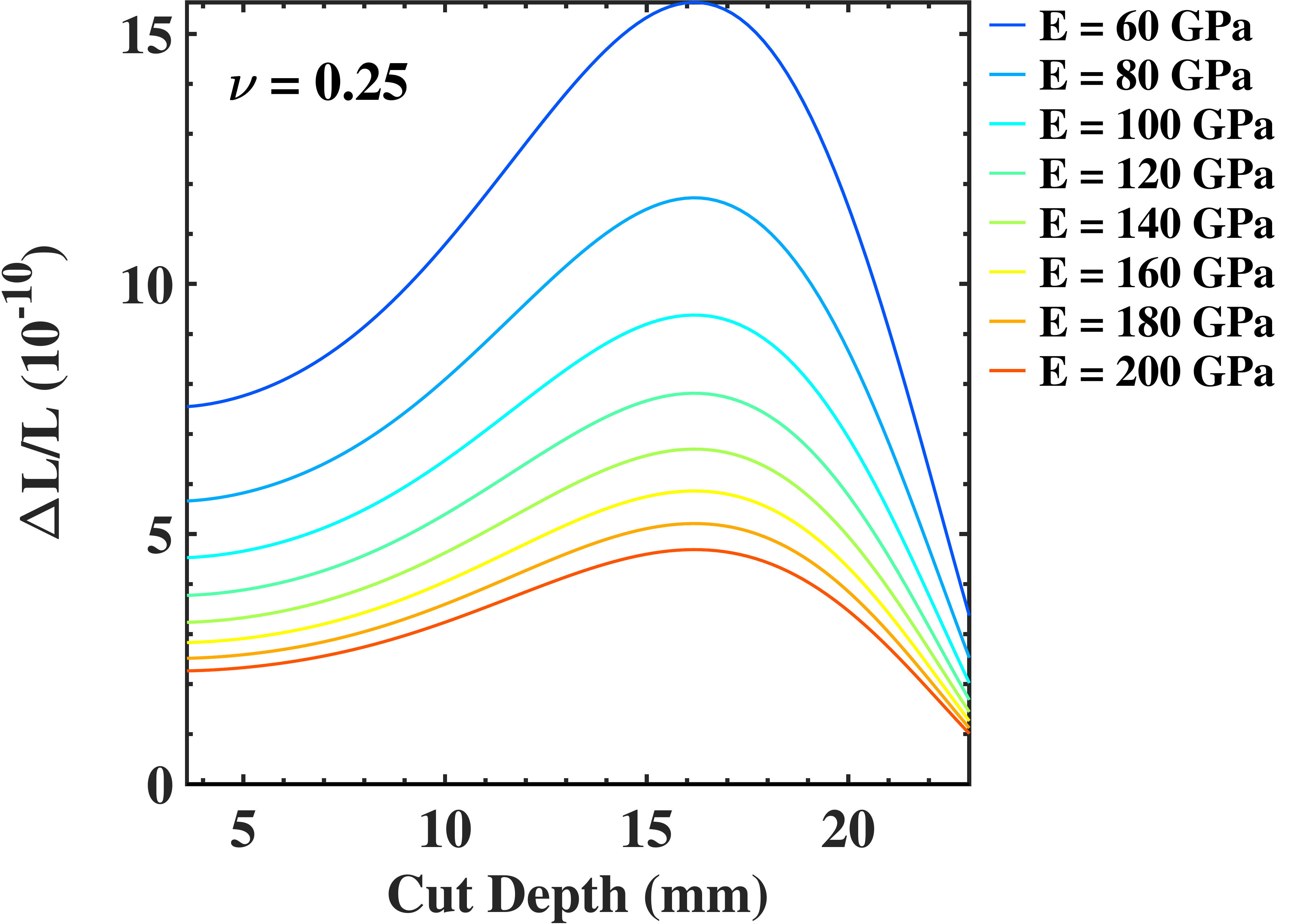}%
\par\end{center}%
\end{centering}
\caption{\label{fig:Cubic Cavity Poisson and Young-1} Top: Left, dependence
of the fractional length change of the resonator on vertex cut depth
and on Poisson's ratio $\nu$, with the Young's modulus held constant.
Right: dependence on Young's modulus with a Poisson's ratio $\nu$
held constant at $0.17$. Bottom: Dependence on Young's modulus with
a Poisson's ratio $\nu$ held constant at $0.11$ (left) and $0.25$
(right).}
\end{figure}

\begin{figure}
\begin{centering}
\begin{center}
\includegraphics[width=0.48\textwidth]{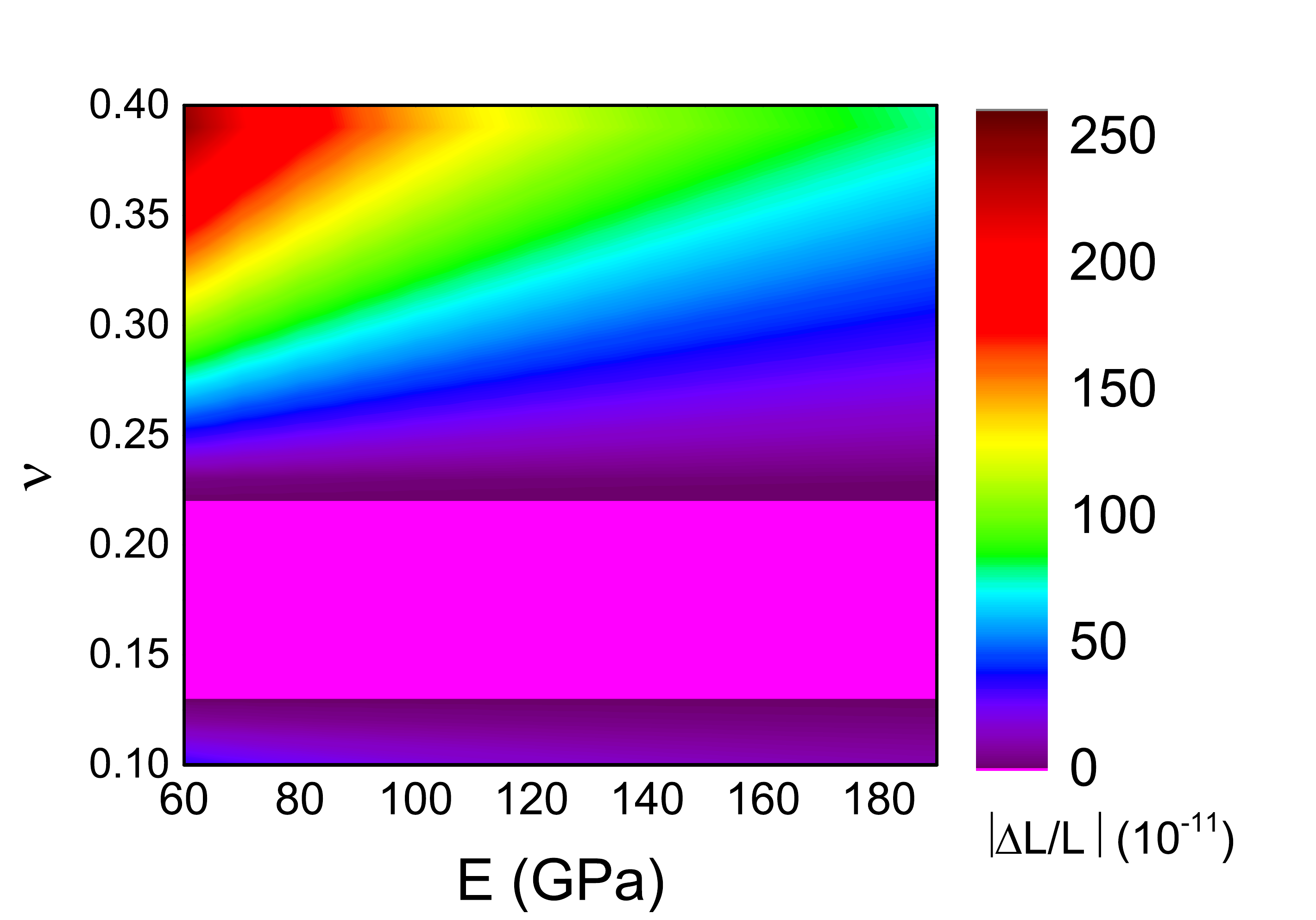}
\quad{}%
\includegraphics[width=0.48\textwidth]{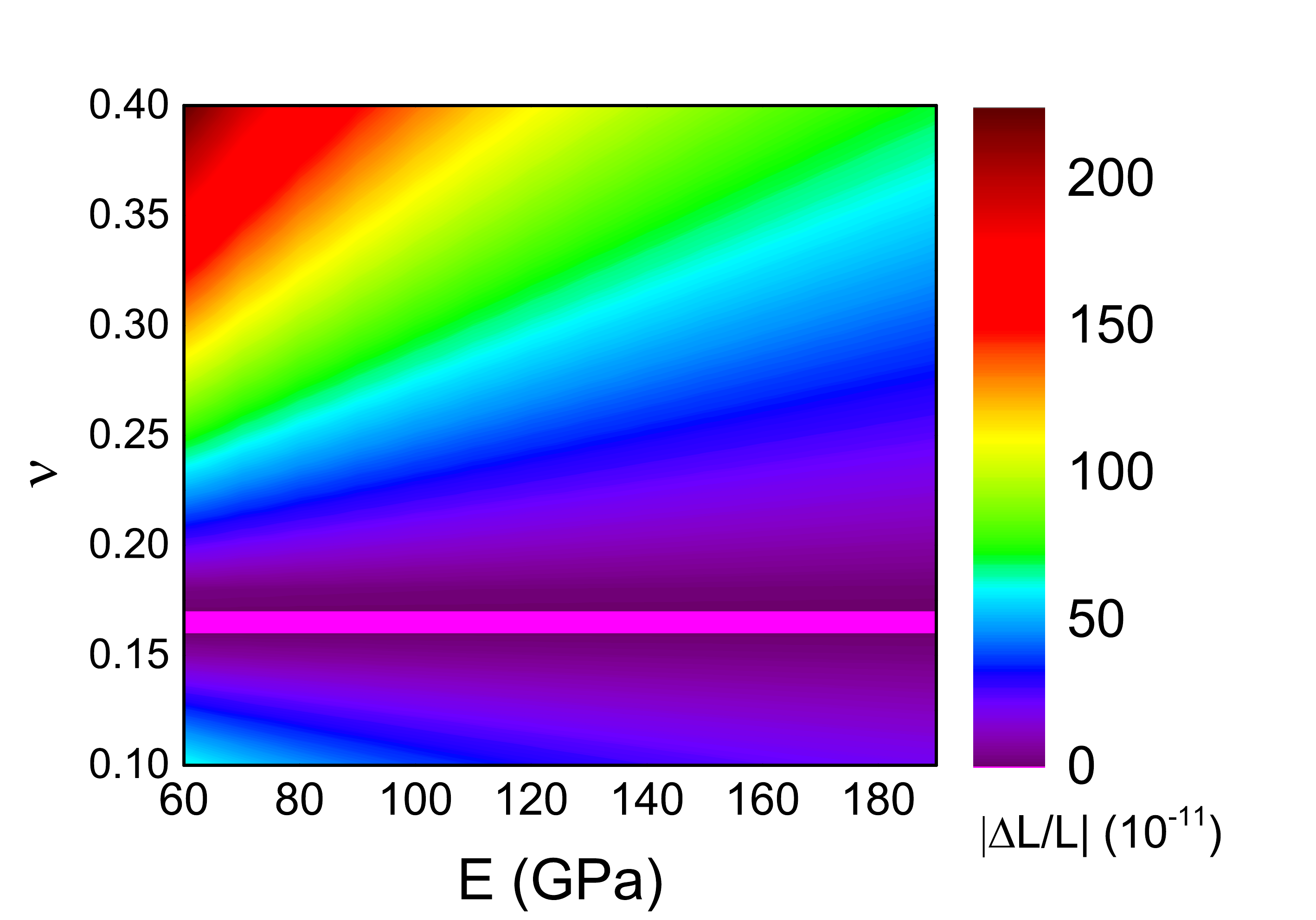}%
\par\end{center}%
\end{centering}
\caption{\label{fig:2-DIM-Map-Poisson-Young}Minimum absolute fractional length
change ${\rm Min}|\Delta L(F_{c})/L|$ among all cut depths between
$3$~mm and $23$~mm (left) and $3$~mm and $9$~mm (right), as
a function of the Young's modulus $E$ and the Poisson's ratio $\nu$.
The region in magenta color indicates those values of $E$ and $\nu$
for which cut depths exist that exhibit zero sensitivity to the support
forces $F_{c}$. }
\end{figure}

To confirm this observation, we assumed a hypothetical material and
varied either the Poisson ratio or the Young's modulus. The result
is shown in Fig.~\ref{fig:Cubic Cavity Poisson and Young-1}. We
find that both the Young's modulus and the Poisson's ratio are critical
parameters. As expected, the deformation and the Young's modulus are inversely proportional to each other. Thus, a low Young's modulus leads to high deformation of the
spacer and to high sensitivity to holding forces. On the other hand,
a high Young's modulus reduces the deformation and thus the sensitivity.
A small Poisson's ratio leads to an overall compression of a spacer,
whereas a high Poisson's ratio effectively redistributes the strain
and leads to an overall expansion of the spacer. Thus, comparing the
sensitivities of two different materials we can generally determine
the material with higher sensitivity by comparing solely their Poisson's
ratio values. If these materials have comparable values of Poisson's
ratio, the Young's modulus must also be taken into account. Substantial
difference in Young's modulus can change the sequence of the sensitivities
based on the Poisson's ratio. This is the case for Invar and Zerodur,
where Invar is the material with higher Poisson's ratio value (see
Tab.~\ref{tab:Isotropic Material Properties}) but lower sensitivity
(see Fig.~\ref{fig:ULE vs Nexcera vs SiC}). In order to have a cut
depth with zero sensitivity, the hypothetical material with the Young's
modulus between $60$~GPa and $200$~GPa must have the Poisson's
ratio within a ``magic'' range $0.13 < \nu < 0.23$ (Fig.~\ref{fig:2-DIM-Map-Poisson-Young},
left). This range is reduced to $0.13 < \nu < 0.18$ for the cut depths
between $3$~mm and $9$~mm (Fig.~\ref{fig:2-DIM-Map-Poisson-Young},
right). 
Note that the density of the material does not play a role in this
consideration. This leaves ULE, fused silica, and polycrystalline
$\beta$-SiC as the only suitable materials among the considered isotropic
ones. 

\subsection{Acceleration sensitivity}
\label{sec:Accelearion-Sensitivity}

The results presented in the previous sections were computed in the
absence of gravity and of acceleration acting at the resonator. Equal
forces $F_{c}$ acting at each of the four tetrahedrically oriented
supports on the block with octahedral symmetry, and pointing towards
the center of the block, preserve the symmetry of arrangement. 

When we include static gravity, which acts as a body force, i.e. on
each volume element of the resonator, the resulting deformation lowers
the resonator's symmetry. Depending on the magnitude of the deformation,
this could make necessary an adjustment of the zero-sensitivity cut
depths obtained in the previous sections. 

\begin{figure}
\begin{centering}
\begin{center}
\includegraphics[width=0.48\textwidth]{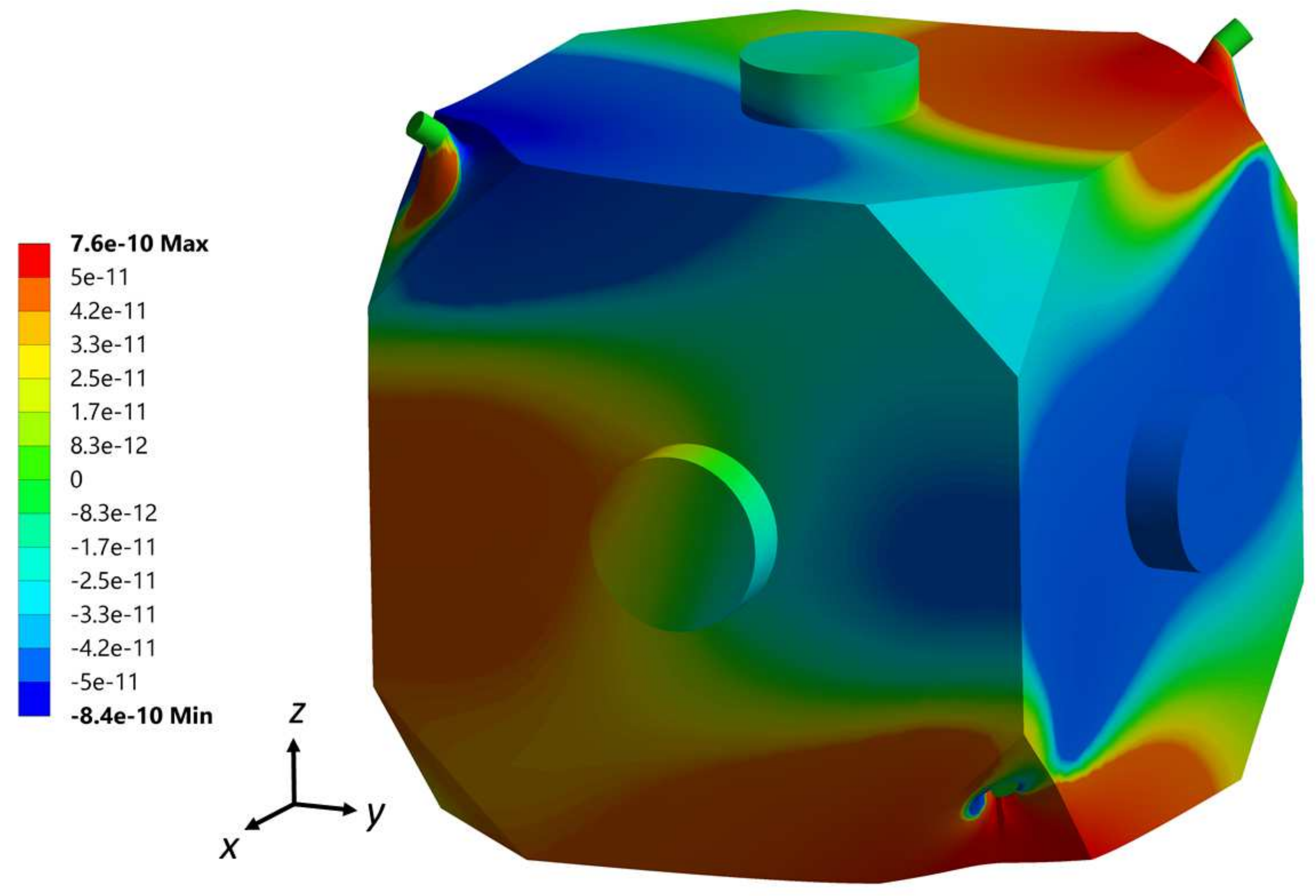}%
\quad{}%
\includegraphics[width=0.48\textwidth]{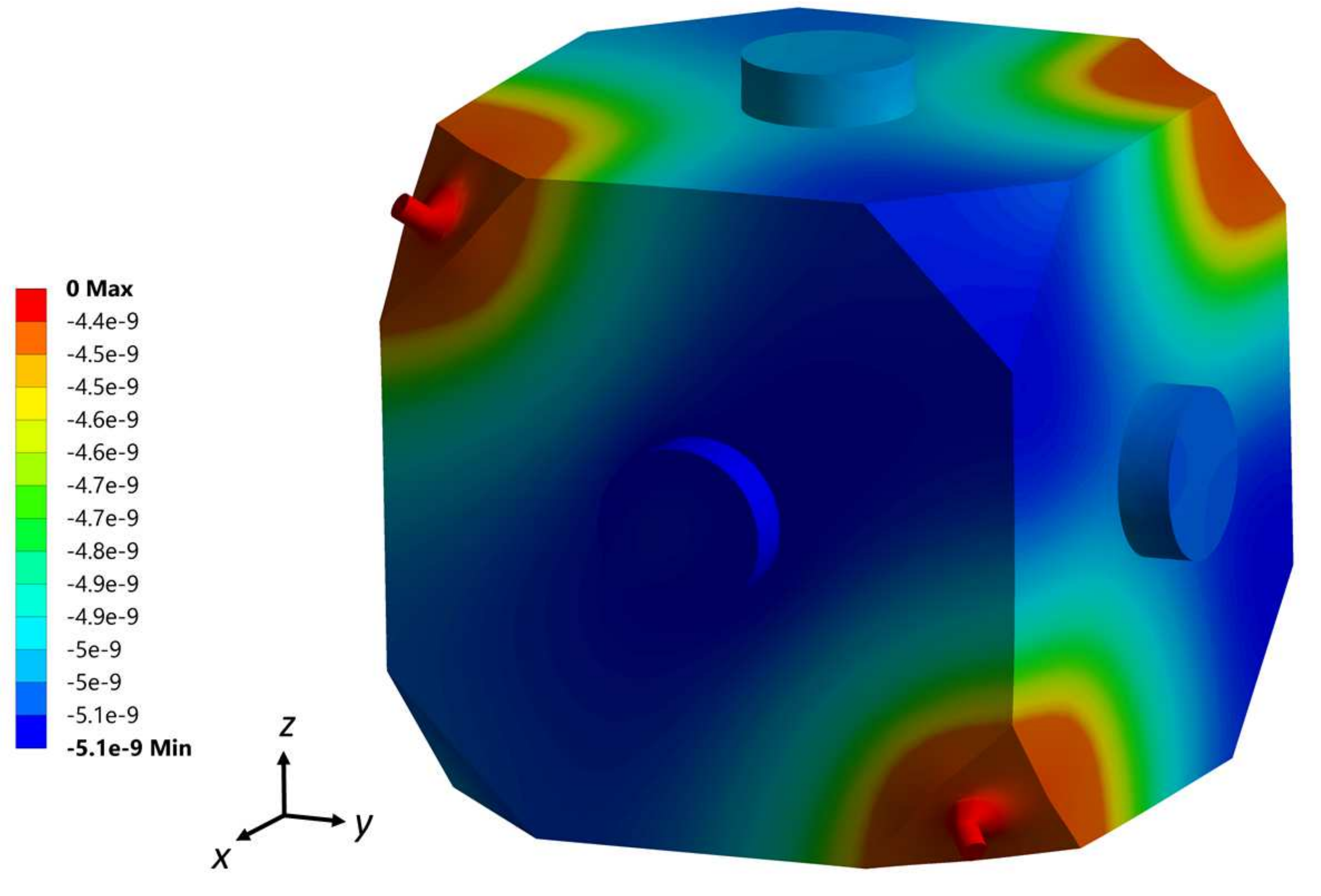}%
\par\end{center}%
\bigskip{}
\begin{center}
\includegraphics[width=0.9\textwidth]{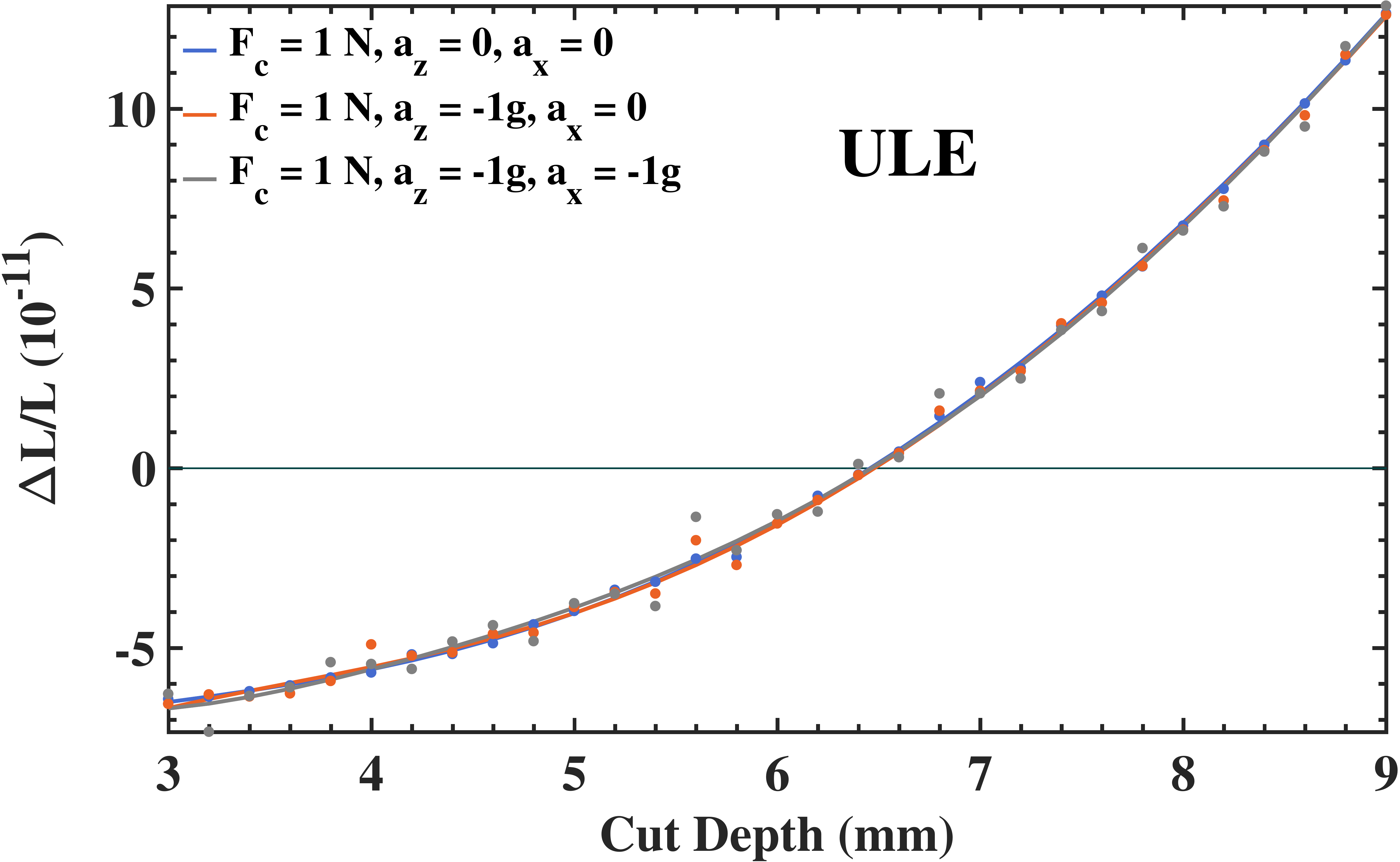}
\par\end{center}%
\par\end{centering}
\caption{\label{fig:ULE-Acceleration-Sensitivity} ULE resonator subjected
to an acceleration of magnitude $|a_{j}|=1\,{\rm g}$. Top left:
acceleration applied along $-z$-axis. The deformation along the $x$-axis
is displayed. Top right: acceleration applied along $-x$-axis. The
deformation along the $x$-axis is displayed. Bottom: Comparison of
(1) sensitivity due to a force of $1$~N on the supports, (2) with
additional application of $-1{\rm g}$ in the direction perpendicular
to the $x$-axis, (3) with additional 1${\rm g}$ accelerations acting
both perpendicular and along the axis of the $x$-cavity. Results presented in this diagram were calculated with an acceleration of $|a_{j}|=100\,{\rm g}$ and scaled to $|a_{j}|=1\,{\rm g}$ afterwards. }
\end{figure}

We computed the effects of acceleration on the cubic ULE resonator
having 6.6~mm cut depth. The resonator was fixed in space by the
supports, a force $F_{c}=1\,$N applied to each support,  and was
additionally subjected to a $1\,{\rm g}$ acceleration perpendicular
or parallel to the $x$-axis (see Fig.~\ref{fig:ULE-Acceleration-Sensitivity}).
For the acceleration along the $-z$-axis, the displacements of the
mirrors' center points along the $x$-axis are zero, see top left
panel in the figure. In contrast, the $1\,{\rm g}$-acceleration along
the $x$-axis generates displacements of both mirrors on the nm scale
($\sim5.1\times10^{-9}$~m), but equal ones, thus leaving the distance
between the mirrors unchanged (see top right panel in the figure).
The cancellation  represents the numerical proof of the concept of
Webster and Gill, which is based on the octahedral symmetry. Further
simulations showed that the above displacements decrease with increasing
Young's modulus, as expected.

We have performed similar simulations for various cut depths. The
results are summarized in Fig.~\ref{fig:ULE-Acceleration-Sensitivity}.
It can be seen that for the different sets of applied forces, the
results are nearly equal. In particular, the optimal cut depth is
not modified in the presence of gravity. We conclude that accelerations
on the order of $1\,{\rm g}$ do not deform the resonator strongly
enough to lower its symmetry so as to destroy the force insensitivity
at the optimal cut depth determined assuming zero acceleration. We
obtain the same results when the acceleration is increased by a factor
100.

\section{Anisotropic materials}
\label{Anisotropic-Materials}

Additional candidate materials for a force-insensitive cubic cavity
might be found among anisotropic materials, where $E$ and $\nu$
depend on the crystallographic direction. Silicon and sapphire are
two crystalline materials of this kind, and they have already been
used successfully for cryogenic optical resonators.

For an anisotropic material the relation between the applied stress
$\sigma$ and the resulting strain $\varepsilon$ is \cite{Nye1964,Wortman1965}:

\begin{equation}
\sigma=C\varepsilon, 
\label{eq:1}
\end{equation}

where  $\sigma$ and $\varepsilon$ are second-rank tensors with
9 elements each and $C$ is the fourth-rank stiffness tensor with
81 elements. For crystals with cubic symmetry (e.g., silicon) both $\sigma$ and $\varepsilon$ tensors contain only six independent elements.  Using a simplified Voigt notation the tensor $C$ can be reduced to the 6$\times$6 symmetric matrix with only three independent elements in the Cartesian coordinate system spanned by the $\vec{e}_{1}=(1,0,0)$, $\vec{e}_{2}=(0,1,0)$, and $\vec{e}_{3}=(0,0,1)$ unit vectors pointing along {[}100{]}, {[}010{]}, and {[}001{]} crystallographic directions. The three independent elements are denoted by $c_{11}$, $c_{12}$, and $c_{44}$ \cite{Hopcroft2010}:

\begin{equation}
C=\left[\begin{array}{cccccc}
c_{11} & c_{12} & c_{12} & 0 & 0 & 0\\
c_{12} & c_{11} & c_{12} & 0 & 0 & 0\\
c_{12} & c_{12} & c_{11} & 0 & 0 & 0\\
0 & 0 & 0 & c_{44} & 0 & 0\\
0 & 0 & 0 & 0 & c_{44} & 0\\
0 & 0 & 0 & 0 & 0 & c_{44}
\end{array}\right].\label{eq:Stiffness matrix-1}
\end{equation}

To transform $C$ to the stiffness matrix $C'$ for any other Cartesian coordinate system, specified by the vectors $\vec{e'}_{1}$, $\vec{e'}_{2}$, and $\vec{e'}_{3}$, we first lay $\vec{e'}_{1}$ along a particular crystallographic direction defined by the Miller
indices {[}hkl{]}. The two vectors $\vec{e'}_{2}$ and $\vec{e'}_{3}$
then necessarily lie in the crystallographic plane (hkl), at right
angles to each other. The transformation of $C$ is done using the algorithm described
in \cite{Zhang2014}. The Young's modulus and the Poisson's ratio can be extracted from the compliance matrix $S'$, the inverse of the stiffness matrix $C'$, as follows
\cite{Zhang2014}: 
\begin{equation}
E_{ii}=\frac{1}{S'_{ii}},
\label{eq:3}
\end{equation}
\begin{equation}
\nu_{ij}=-\frac{S'_{ij}}{S'_{ii}},
\label{eq:4}
\end{equation}
with $i,\,j=1,\,2,\,3$ and $i\neq j$, where $i$ and $j$ denote the
three orthogonal directions in the new coordinate system. Thus, $E_{11}$
and ($E_{22}$, $E_{33}$) are the Young's moduli along the $\vec{e'}_{1}$ axis
and perpendicular to it, respectively. 
($\nu_{12}$, $\nu_{13}$) and $\nu_{23}$ are the Poisson's
ratios for the directions along the $\vec{e'}_{1}$ axis and perpendicular
to it, respectively. 

\subsection{Silicon}
\label{sec:Silicon}

Silicon is an anisotropic material which enjoys increasing popularity
as a material for cryogenic optical resonators \cite{Zhang2017,Parker2014,Millo2014,Matei2017,Kessler2012a,Hagemann2014}
due to the high thermal conductivity \cite{Glassbrenner1964}, the
ultra-low expansion coefficient at cryogenic temperatures \cite{Lyon1977,Richard1991,Wiens2014,Wiens2015}
and the ultra-low length drift \cite{Hagemann2014,Wiens2016}. Three
independent elements of the stiffness matrix $C$ from eq.~(\ref{eq:Stiffness matrix-1})
are $(c_{11},c_{12},c_{44})=(165.7,\,63.9,\,79.6)$~GPa \cite{Mason1958}. 

\begin{figure}
	\noindent\begin{minipage}[t]{1\columnwidth}%
		\begin{center}
			\includegraphics[width=0.48\textwidth]{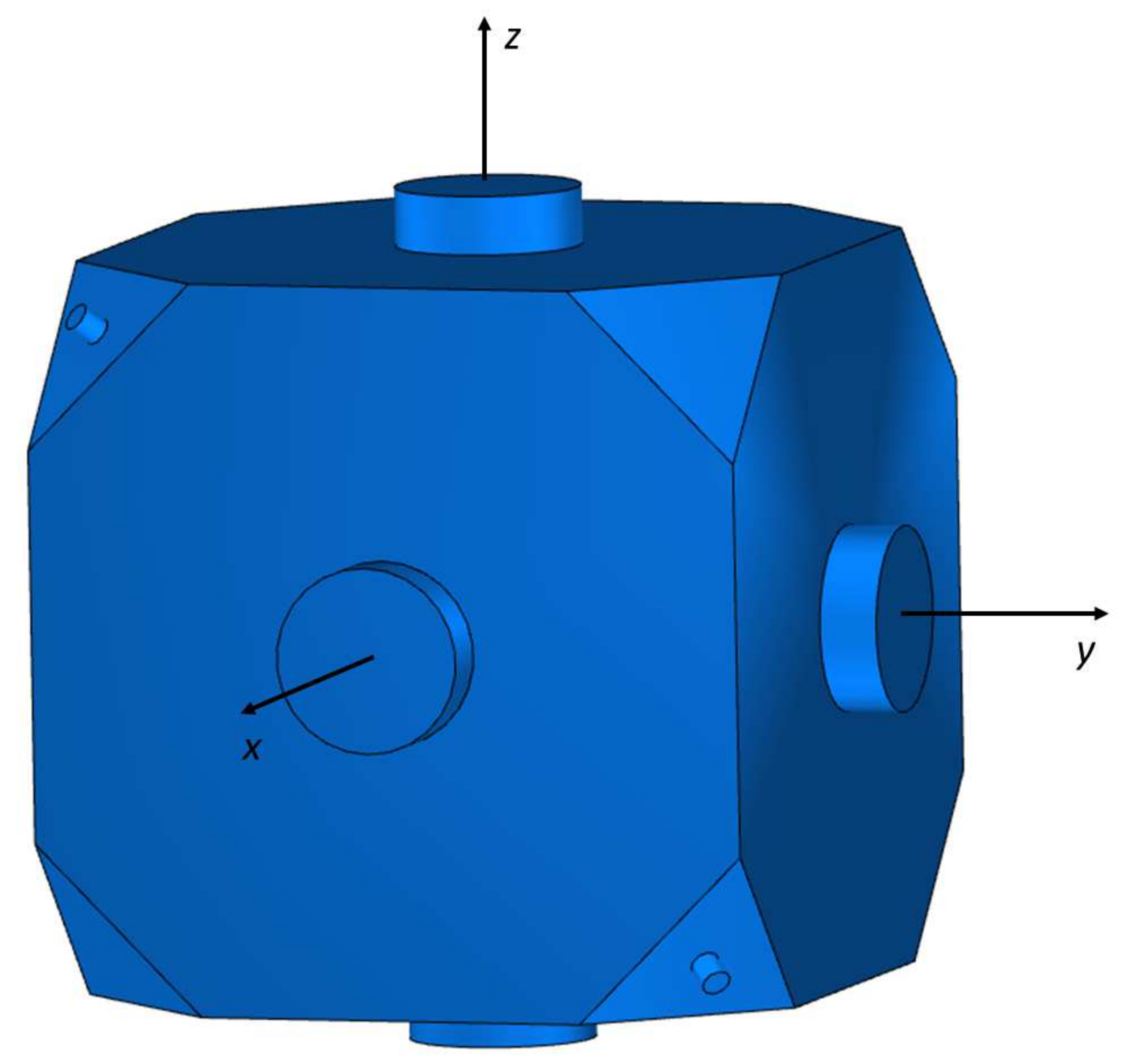}
			\quad{}%
			\includegraphics[width=0.48\textwidth]{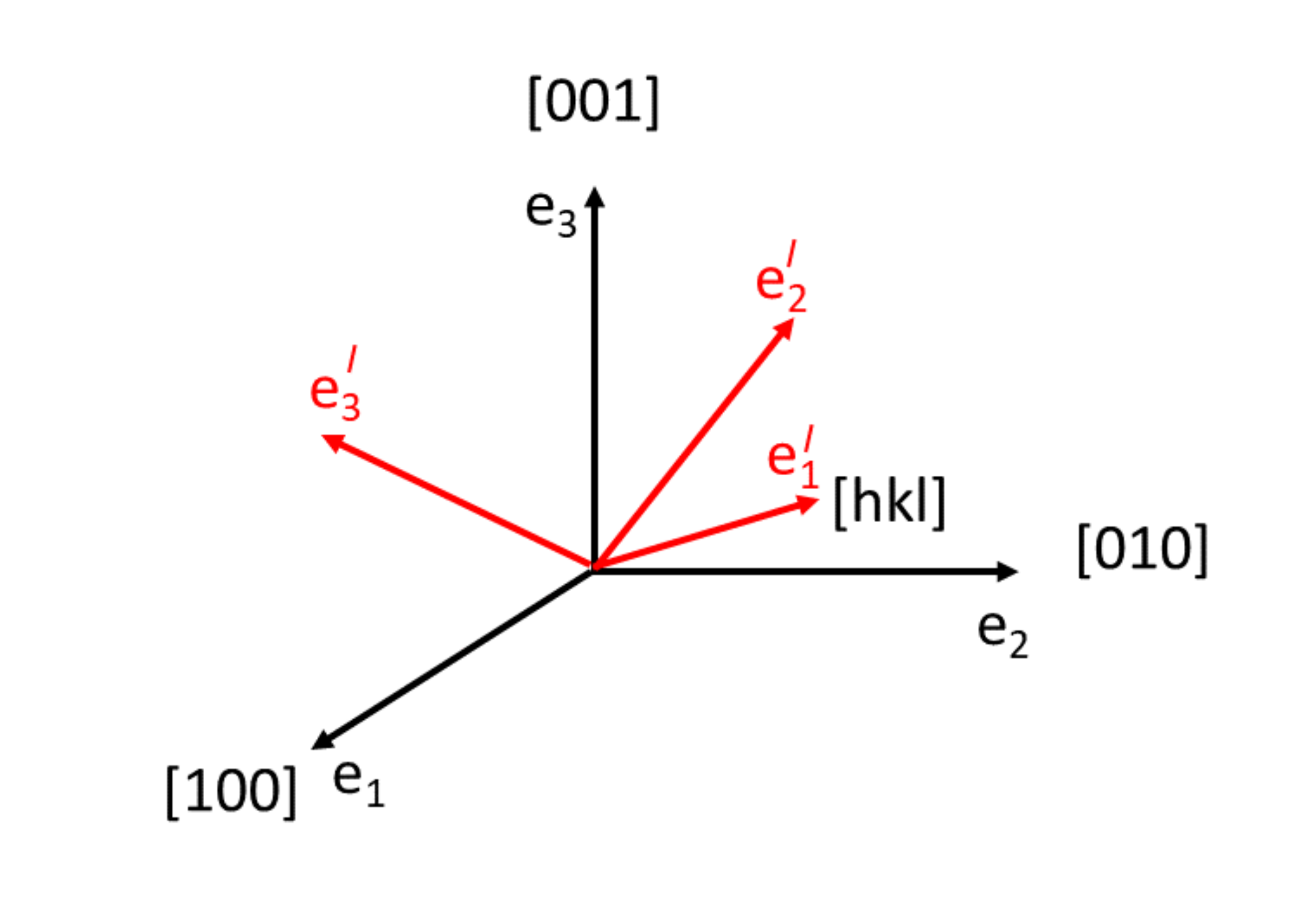}
		\par\end{center}%
		\bigskip{}
		\begin{center}
			\includegraphics[width=0.48\textwidth]{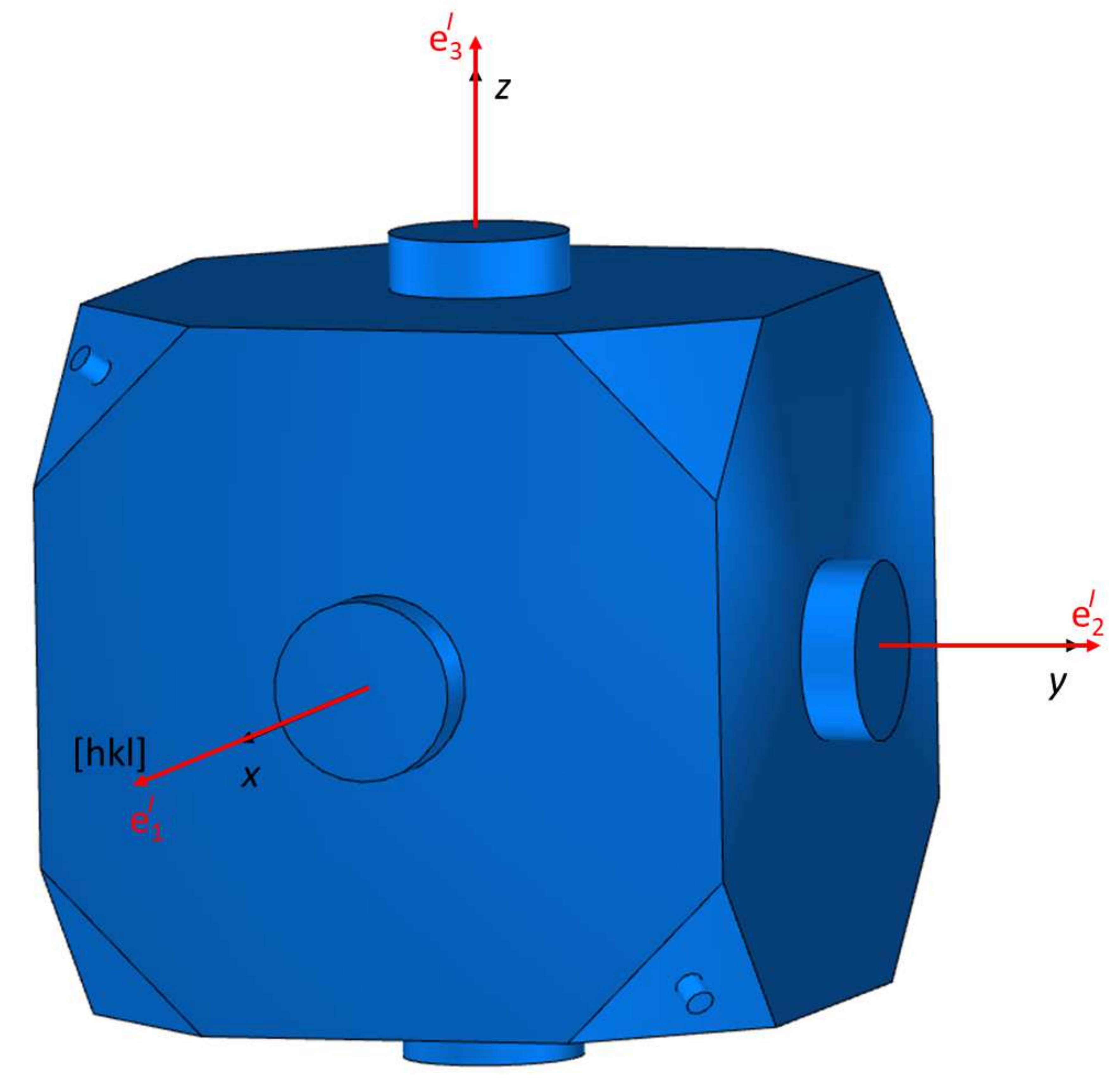}
			\quad{}%
			\includegraphics[width=0.48\textwidth]{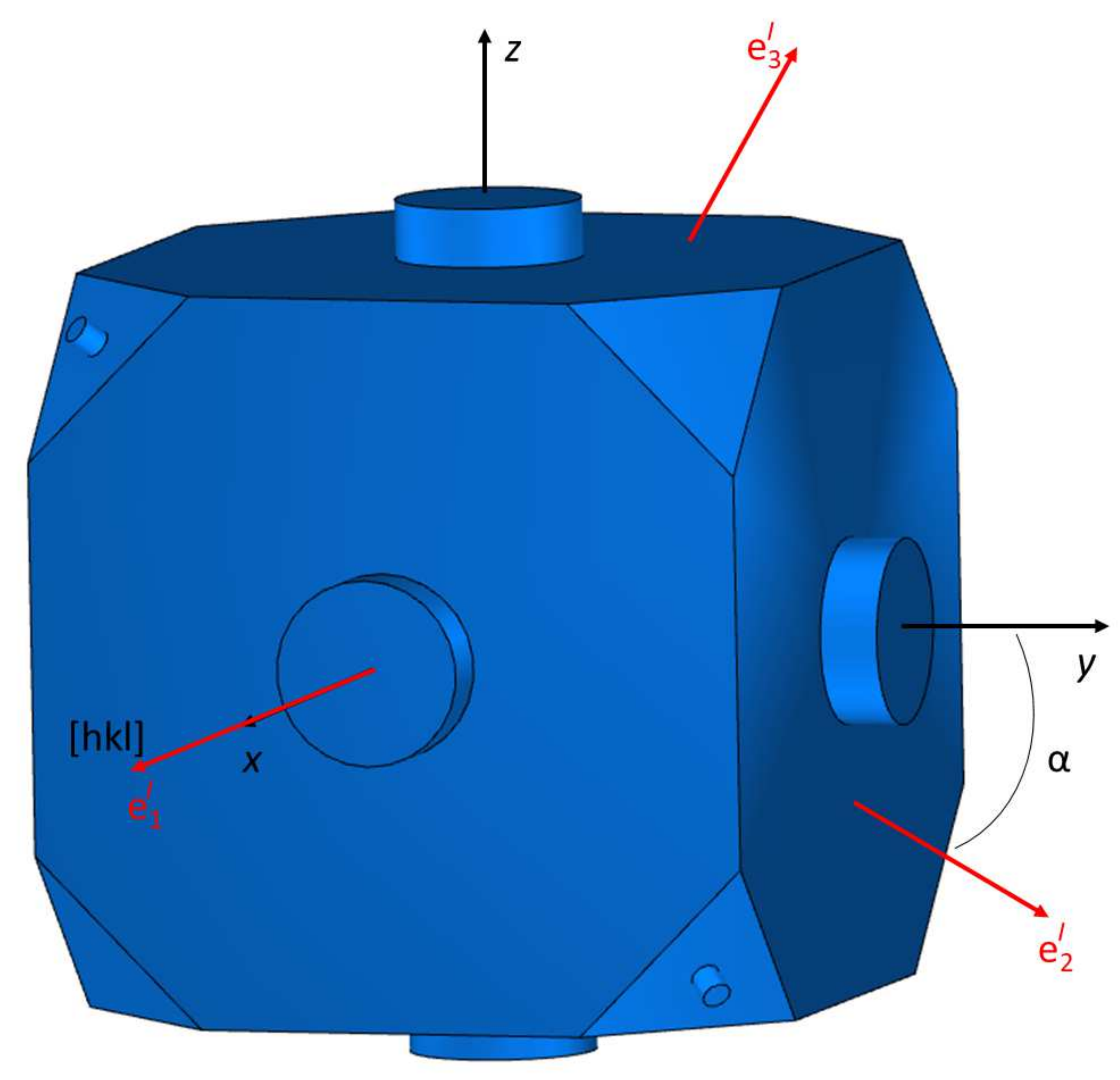}
			\par\end{center}%
		\bigskip{}
	\end{minipage}
\caption{\label{fig:Transformation-Coordinate-System} Orientation of the crystal lattice with respect to the cavity body. Top left: The silicon resonator is always oriented with its cavities aligned with the $(x,\,y,\,z)$ coordinate axes of the laboratory reference frame. Top right: Orientation of the crystal-lattice-fixed coordinate system of the silicon crystal ($\vec{e'}_{1}$, $\vec{e'}_{2}$, $\vec{e'}_{3}$) with vector $\vec{e'}_{1}$ coincident with a selected crystallographic direction {[}hkl{]}, relative to the coordinate system defined by the ($\vec{e}_{1}$, $\vec{e}_{2}$, $\vec{e}_{3}$) unit vectors oriented along the [100], [010], and [001] crystallographic directions, respectively. Bottom left: Silicon crystal is oriented with the selected crystallographic direction [hkl] coincident with the $x$-axis of the laboratory frame and $\vec{e'}_{3}$ along z. Bottom right: Rotation of the silicon crystal around the [hkl] direction by an angle $\alpha$. For this orientation the stiffness matrix $C'$ is calculated.}
\end{figure}

In order to set up the simulation for any desired crystallographic
direction [hkl], we first orient the resonator with the optical axes of the three cavities laying parallel to the $(x,\,y,\,z)$ coordinate axes of the fixed laboratory reference frame, as shown in Fig.~\ref{fig:Transformation-Coordinate-System}, top left panel. Then, we define the new coordinate system by pointing $\vec{e'}_{1}$
along the [hkl] crystallographic direction and by defining the two vectors $\vec{e'}_{2}$ and $\vec{e'}_{3}$ in the crystallographic plane (hkl), at right
angles to each other (see Fig.~\ref{fig:Transformation-Coordinate-System}, top right panel). Because of the cubic symmetry of the silicon lattice, we
only need to consider crystallographic directions that lie inside
the unit stereographic triangle whose corners are defined by the {[}100{]},
{[}110{]}, and {[}111{]} directions. In the next step, we orient the crystal structure with the chosen crystallographic direction {[}hkl{]} along the $x-$axis
of the cube. Two other unit vectors $\vec{e'}_{2}$ and $\vec{e'}_{3}$ are laid along the $y$ and $z$ axes, respectively (see Fig.~\ref{fig:Transformation-Coordinate-System}, bottom left panel). To find all possible orientations of interest we can introduce an additional degree of freedom by rotating the crystal counterclockwise around the {[}hkl{]} direction, as seen along the x-axis (see Fig.~\ref{fig:Transformation-Coordinate-System}, bottom right panel). This is done first by rotating the vectors $\vec{e'}_{2}$ and $\vec{e'}_{3}$
in the crystallographic plane (hkl) by an angle $\alpha$ around the
$\vec{e'}_{1}$ axis using Rodrigues' rotation formula \cite{Battin1999} (see Fig.~\ref{fig:Transformation-Coordinate-System}, bottom right panel).
After rotation, the algorithm from \cite{Zhang2014} is applied again
to obtain the stiffness matrix $C'$, the Poisson's ratio $\nu_{12}$ 
and $\nu_{23}$, and the Young's modulus $E_{11}$
and $E_{22}$. This procedure is repeated for different values of
$\alpha$ until one full circle of rotation is completed. We note, that due to rotation,  we only need to consider $E_{22}$ and $\nu_{12}$ as the Young's modulus and the Poisson's ratio for the direction perpendicular to the $x$-axis, respectively, as they contain all necessary information. The $E_{33}$ and $\nu_{13}$ can be ignored.   

\begin{figure}[tb]
\begin{centering}
\begin{center}
\includegraphics[width=0.48\textwidth]{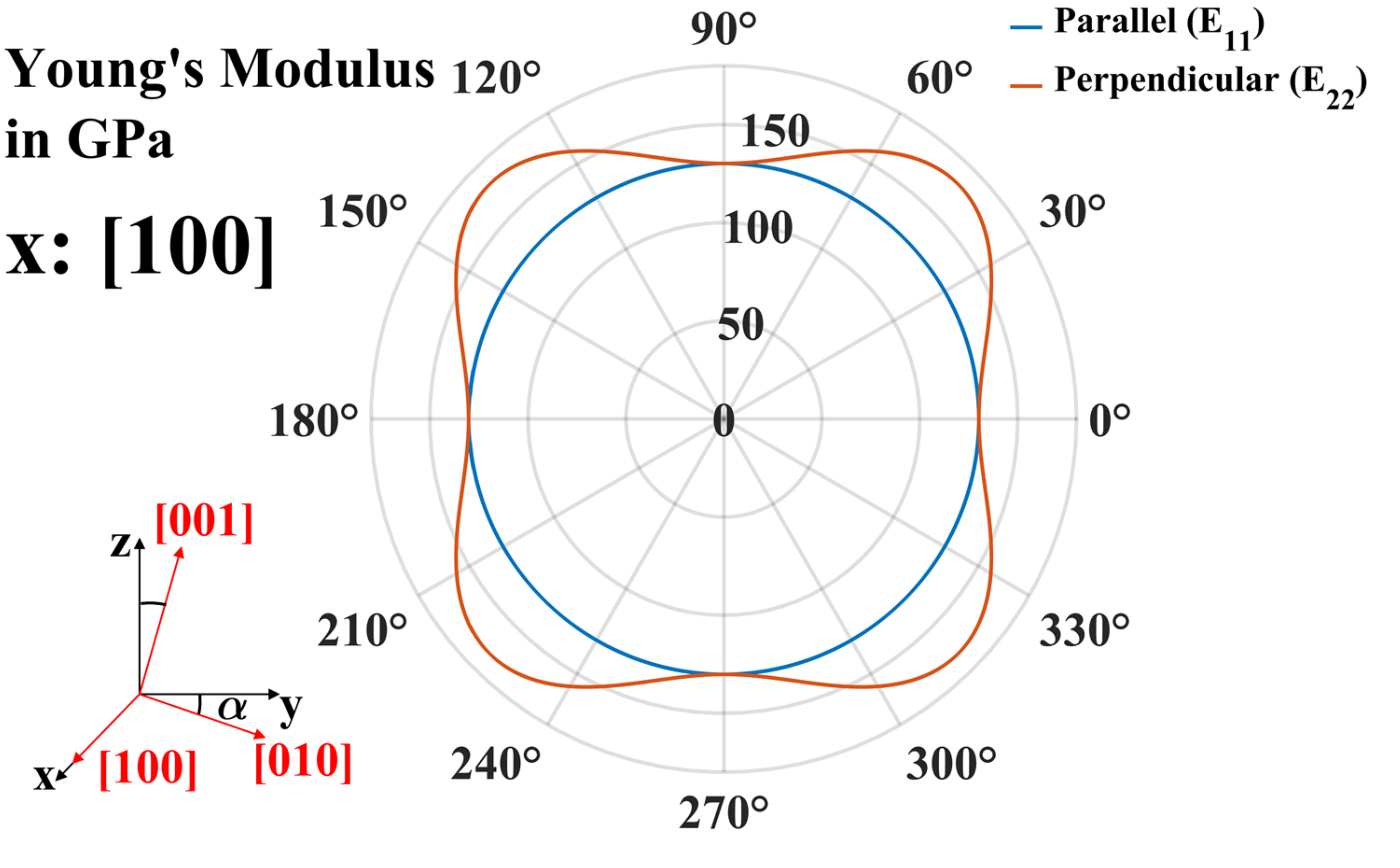}
\quad{}%
\includegraphics[width=0.48\textwidth]{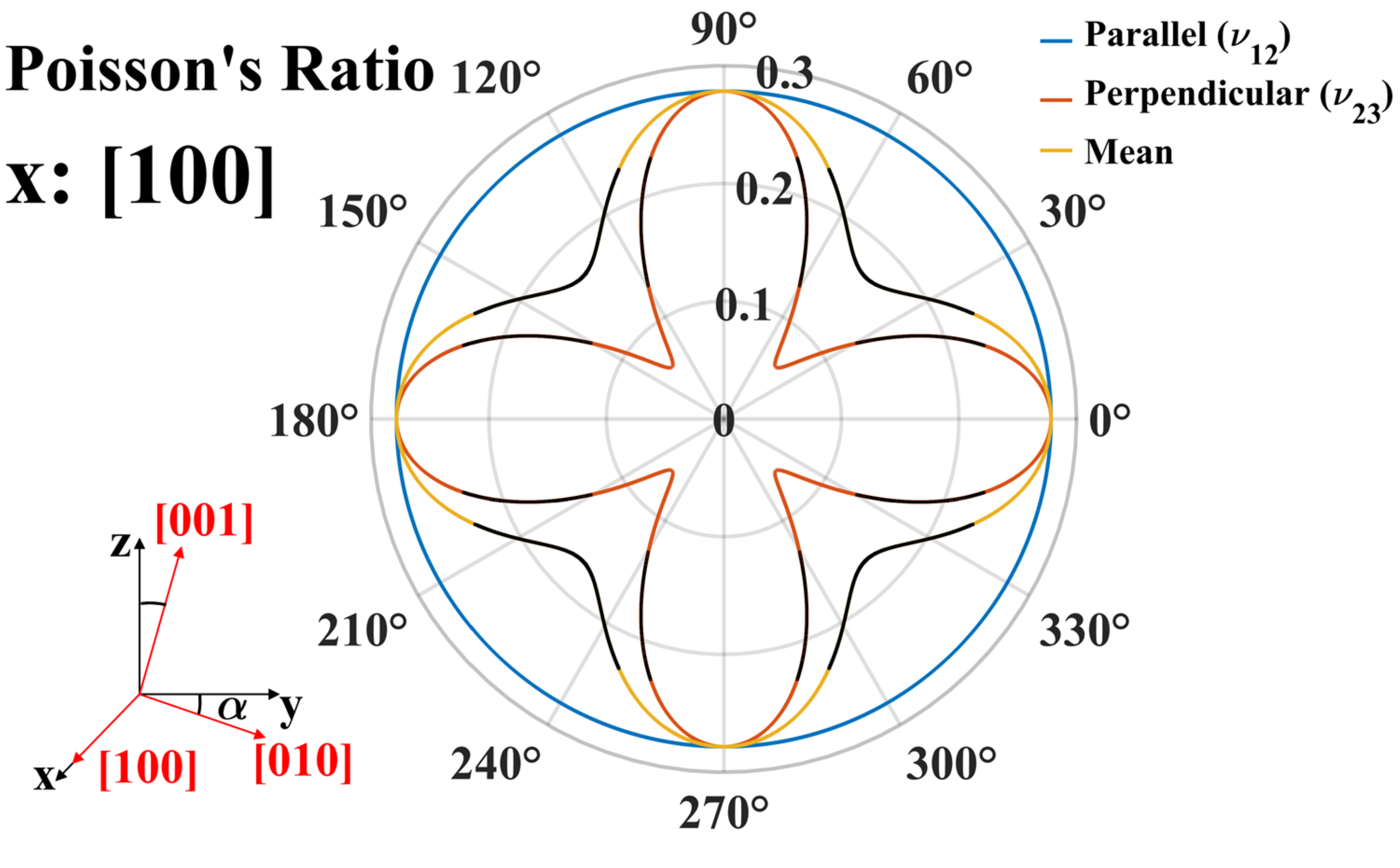}
\par\end{center}%
\bigskip{}
\begin{center}
\includegraphics[width=0.48\textwidth]{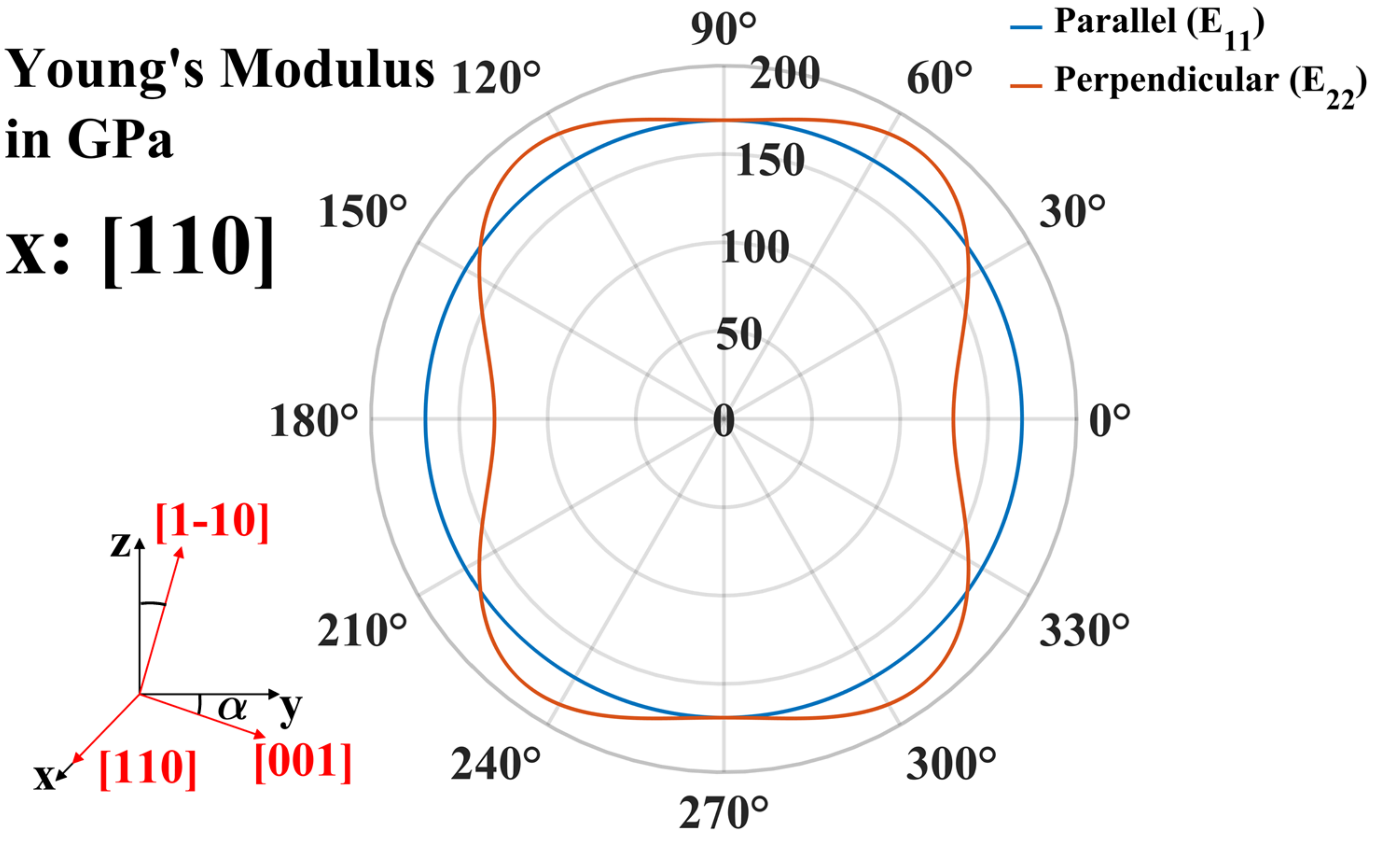}
\quad{}%
\includegraphics[width=0.48\textwidth]{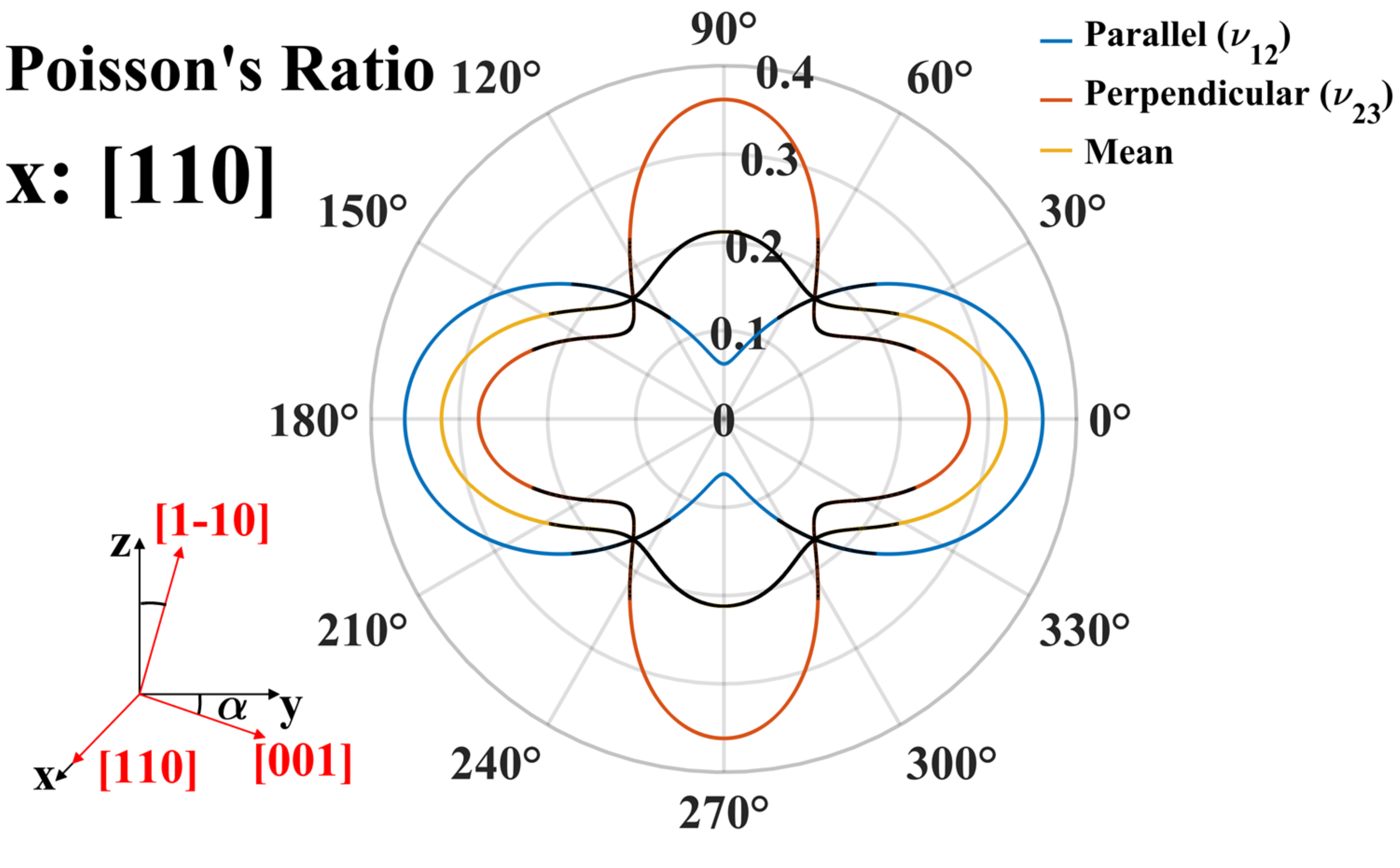}
\par\end{center}%
\bigskip{}
\begin{center}
\includegraphics[width=0.48\textwidth]{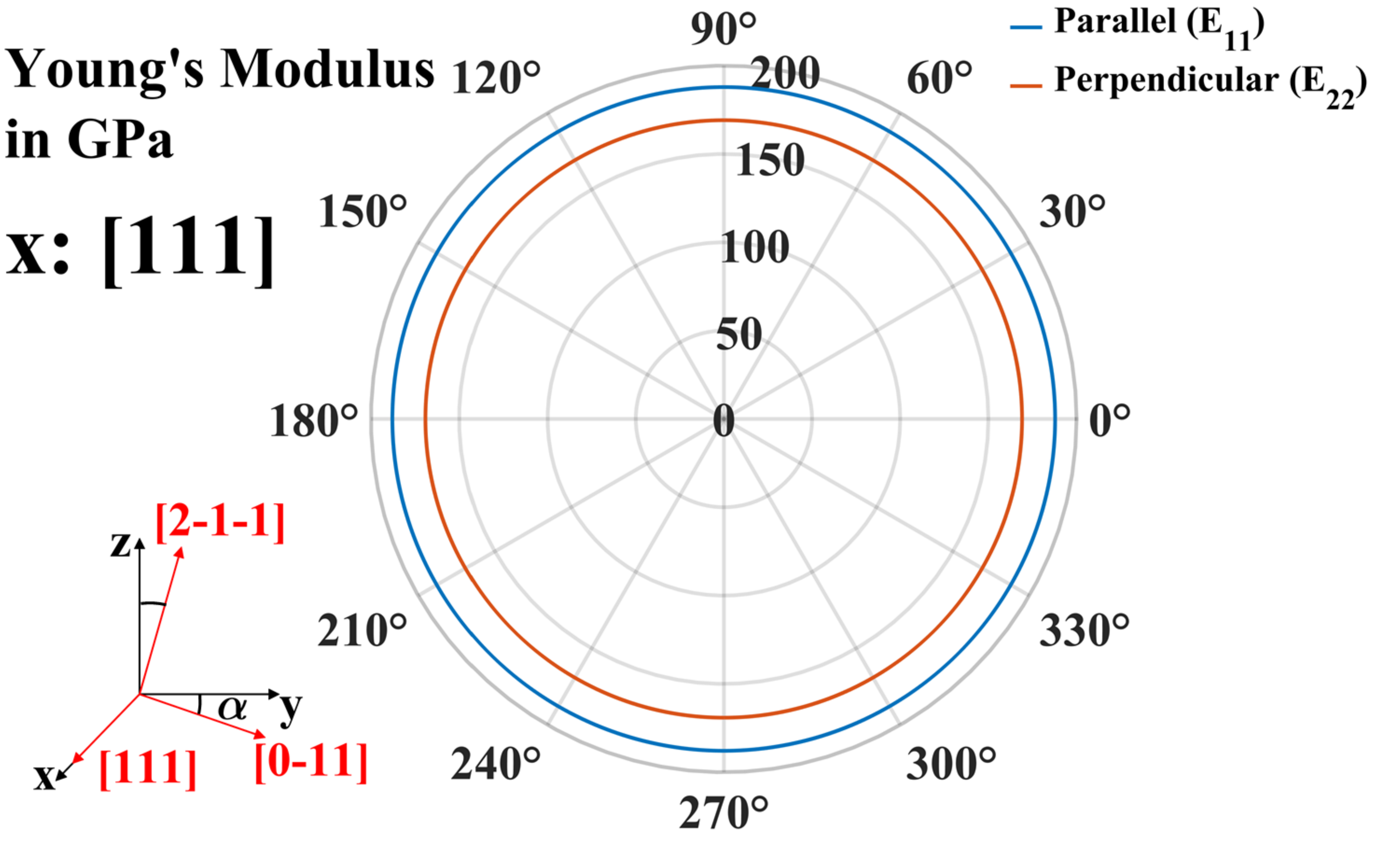}
\quad{}%
\includegraphics[width=0.48\textwidth]{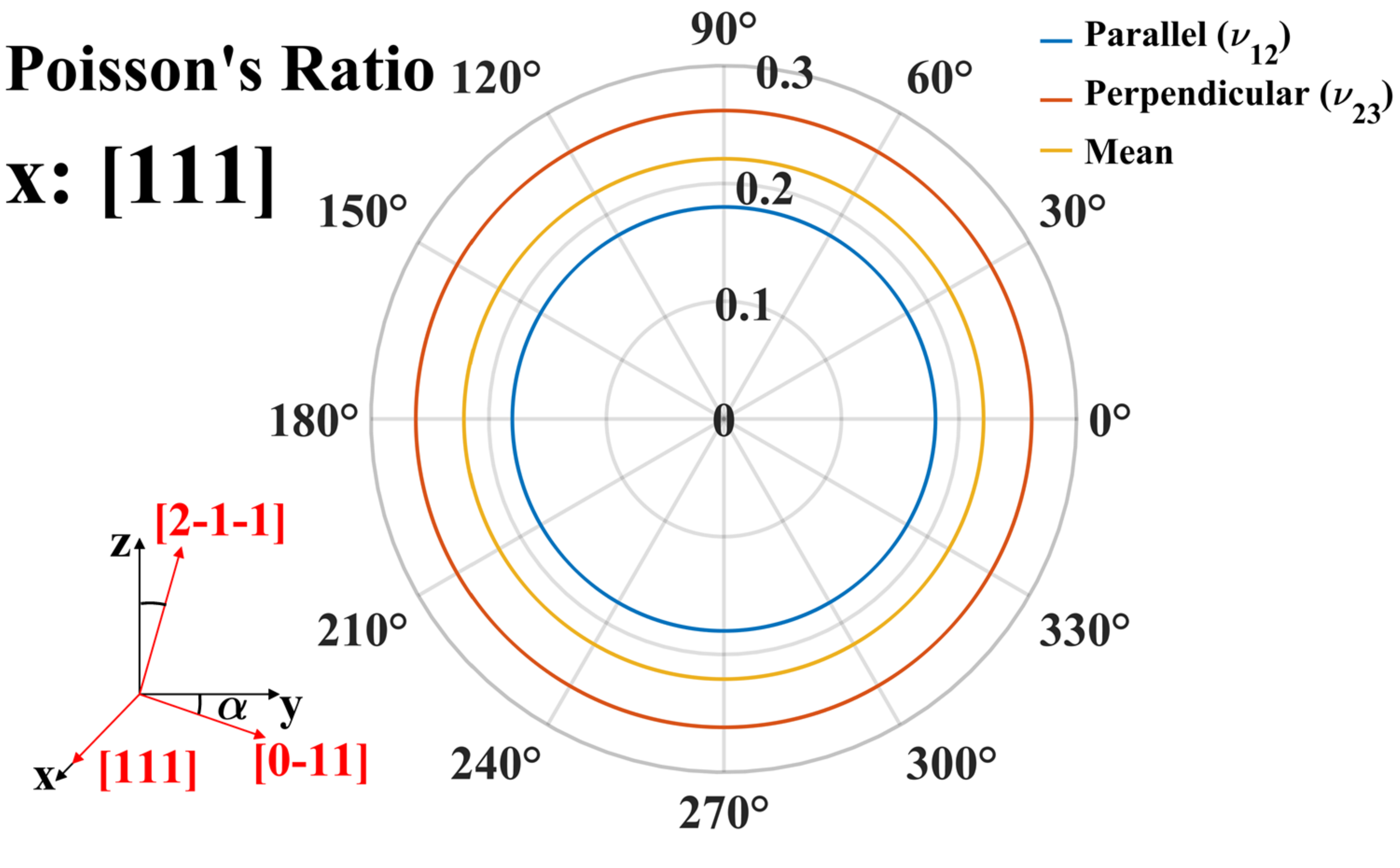}
\par\end{center}%
\end{centering}
\caption{\label{fig:Si100-Young-Poisson}Silicon's Young's modulus (left column) and Poisson's ratio (right column) calculated for the {[}100{]}, {[}110{]}, and {[}111{]} crystallographic directions (blue lines) and for directions perpendicular to them (red lines), for different values of angle $\alpha$. Values of the Poisson's ratio that lie within a ``magic'' range $0.13 < \nu < 0.23$ are marked with black color. The definition of the angle $\alpha$ is shown in the small panels.}
\end{figure}

The values of $E,\,\nu$ for
rotation around the  {[}100{]}, {[}110{]}, and {[}111{]} characteristic directions are visualized in Fig.~\ref{fig:Si100-Young-Poisson}, where the values of the Poisson's
ratio that correspond to the ``magic'' range $0.13 < \nu < 0.23$
were colored green. 

\begin{figure}[tb]
\begin{centering}
\begin{center}
\includegraphics[width=0.48\textwidth]{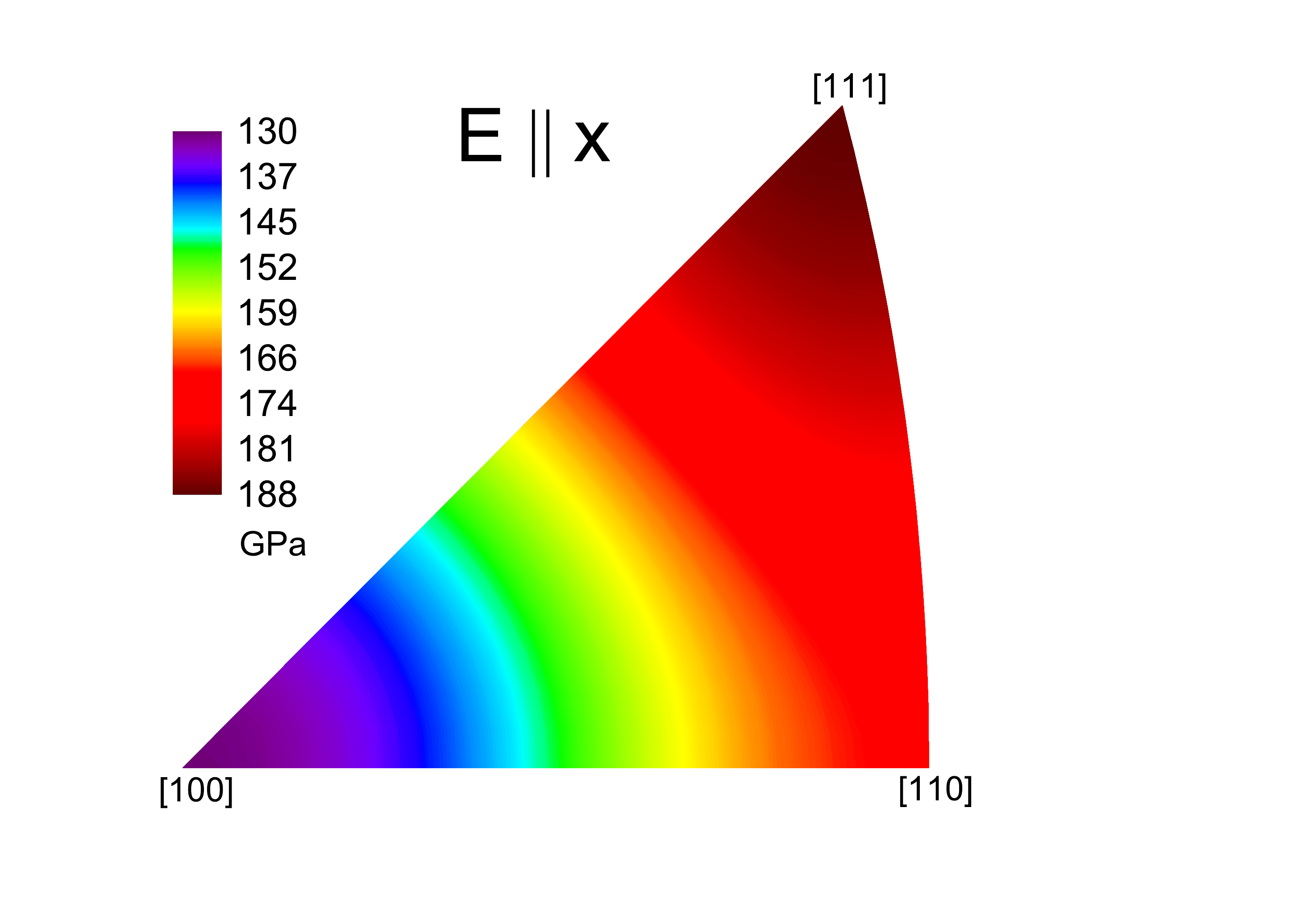}
\quad{}%
\includegraphics[width=0.48\textwidth]{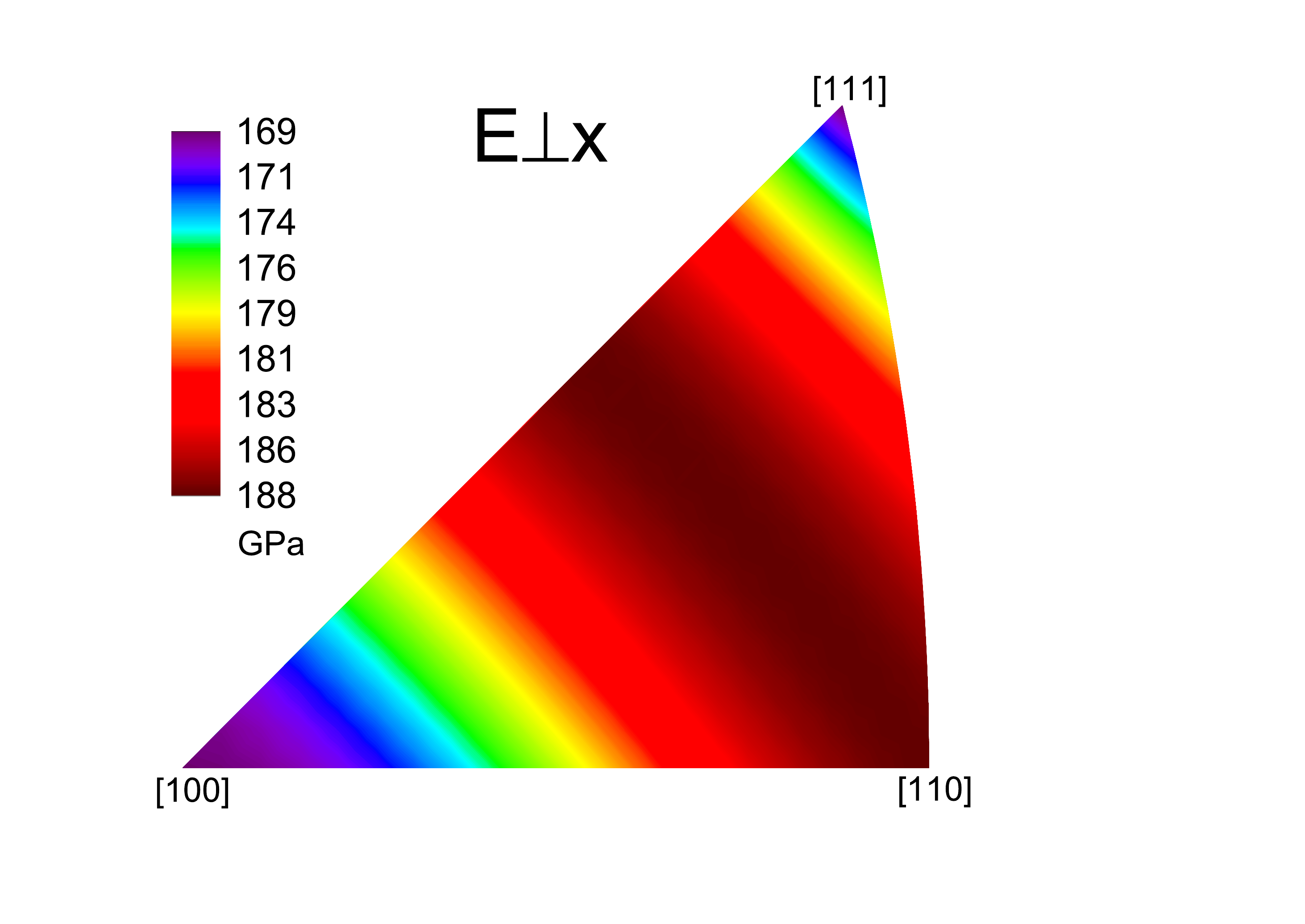}
\par\end{center}%
\begin{center}
\includegraphics[width=0.48\textwidth]{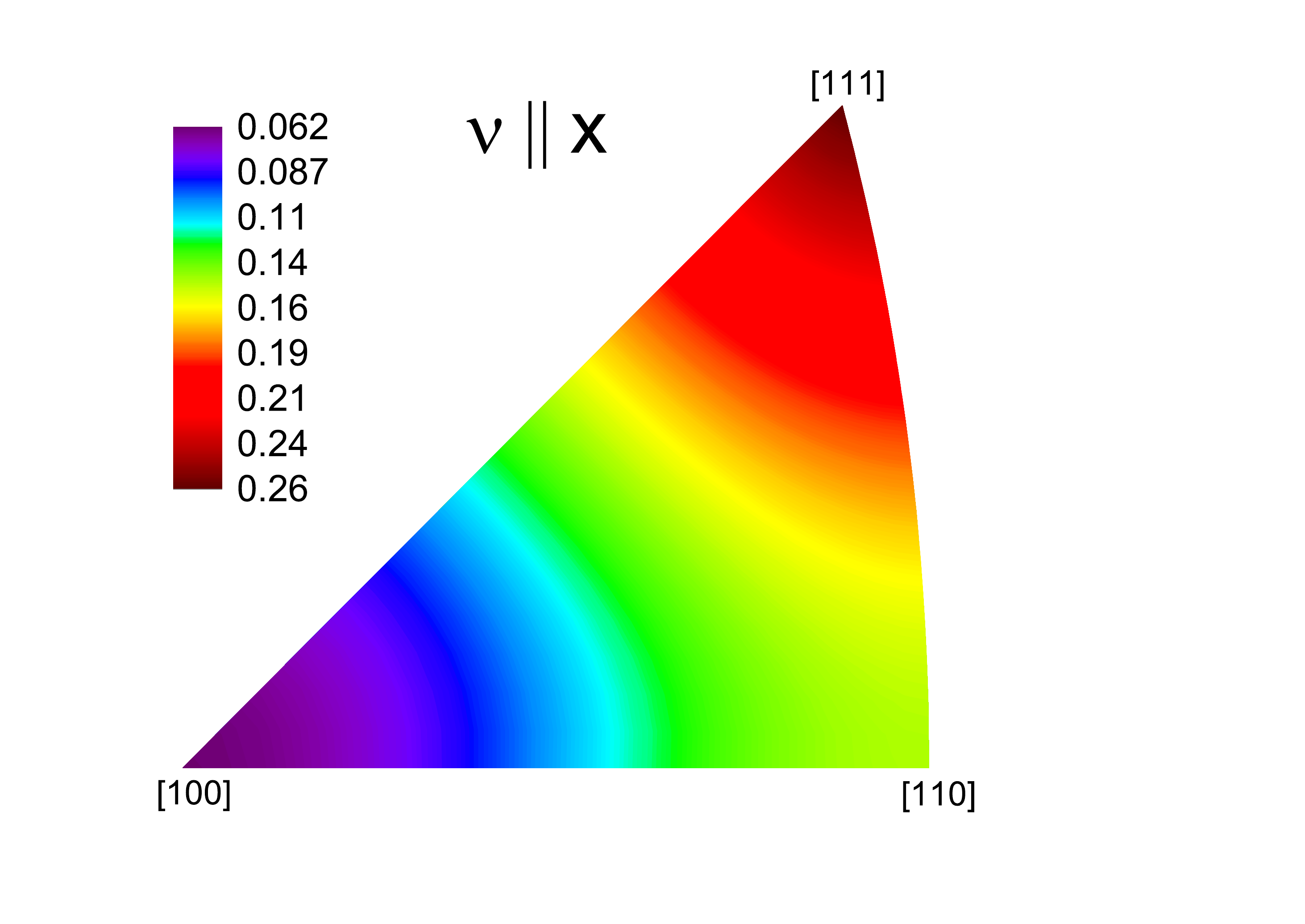}
\quad{}%
\includegraphics[width=0.48\textwidth]{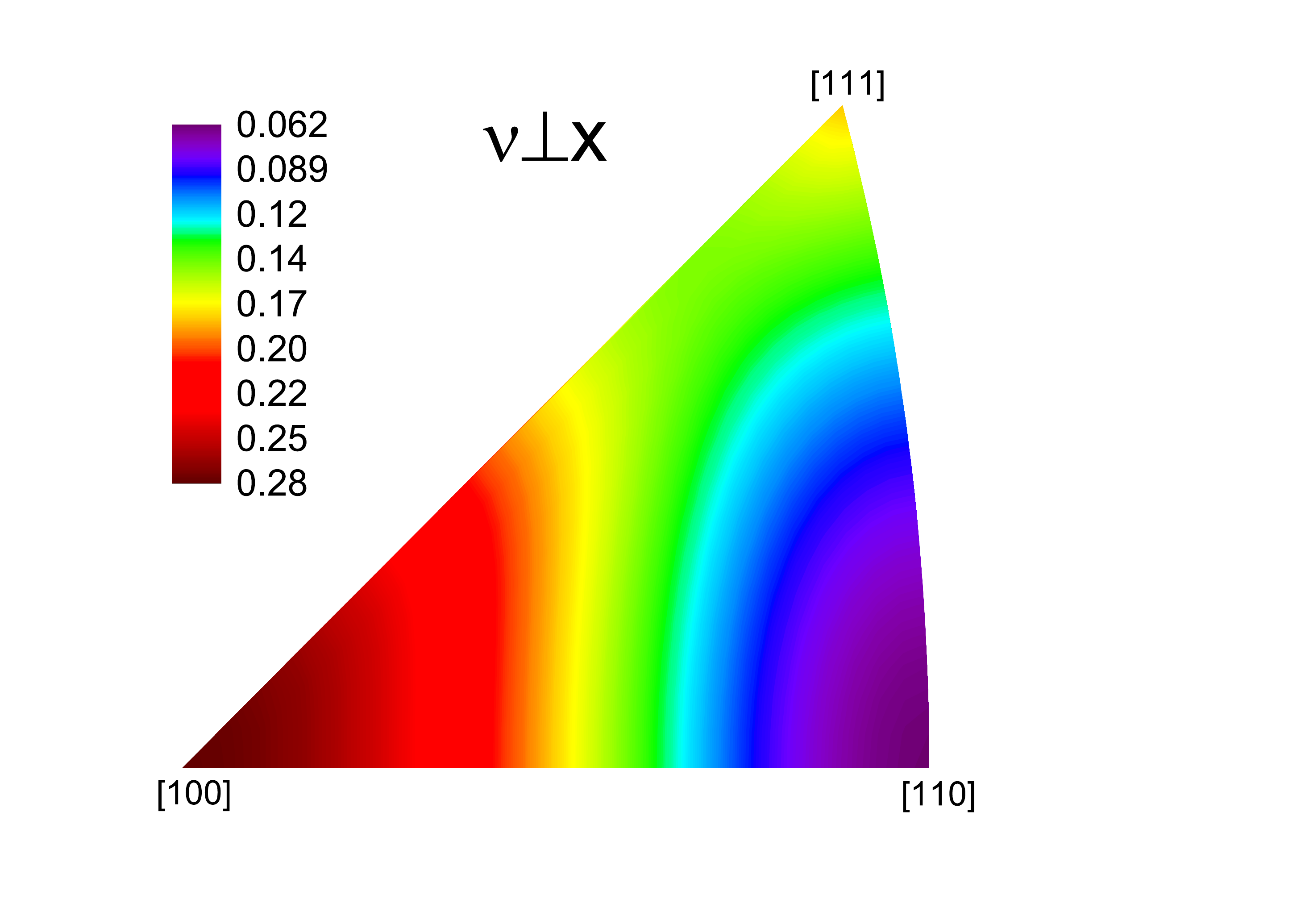}
\par\end{center}%
\par\end{centering}
\caption{\label{fig:SST-Poisson_Young}Maximum of the Young's modulus and minimum of the Poisson's ratio for different crystallographic orientations
of Si inside the unit stereographic triangle oriented along the $x$-axis of the resonator.}
\end{figure}

The \textit{maximum} of the Young's modulus and the \textit{minimum}
of the Poisson's ratio for all angles of rotation for any given direction
{[}hkl{]} inside the unit stereographic triangle are presented in
Fig.~\ref{fig:SST-Poisson_Young}. The Young's modulus varies from
$130.1$~GPa to $187.9$~GPa for the directions parallel to {[}hkl{]}
(top, left) and from $169.1$~GPa to $187.9$~GPa for the perpendicular
direction (top, right). Directions which provide the highest stiffness
are the {[}111{]} and {[}110{]} directions, respectively. 

As we know from the foregoing discussion on isotropic materials, the
Poisson's ratio plays the crucial role. In order to have zero sensitivity
it should lie within a ``magic range'' $0.13 < \nu < 0.23$. The
Poisson's ratio varies from 0.062 to 0.26 for the parallel direction
(bottom, left) with a minimum along {[}100{]} and a maximum along
{[}111{]}. The variation in the perpendicular direction is from 0.062
to 0.28 with the minimum along {[}110{]} and the maximum along {[}100{]}.
This large difference makes it impossible to predict which directions
will have zero sensitivity and makes extensive simulations necessary.

Since the material is anisotropic, we now calculate the length changes
of all three cavities within the cube, i.e. $\Delta L_{i}(F_{c})/L$
for $i=x,\,y,\,z$, which may differ. We repeat the simulations for
different crystallographic directions inside the unit stereographic
triangle, which is oriented along the $x$-axis. 

\begin{figure}
\begin{centering}
\begin{center}
\begin{minipage}[t]{0.48\columnwidth}%
\includegraphics[width=0.98\textwidth]{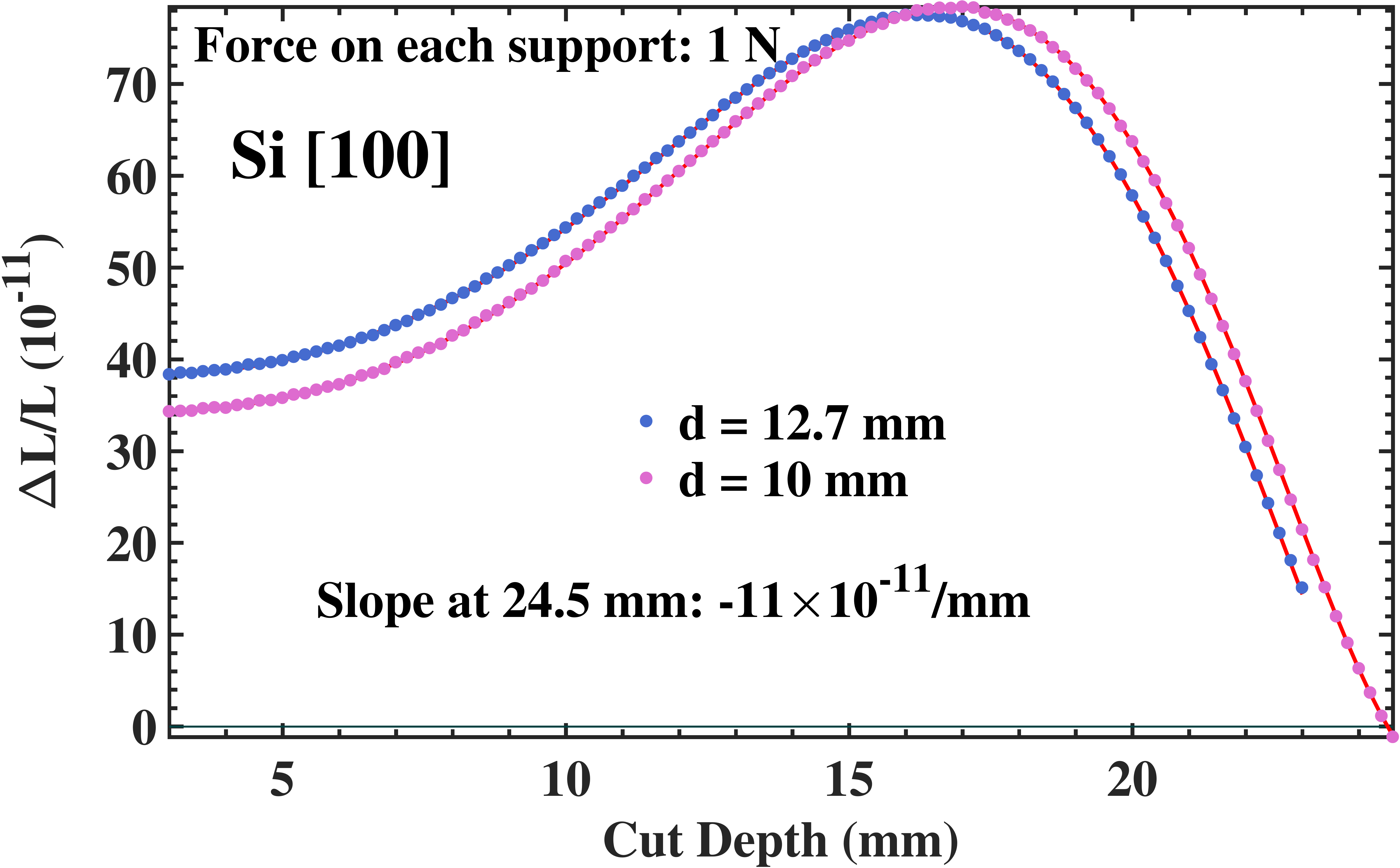}
\end{minipage}
\quad{}%
\begin{minipage}[t]{0.48\columnwidth}%
\begin{center}
\includegraphics[width=1.0\textwidth]{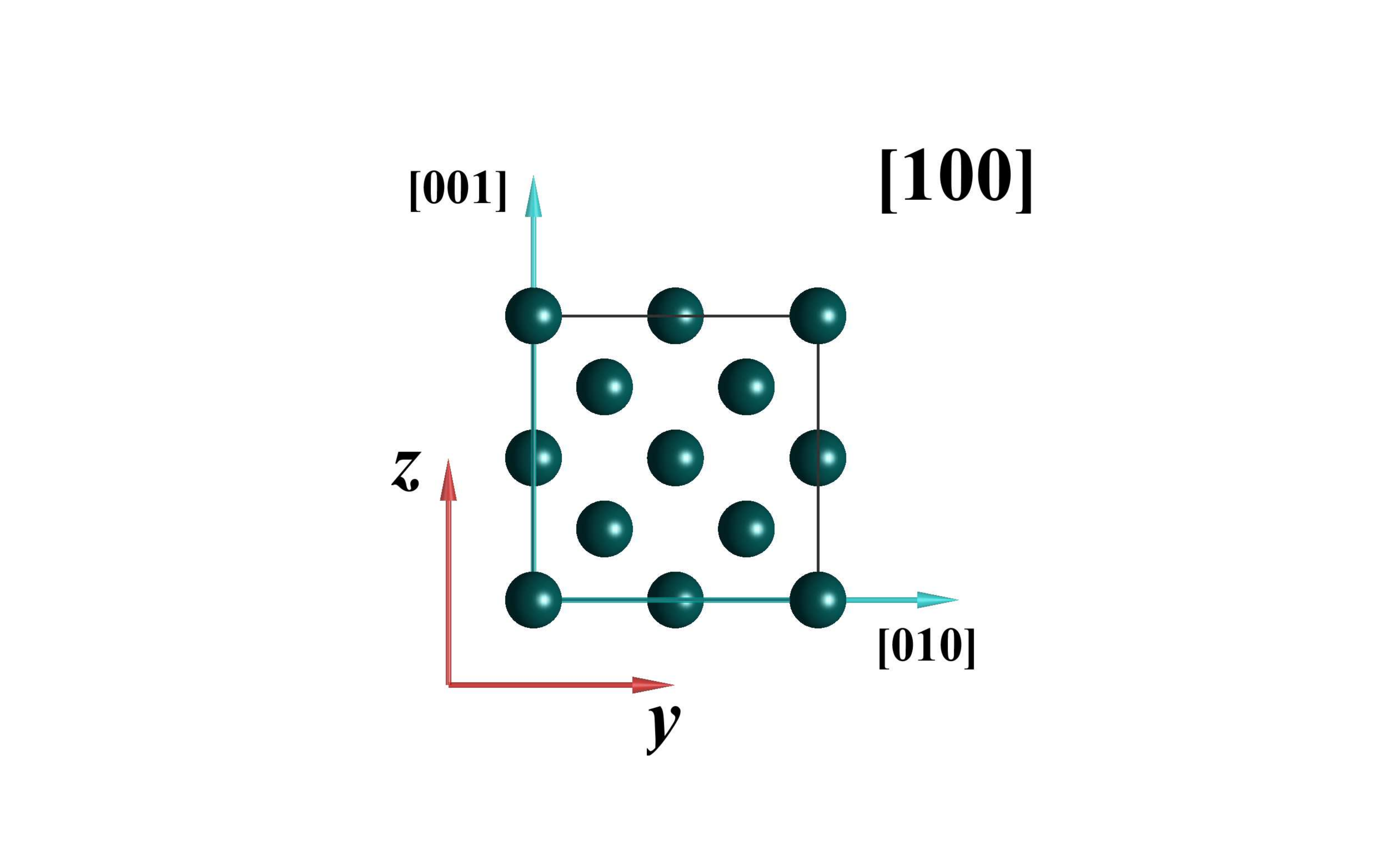}
\par\end{center}%
\end{minipage}
\par\end{center}%
\medskip{}
\begin{center}
\begin{minipage}[t]{0.48\columnwidth}%
\includegraphics[width=0.98\textwidth]{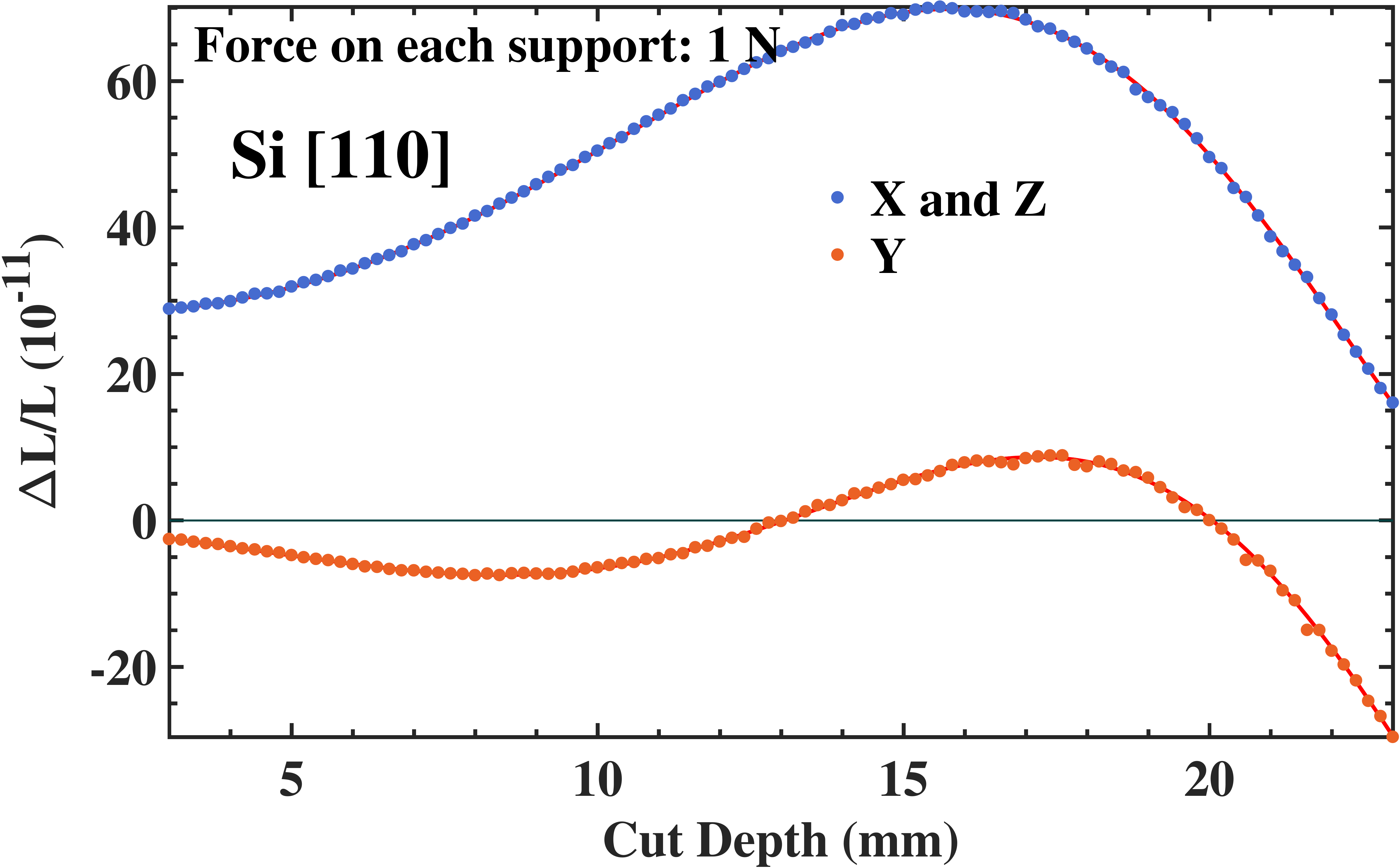}
\end{minipage}
\quad{}%
\begin{minipage}[t]{0.48\columnwidth}%
\begin{center}
\includegraphics[width=1.0\textwidth]{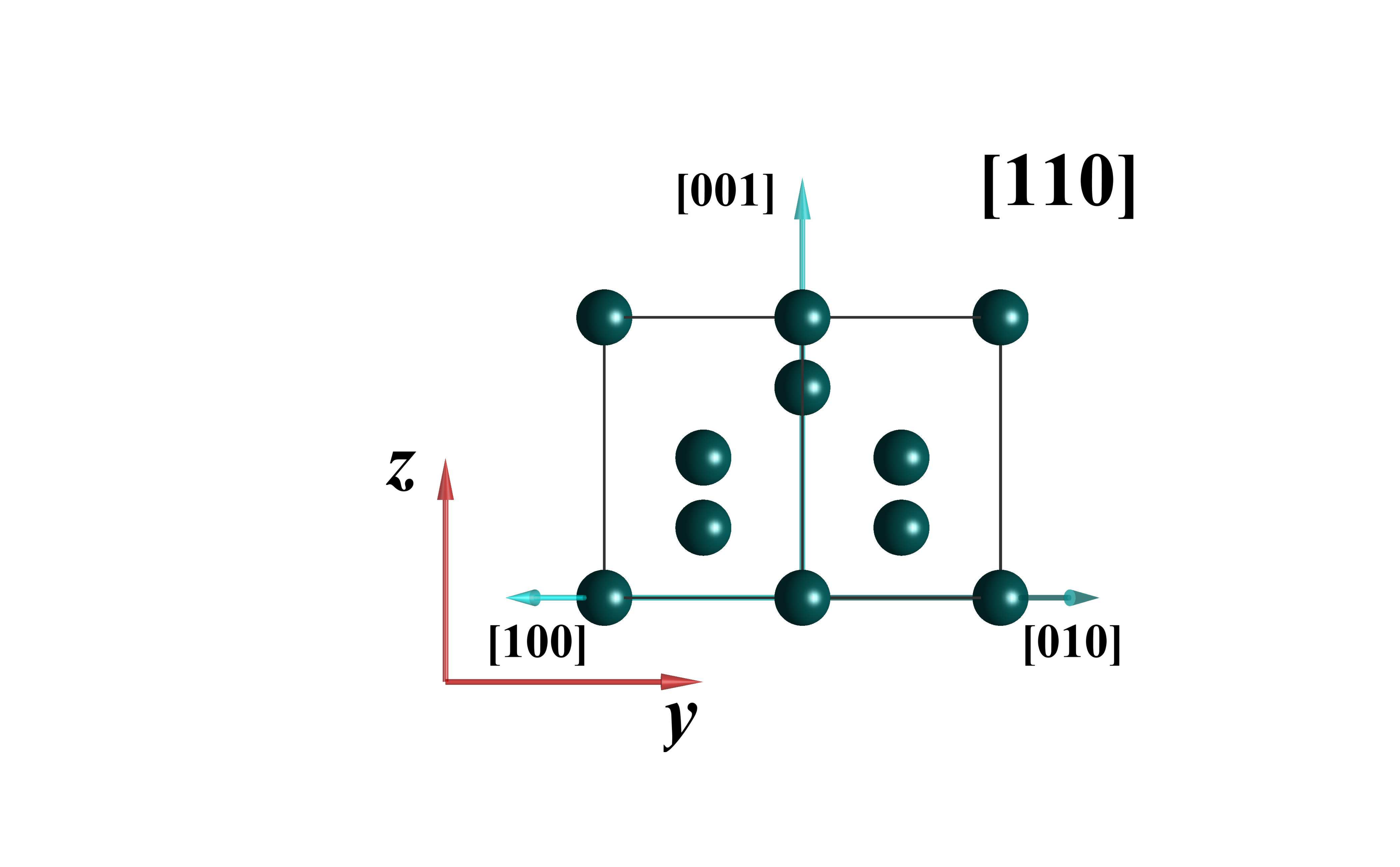}
\par\end{center}%
\end{minipage}
\par\end{center}%
%
%
\begin{center}
\begin{minipage}[t]{0.48\columnwidth}%
\includegraphics[width=0.98\textwidth]{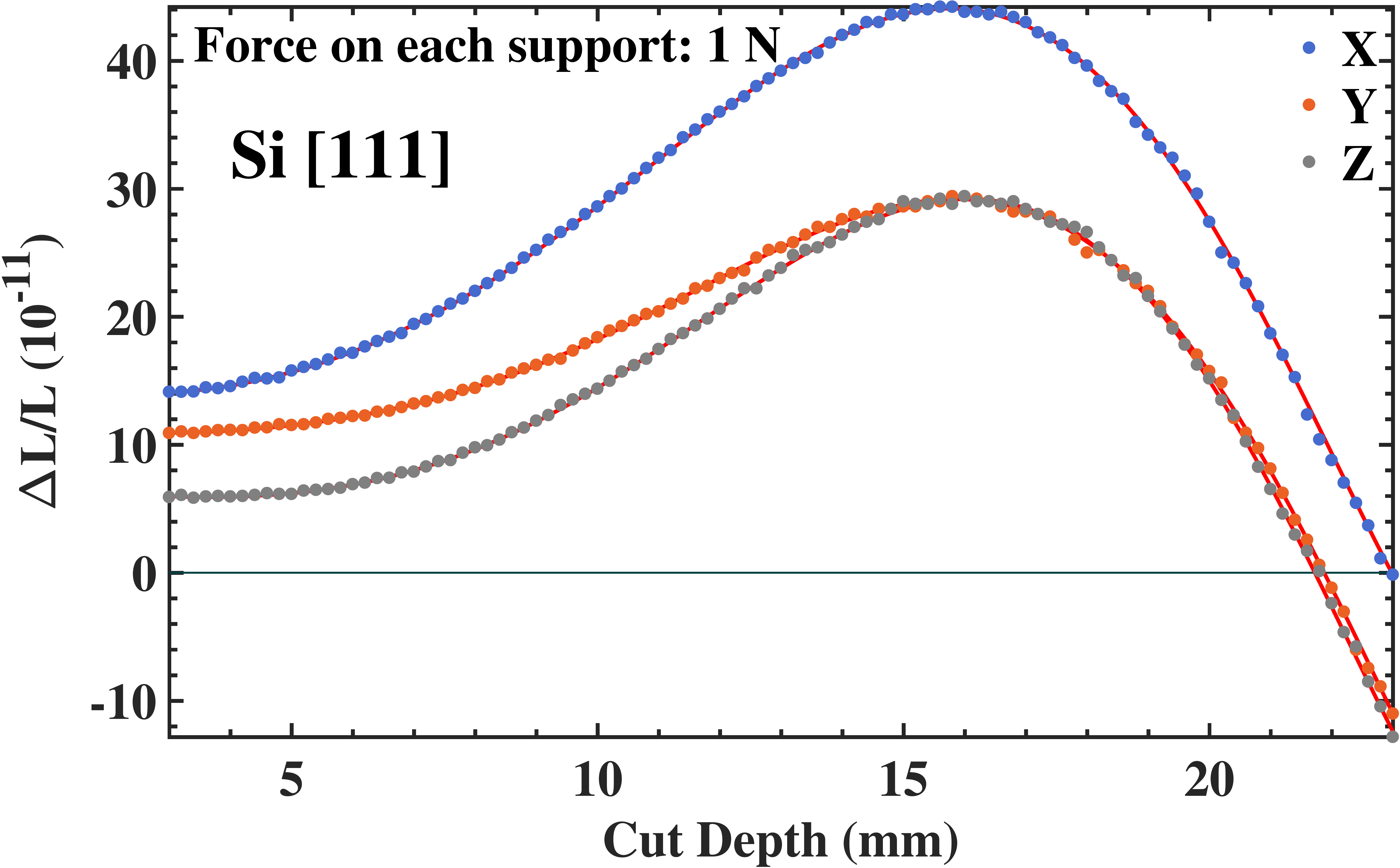}
\end{minipage}
\quad{}%
\begin{minipage}[t]{0.48\columnwidth}%
\begin{center}
\includegraphics[width=1.0\textwidth]{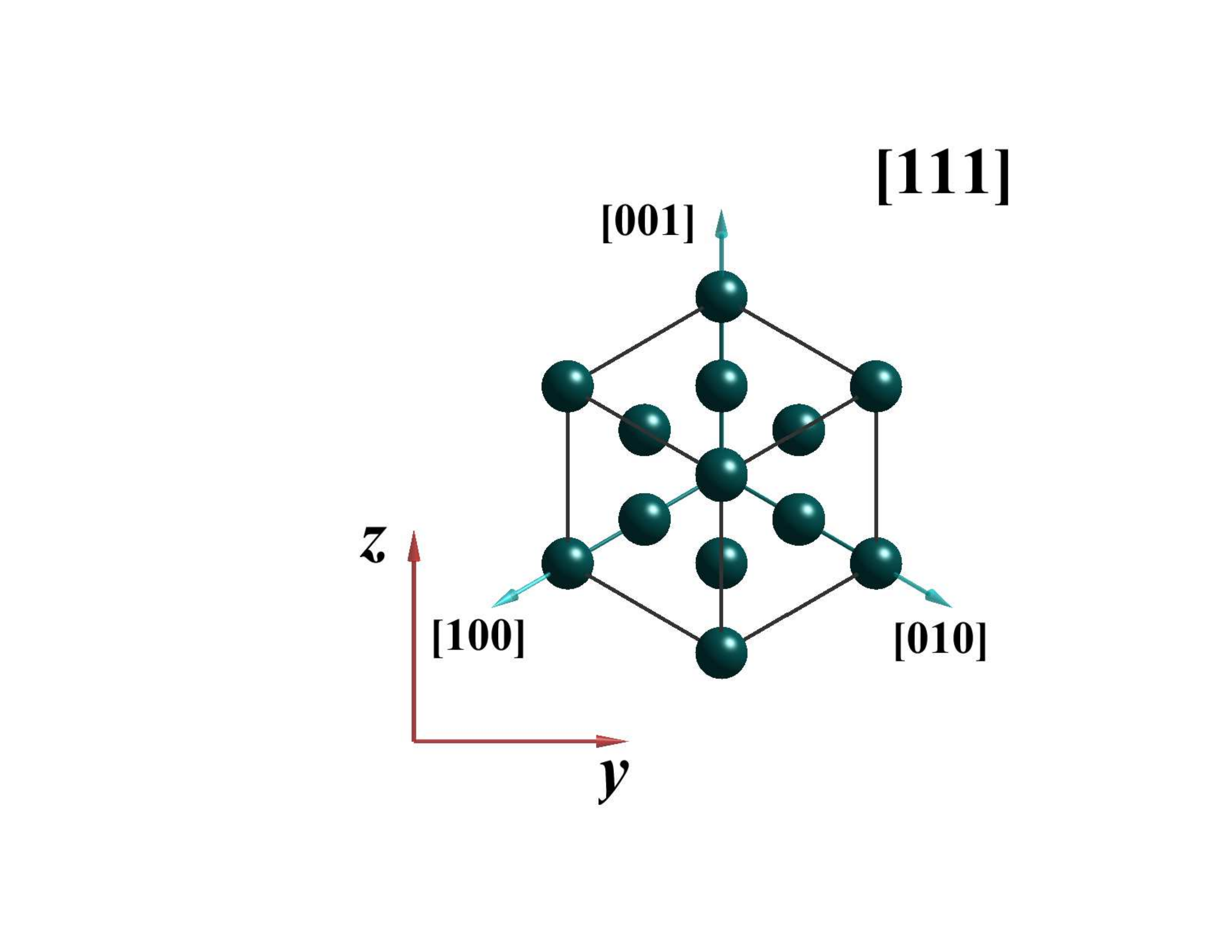}
\par\end{center}%
\end{minipage}
\par\end{center}%
\par\end{centering}
\caption{\label{fig:CutDepth-Silicon}Sensitivity to $F_{c}$ for the three
cavities when the Si crystal is oriented in particular crystallographic
directions. The corresponding view on the crystal along the direction
in question is shown on the right, each ball representing the top
Si atom in the plane perpendicular to the observation direction.}
\end{figure}

A first result of the FEA simulation is shown in Fig.~\ref{fig:CutDepth-Silicon},
left column, where only the three characteristic directions {[}100{]},
{[}110{]}, and {[}111{]} are considered. The right column visualizes
the corresponding crystal structure seen along these directions. 

\begin{figure}
\begin{centering}
\begin{minipage}[t]{0.98\columnwidth}%
\begin{center}
\includegraphics[scale=0.3]{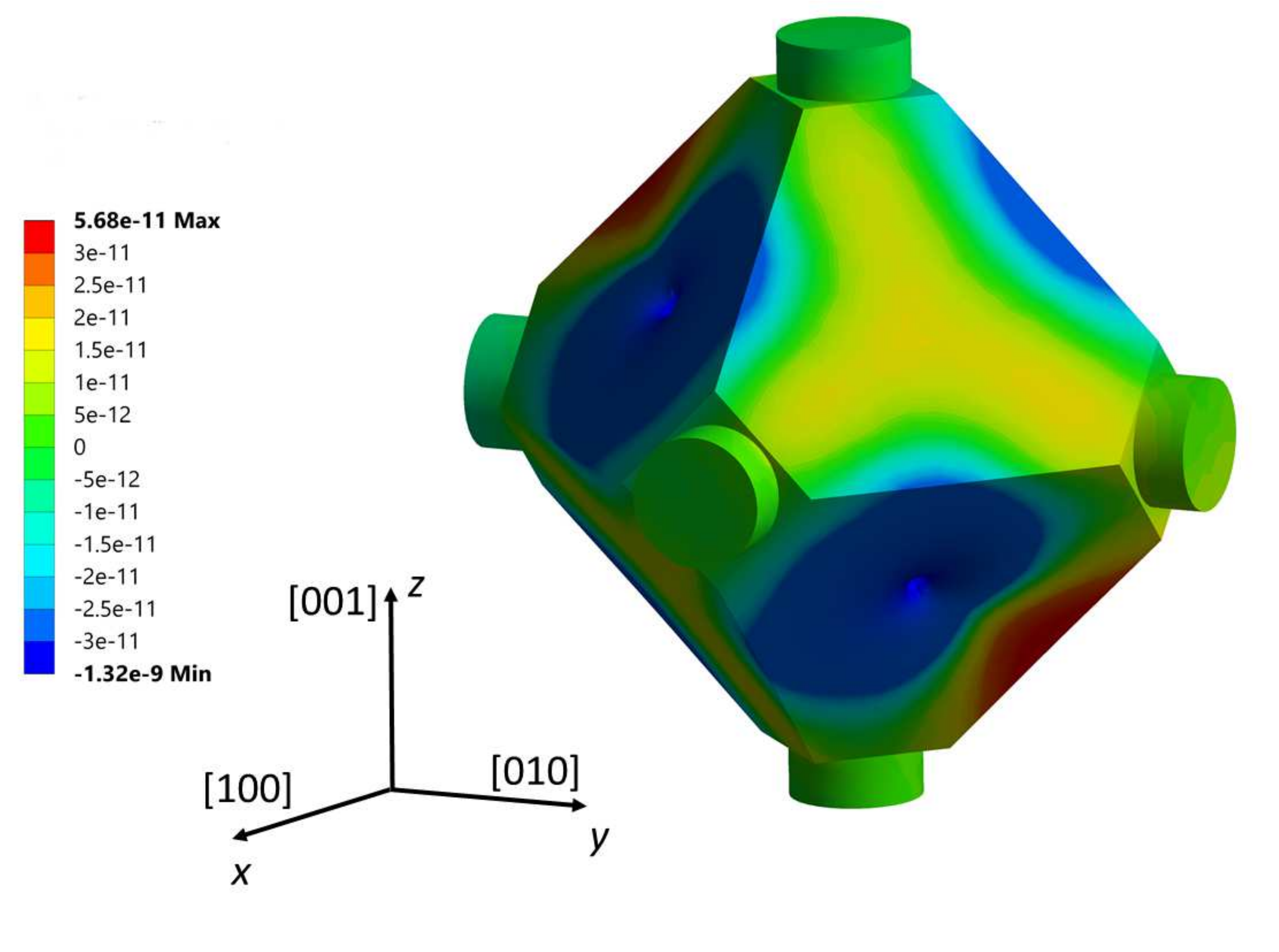}%
\par\end{center}
\end{minipage}\hfill{}
\par\end{centering}
\caption{\label{fig:Si100-Reduced-Mirror-Diameter-Ansys}Si {[}100{]} resonator
structure having the special geometry Si-I:  cut depth of $24.5$~mm,
with reduced mirror diameter of $d=10$~mm. For an angle of material rotation $\alpha = 0$ all three cavities exhibit zero length change upon application of the four support forces $F_{c}=1$~N. Shown are the displacements of the body along the $x$-axis. The scale
is in meter. }
\end{figure}

The {[}100{]} direction (top left panel) is the only one, for which
all three cavities exhibit equal sensitivity to the holding force.
For this direction, the FEA is performed directly with $C'=C$, eq.~(\ref{eq:Stiffness matrix-1}).
However, no zero sensitivity is possible with the half-inch mirror
substrates (blue points). Reduction of the mirror diameter to $d=10$~mm
(magenta points) allows achieving a zero sensitivity for all three
cavities simultaneously at a cut depth of $24.5$~mm with a slope
of $11\times10^{-11}$/mm. The corresponding simulation results are
depicted in more detail in Fig.~\ref{fig:Si100-Reduced-Mirror-Diameter-Ansys}.
We designate this geometry as Si-I. 

The {[}110{]} direction (middle left panel) displays identical sensitivity
for two cavities but without zero sensitivity cut depth. The third
cavity, the $y$-cavity, has zero sensitivity for two appropriate
cut depths (see Fig.~\ref{fig:CutDepth-Silicon}, middle left panel).
The slopes at $13$~mm and at $20$~mm cut depth are $2.9\times10^{-11}$/mm
and $6.5\times10^{-11}$/mm, respectively. They are smaller than for
the ULE case. 

The {[}111{]} direction (bottom left panel) has different sensitivities
for all three cavities (see Fig.~\ref{fig:CutDepth-Silicon}, bottom
left panel). They all have  zero sensitivity at large cut depths.
The difference in optimum cut depth for the $y$- and $z$-cavities
is particularly low, $0.16$~mm. This value is comparable with the
typical manufacturing precision of $0.1$~mm. Thus, the {[}111{]}
orientation makes it possible to access two orthogonal cavities having
small sensitivities. If the cut depth is chosen such that the sensitivity
of one cavity is zero,  the sensitivity of the second cavity is then
approximately $1.5\times10^{-11}$/mm. 

The foregoing discussion makes it clear that  there must be multiple
orientations inside the unit triangle which yield zero sensitivity
for at least one cavity. However, we are only interested in orientations
which have the effective Young's modulus along the {[}hkl{]} direction
and perpendicular to it as high as possible, in order to relax the
requirement on the manufacturing precision of the vertices' cut depth.
The effective Poisson's ratio for these directions must be in the
range which allows the cavities to have zero sensitivity. To identify
these directions an extensive simulation was carried out, which is
described in the next section.

\subsection{Simulation procedure}
\label{sec:Simulation-Procedure}

We performed simulations for more than 100 different directions inside
the unit triangle. The stiffness matrix $C'$ for each direction was
input into the simulation software. The chosen crystallographic direction was oriented along the $x$-axis of the laboratory reference frame. 
This orientation together with the stiffness matrix $C'$ defines
the orientation of the crystal along the $y$- and $z$-axes. Then the crystal was turned in 10 degree steps around the $x$-axis of the laboratory reference frame.  (see Fig.~\ref{fig:Transformation-Coordinate-System}). At each angle the force $F_{c}=$1~N was applied at each of the supports and pointing to the center of the cube, and the deformations $\Delta L_{j}$ of the three cavities along their axes $j=x,\,y,\,z$
calculated. 

The cut depth of the resonator was held constant at one particular
value, since it would have been too time-consuming to vary this parameter
as well. Its value was chosen based on the foregoing discussion, which
made clear that the slope of sensitivity is lower when the resonator
has the shape of a truncated cube. At a cut depth of $14.47$~mm
the shape of the resonator changes to a truncated octahedron, which
always has a higher sensitivity slope. For that reason, the cut depth
was fixed near the mean of the values corresponding to a truncated
cube, $7.27$~mm. 

\subsection{The support force sensitivity}
\label{sec:Support-Force-Sensitivity}

\begin{figure}
\begin{centering}
\begin{center}
\includegraphics[width=0.31\textwidth]{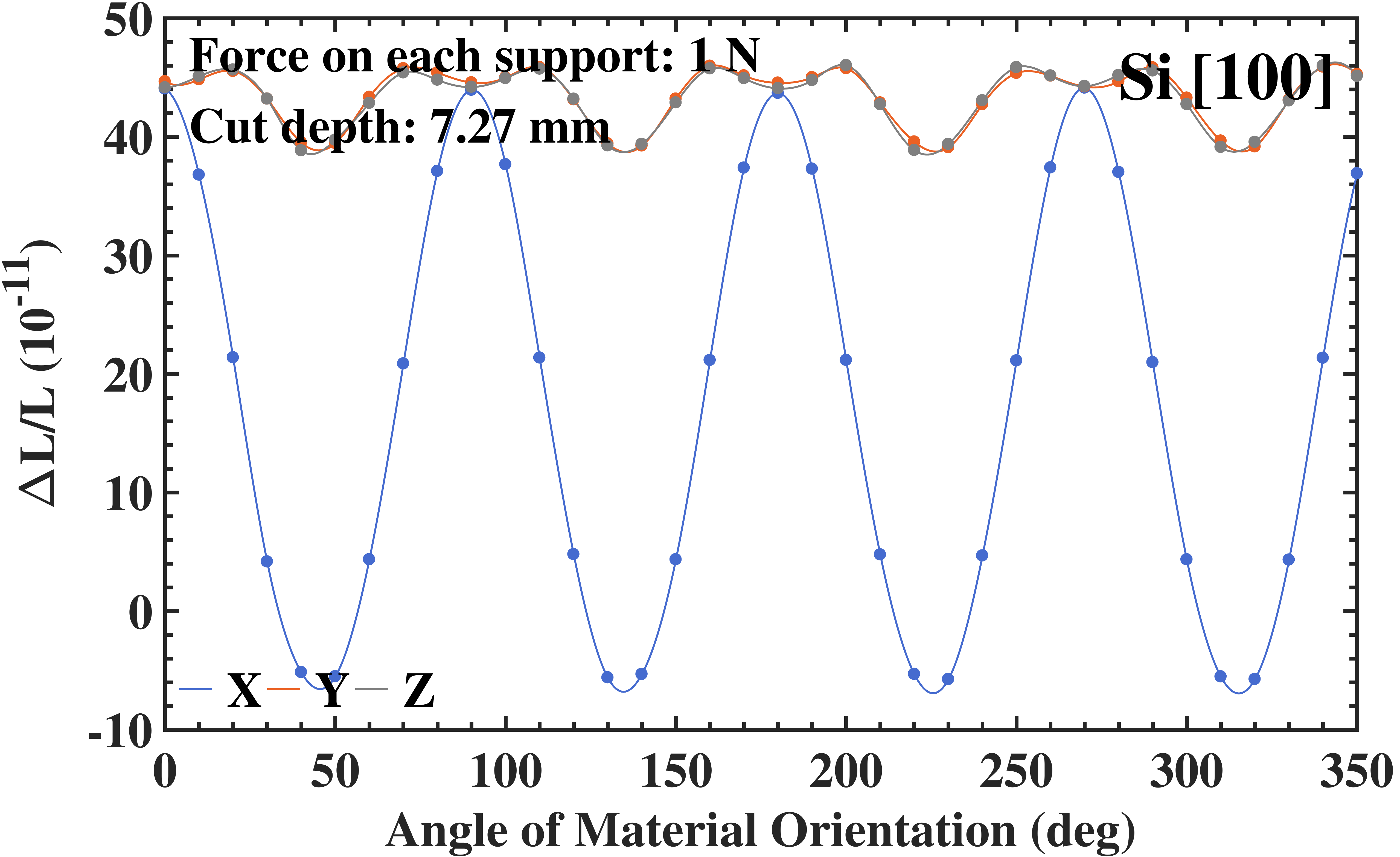}%
\quad{}%
\includegraphics[width=0.31\textwidth]{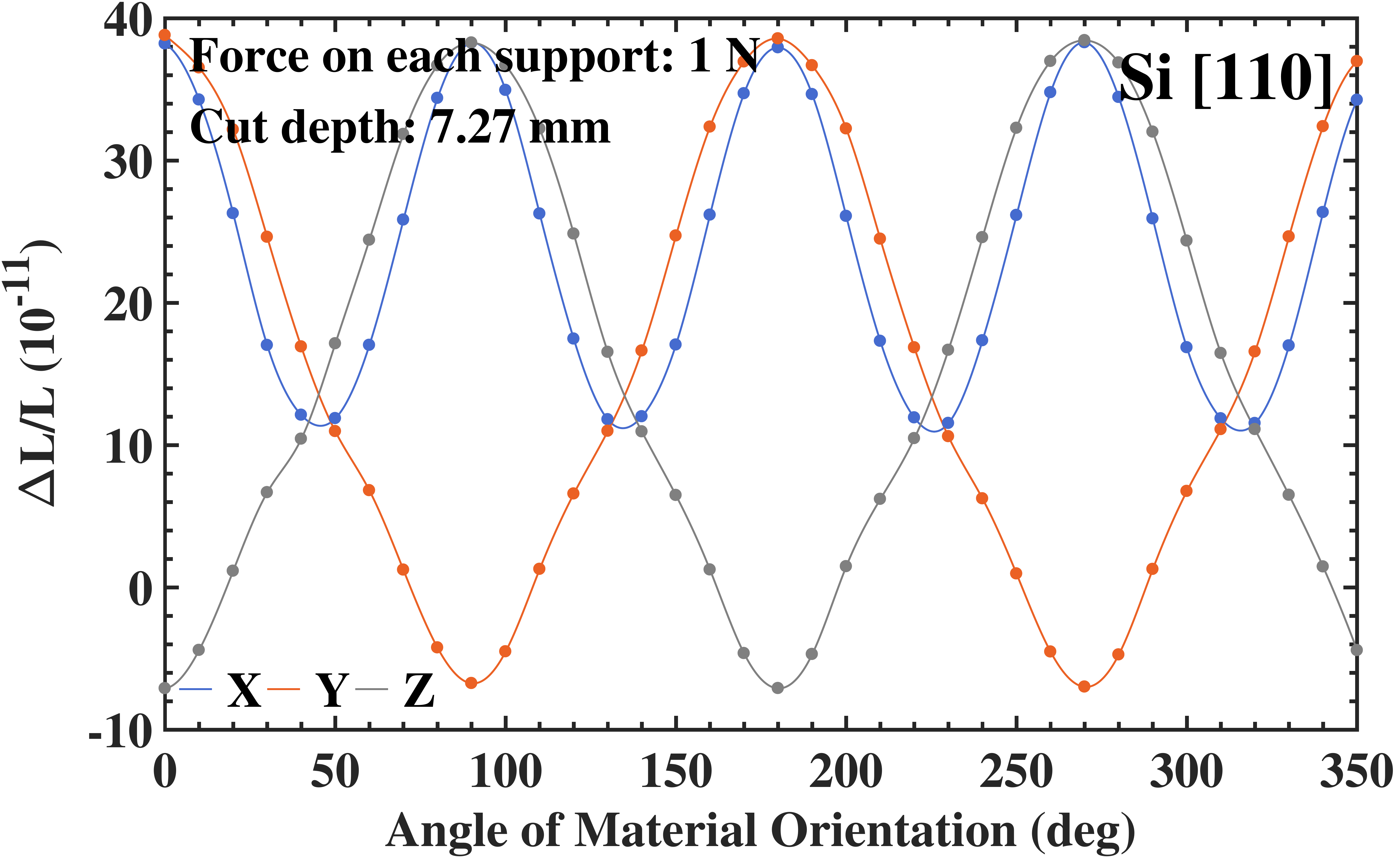}%
\quad{}%
\includegraphics[width=0.31\textwidth]{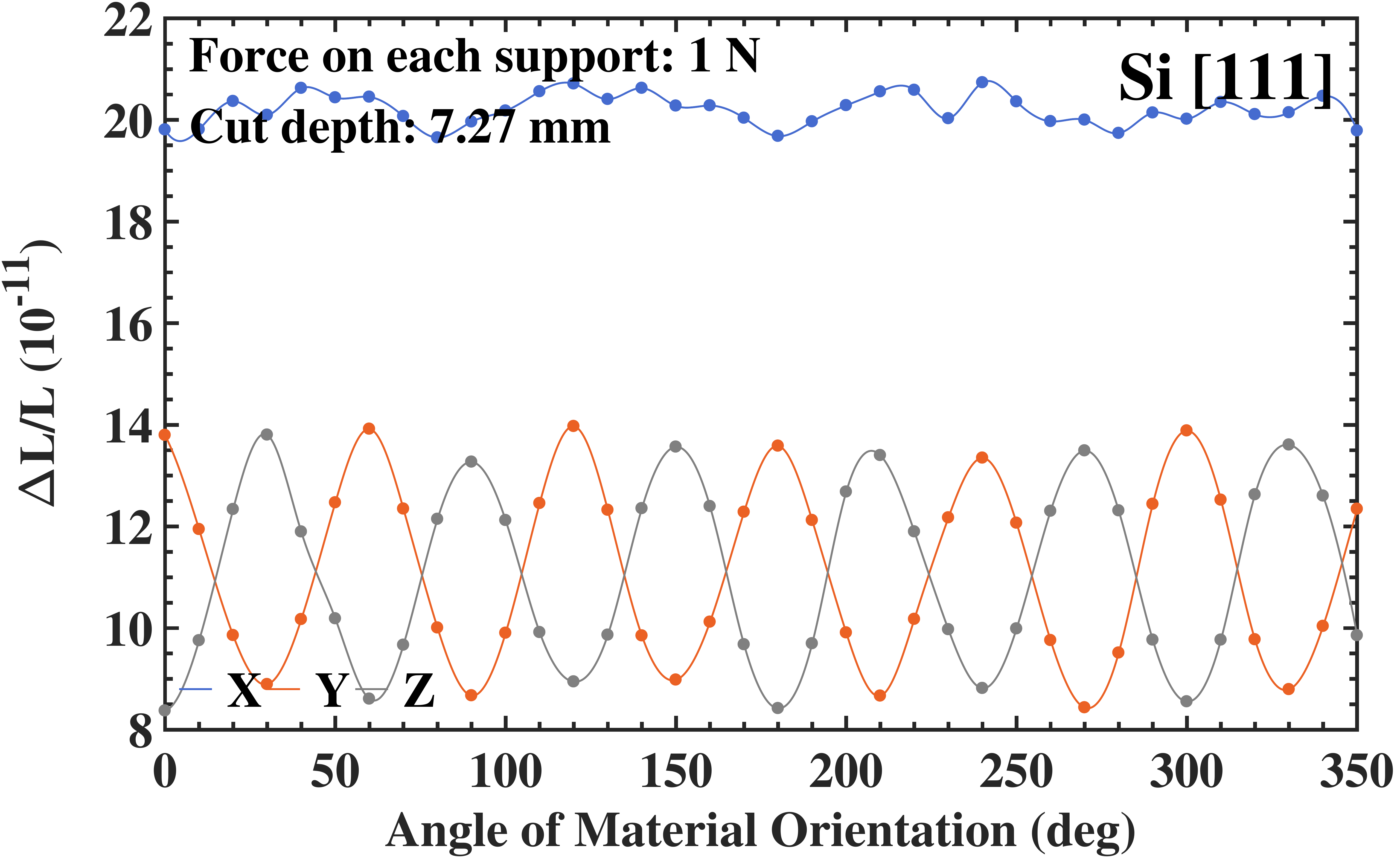}%
\par\end{center}
\medskip{}
\end{centering}
\caption{\label{fig:Resonator rotation} The three plots show the cavity length
changes upon  rotation around the {[}100{]}, {[}110{]}, and {[}111{]}
crystallographic directions of silicon, respectively. }
\end{figure}

The results for the three corner directions of the unity triangle,
{[}100{]}, {[}110{]}, {[}111{]}, are presented in Fig.~\ref{fig:Resonator rotation}.
As the top right panel shows, the fractional length changes of the
three cavities are equal only for the $x:$~{[}100{]} crystallographic
direction and $\alpha=0$ rotational angle. For all other orientations
and angles (all panels), at least two of the three cavities display
different fractional length changes. This is due to the differences
in the lattice structure along the cavity axes. The $x$-cavity of
the {[}100{]} orientation crosses zero fractional length change twice
in the $\alpha$ angle interval between 0 and 90 degrees. Two other
cavities have an equal sensitivity at all angles with a minimum of
$4\times10^{-10}$. The cube with the {[}110{]} material orientation
(bottom left panel) has zero sensitivity crossings for the $y$- and
$z$-cavities and no crossing for the $x$-cavity. All cavities of
the resonator with the {[}111{]} orientation (bottom right panel in
the figure) have no zero sensitivity. 

\begin{figure}[tb]
\begin{centering}
\begin{center}
\includegraphics[width=0.48\textwidth]{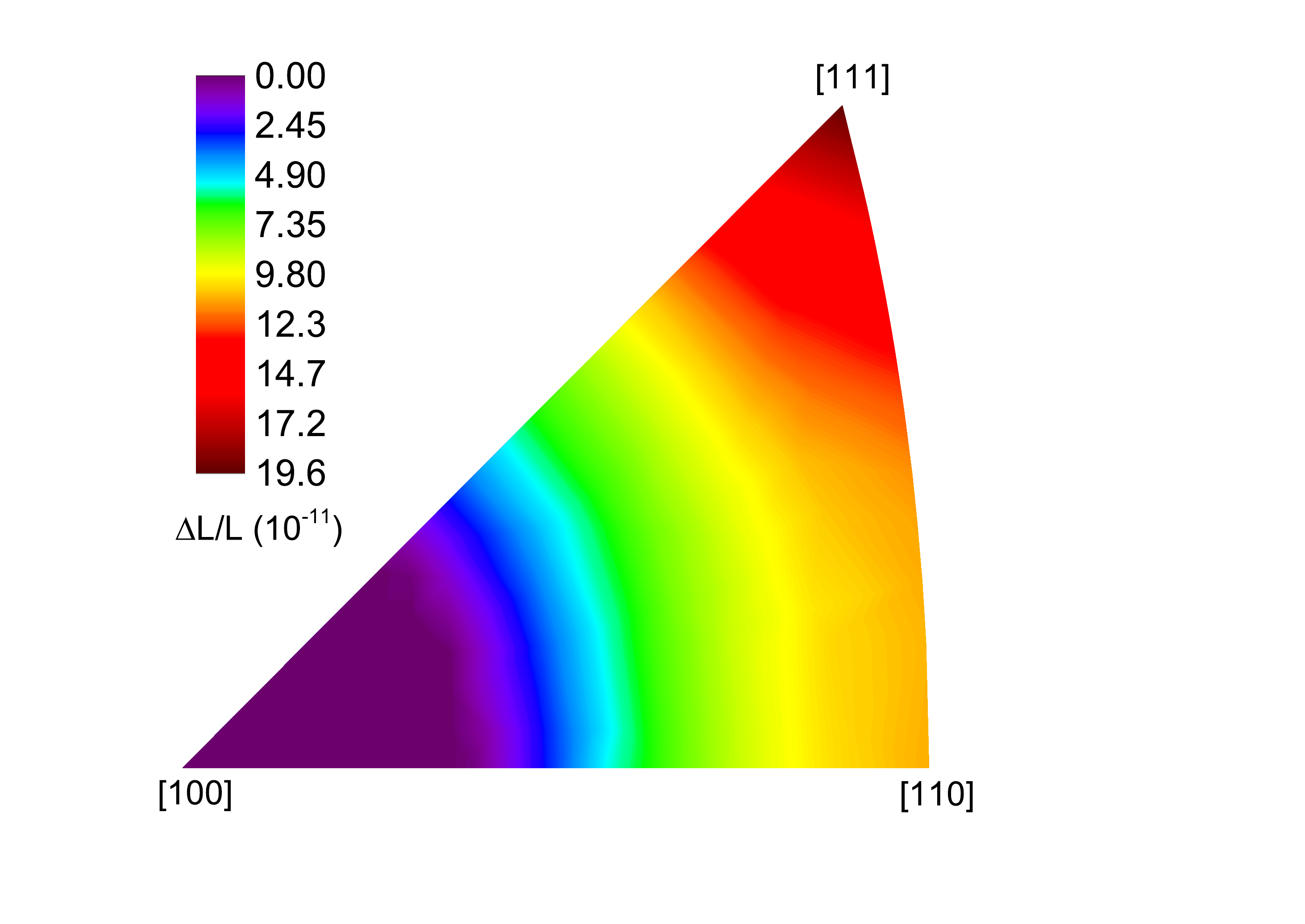}
\quad{}%
\includegraphics[width=0.48\textwidth]{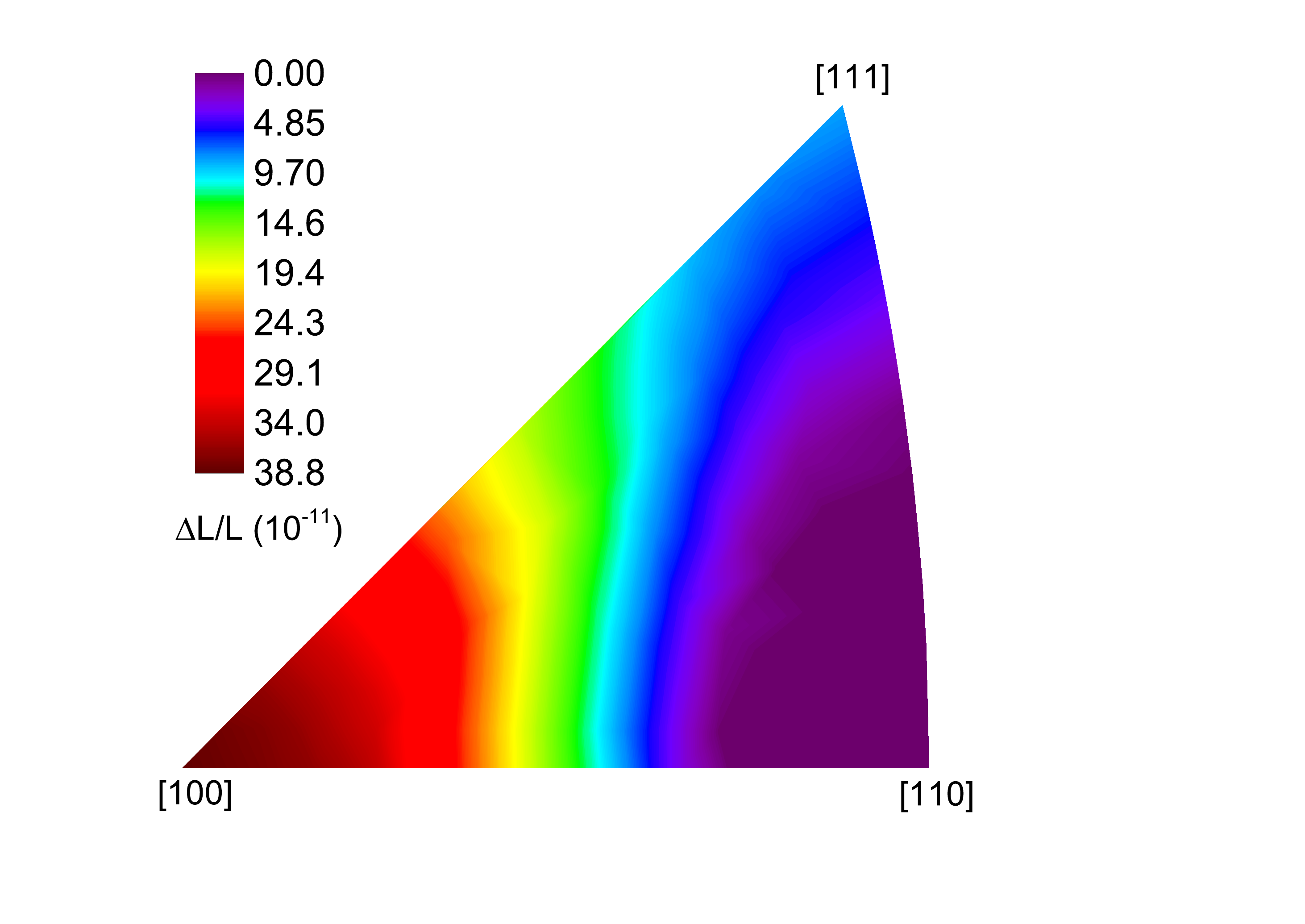}
\par\end{center}%
\bigskip{}
\begin{center}
\includegraphics[width=0.48\textwidth]{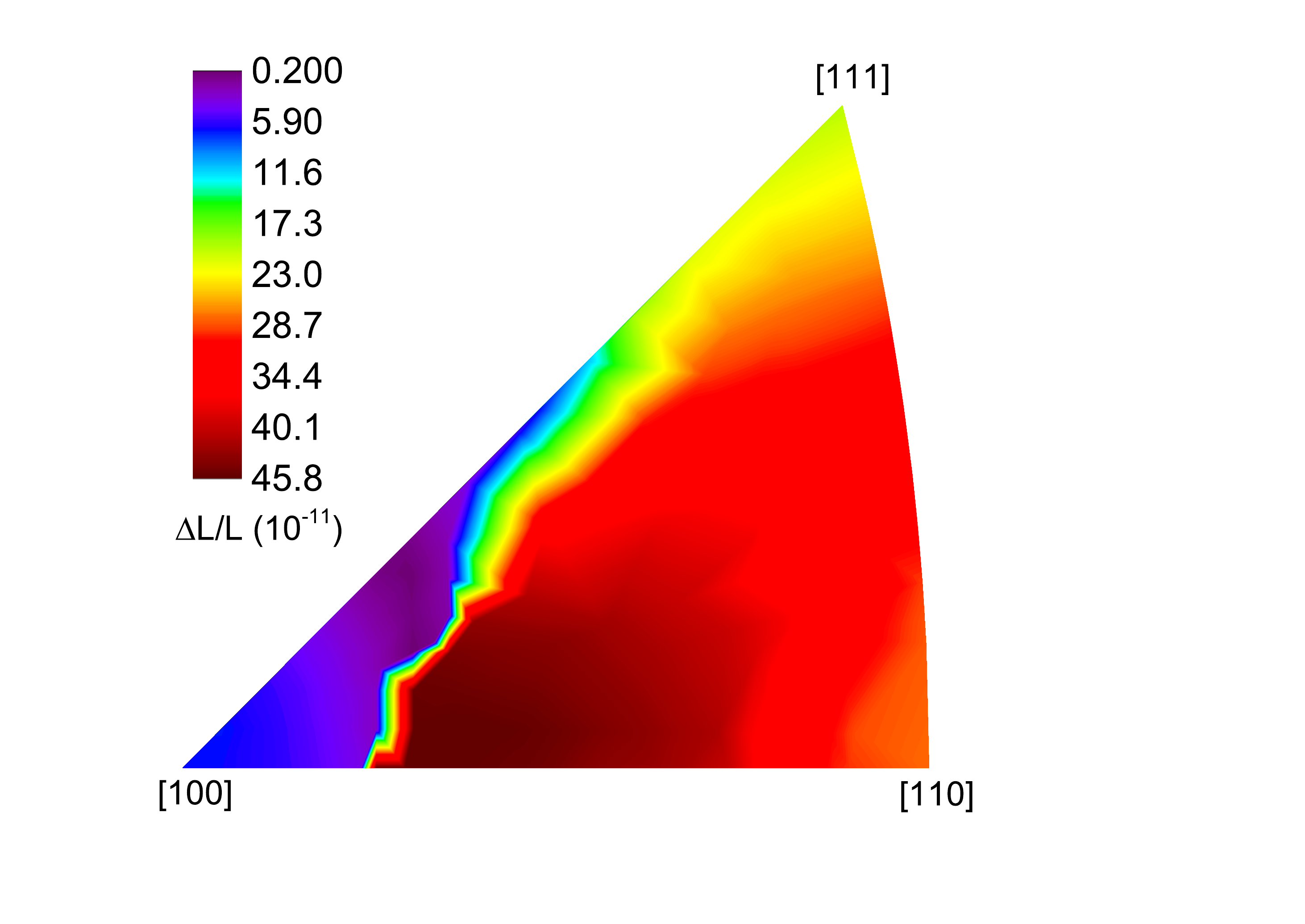}
\quad{}%
\includegraphics[width=0.48\textwidth]{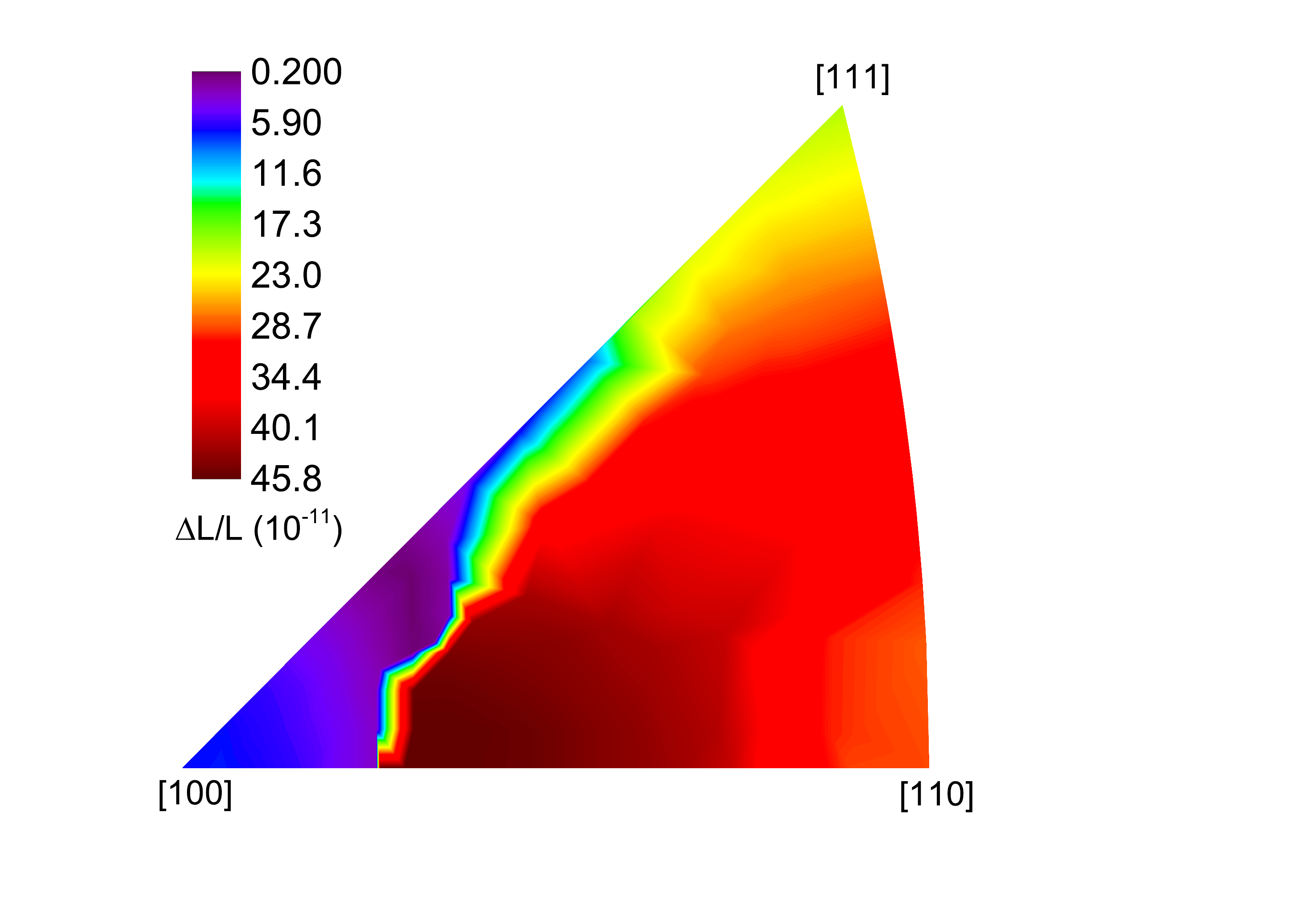}
\par\end{center}%
\bigskip{}
\begin{center}
\includegraphics[width=0.48\textwidth]{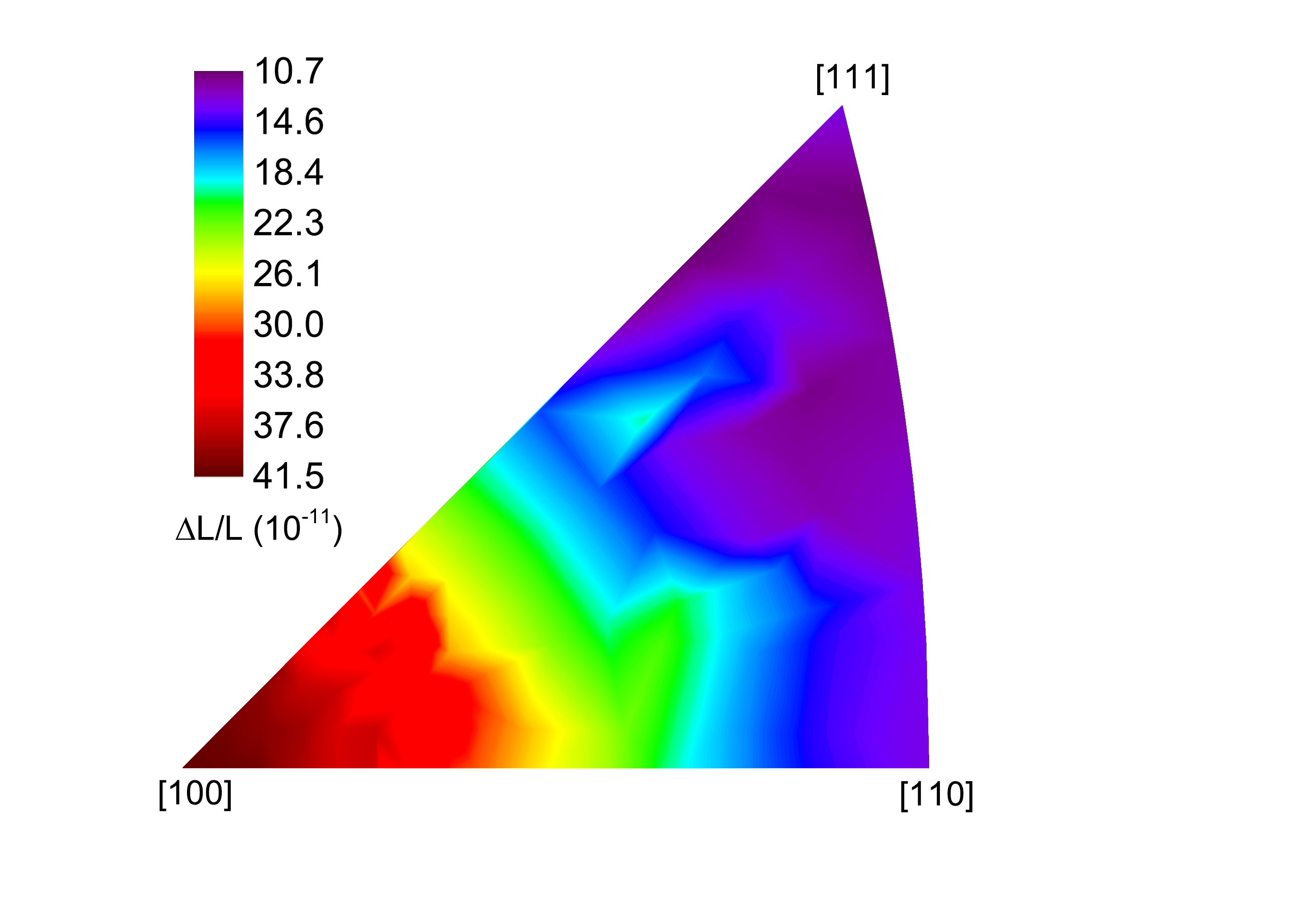}
\quad{}%
\includegraphics[width=0.48\textwidth]{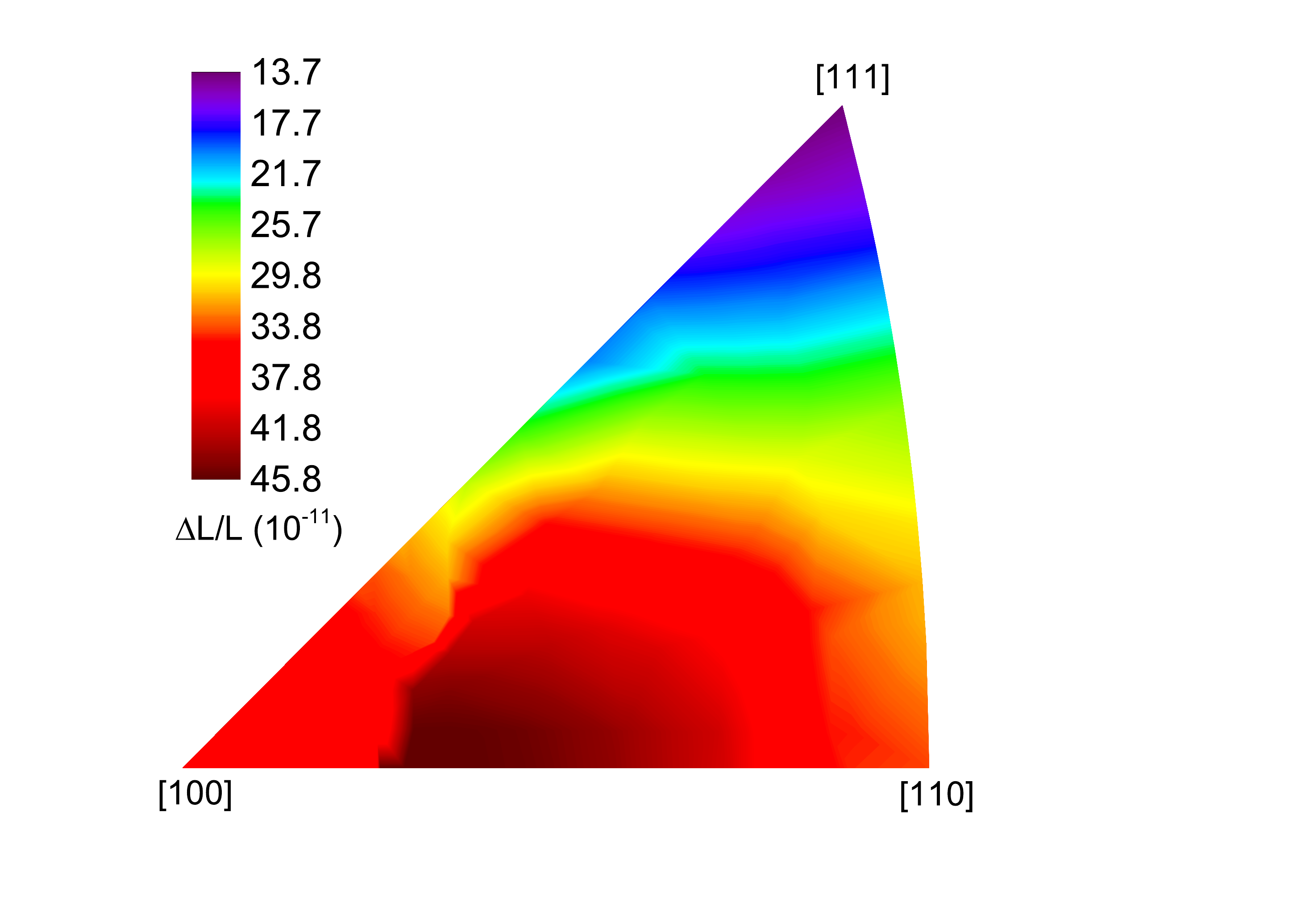}
\par\end{center}%
\end{centering}
\caption{\label{fig:Unity-Triangle-Sensitivity-Poisson-Young-1}Sensitivities
of cavity lengths to the application of the four support forces $F_{c}=1\,{\rm N}$,
for all crystallographic orientations. Top row: maps of minimum fractional
length change for the $x$-cavity (left) and for the $y/z$-cavities
(right); Middle row: maps of the $y$-cavity sensitivity (left) and
of the $z$-cavity sensitivity (right), both at the angle of smallest
$x$-sensitivity; Bottom row, left: maps of the $x/z$-cavity sensitivity
at the angle of smallest $y$-cavity sensitivity; bottom row, right:
maps of the $x/y$-cavity sensitivity at the angle of smallest $z$-cavity
sensitivity.}
\end{figure}

The minimum fractional length changes ocurring over a full turn around
all orientations ($\alpha$ varies between 0 and $360$~deg.) inside
the unity triangle are displayed in Fig.~\ref{fig:Unity-Triangle-Sensitivity-Poisson-Young-1}.
As we can see, there is only one favorable orientation for each cavity,
shown in purple (top row). To have zero sensitivity for the $x$-cavity
the resonator must be oriented along the {[}100{]} direction, see
top left panel. Zero sensitivity for the $y$- and $z$-cavities
is only possible if resonator is oriented along the {[}110{]} direction,
but not at the same angle. As Fig.~\ref{fig:Resonator rotation},
bottom left panel, suggests, there is an angle shift of 90~degrees
between them. Our results rule out the possibility of having zero
sensitivity for more then one cavity simultaneously, for the considered
vertex cut depth. In Fig.~\ref{fig:Unity-Triangle-Sensitivity-Poisson-Young-1}
we display sensitivities along the two axes at an angle of minimum
sensitivity for the one of the three axis. As in the case of isotropic
materials we compare our results with the Poisson's ratio (see Fig.~\ref{fig:SST-Poisson_Young}
and Fig.~\ref{fig:Unity-Triangle-Sensitivity-Poisson-Young-1}).
Both the $x$-cavity sensitivity and the Poisson's ratio have a minimum
in the vicinity of the {[}100{]} direction. The minimum of the sensitivity
for the $y$- and $z$-cavities and for {[}110{]} silicon orientation
(see Fig.~\ref{fig:Unity-Triangle-Sensitivity-Poisson-Young-1},
top right panel) corresponds to the minimum of the Poisson's ratio
calculated for the direction perpendicular to {[}110{]} (see Fig.~\ref{fig:SST-Poisson_Young},
bottom right panel). 

The evaluation of the Young's modulus for the direction parallel to
the crystallographic orientation and perpendicular to it is presented
in Fig.~\ref{fig:Unity-Triangle-Sensitivity-Poisson-Young-1}, top
row panels. For example, we find that the direction {[}110{]}, with
the Young's modulus of $169.1$~GPa along the $x$-axis and $187.9$~GPa
perpendicular to it, is more favorable than the {[}100{]} direction,
for which the values are $130.1$~GPa and $169.1$~GPa, respectively.

\begin{figure}[tb]
\begin{centering}
\begin{center}
\begin{minipage}[t]{0.48\columnwidth}%
\begin{center}
\includegraphics[width=0.98\textwidth]{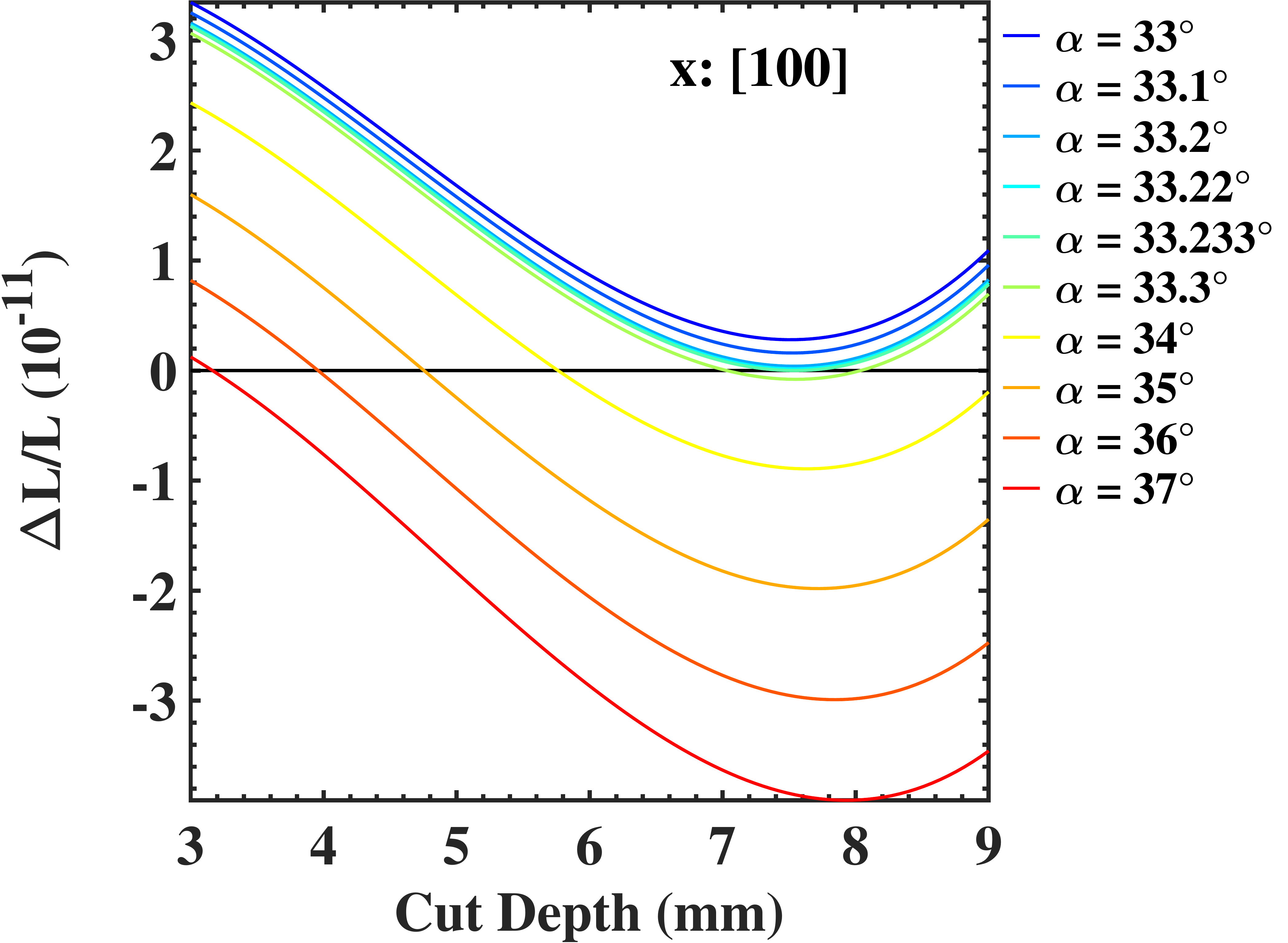}
\par\end{center}%
\end{minipage}
\quad{}%
\begin{minipage}[t]{0.48\columnwidth}%
\begin{center}
\includegraphics[width=0.98\textwidth]{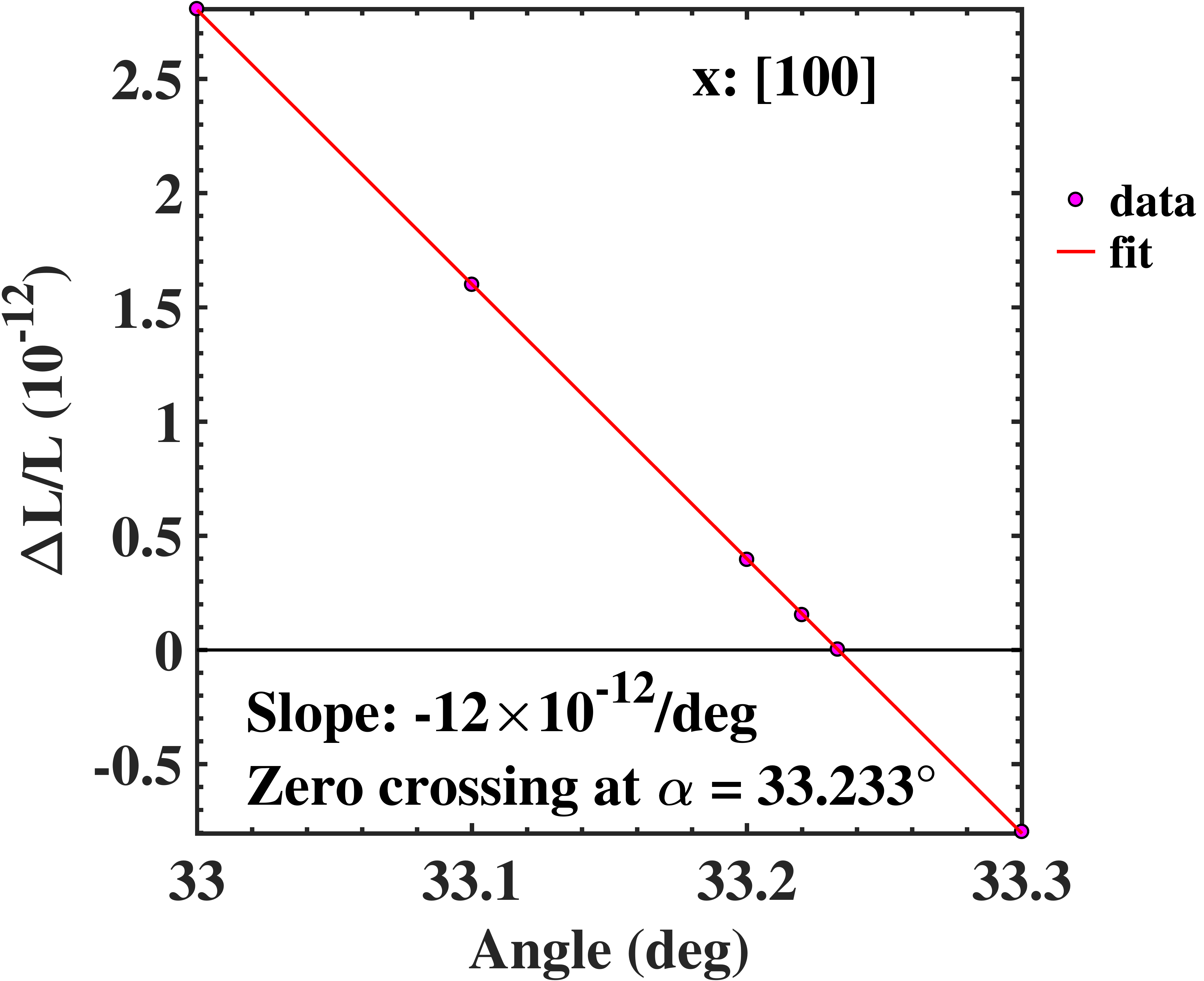}
\par\end{center}%
\end{minipage}
\par\end{center}%
\bigskip{}
\begin{center}
\begin{minipage}[t]{0.48\columnwidth}%
\begin{center}
\includegraphics[width=0.98\textwidth]{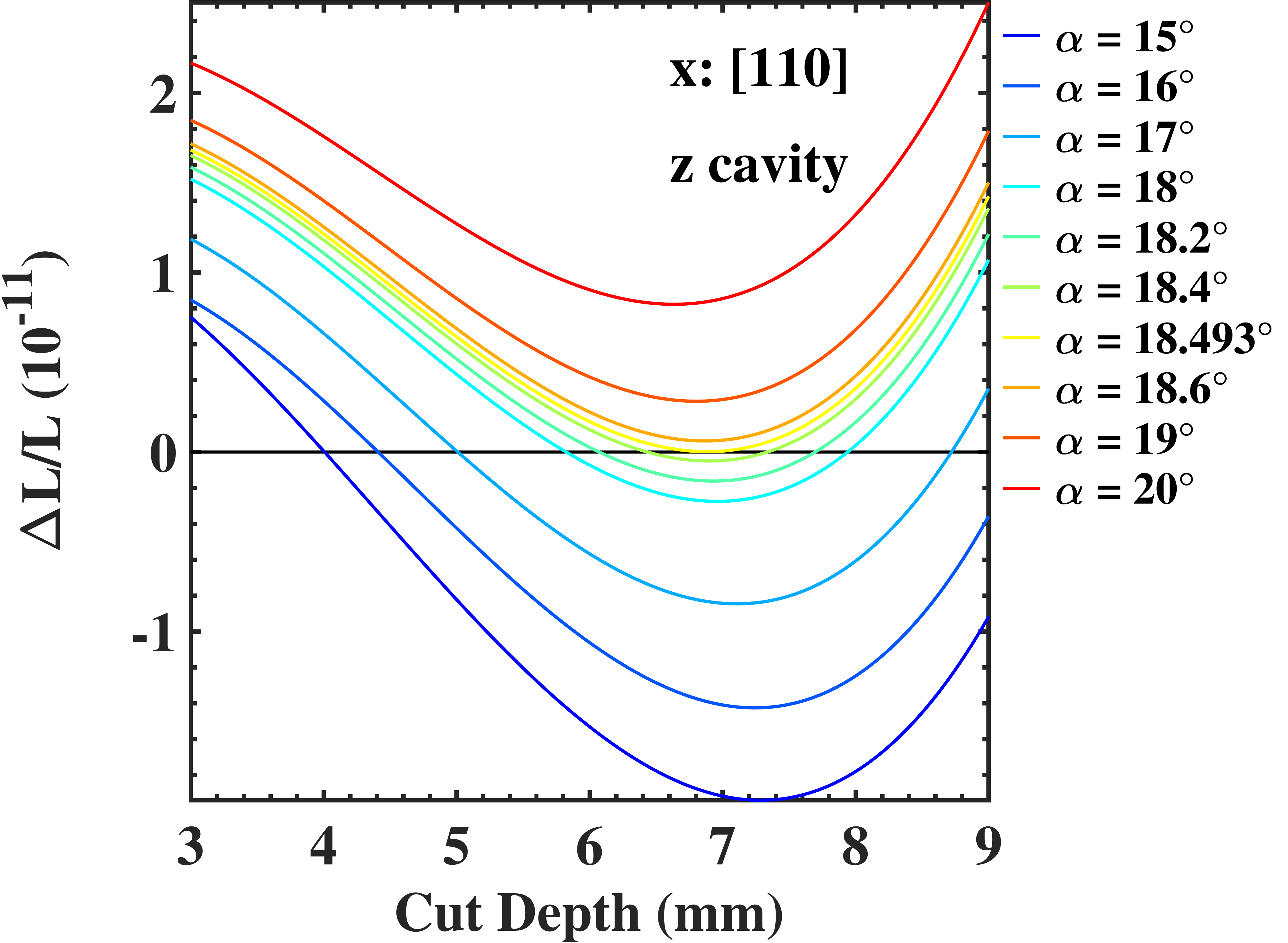}
\par\end{center}%
\end{minipage}
\quad{}%
\begin{minipage}[t]{0.48\columnwidth}%
\begin{center}
\includegraphics[width=0.98\textwidth]{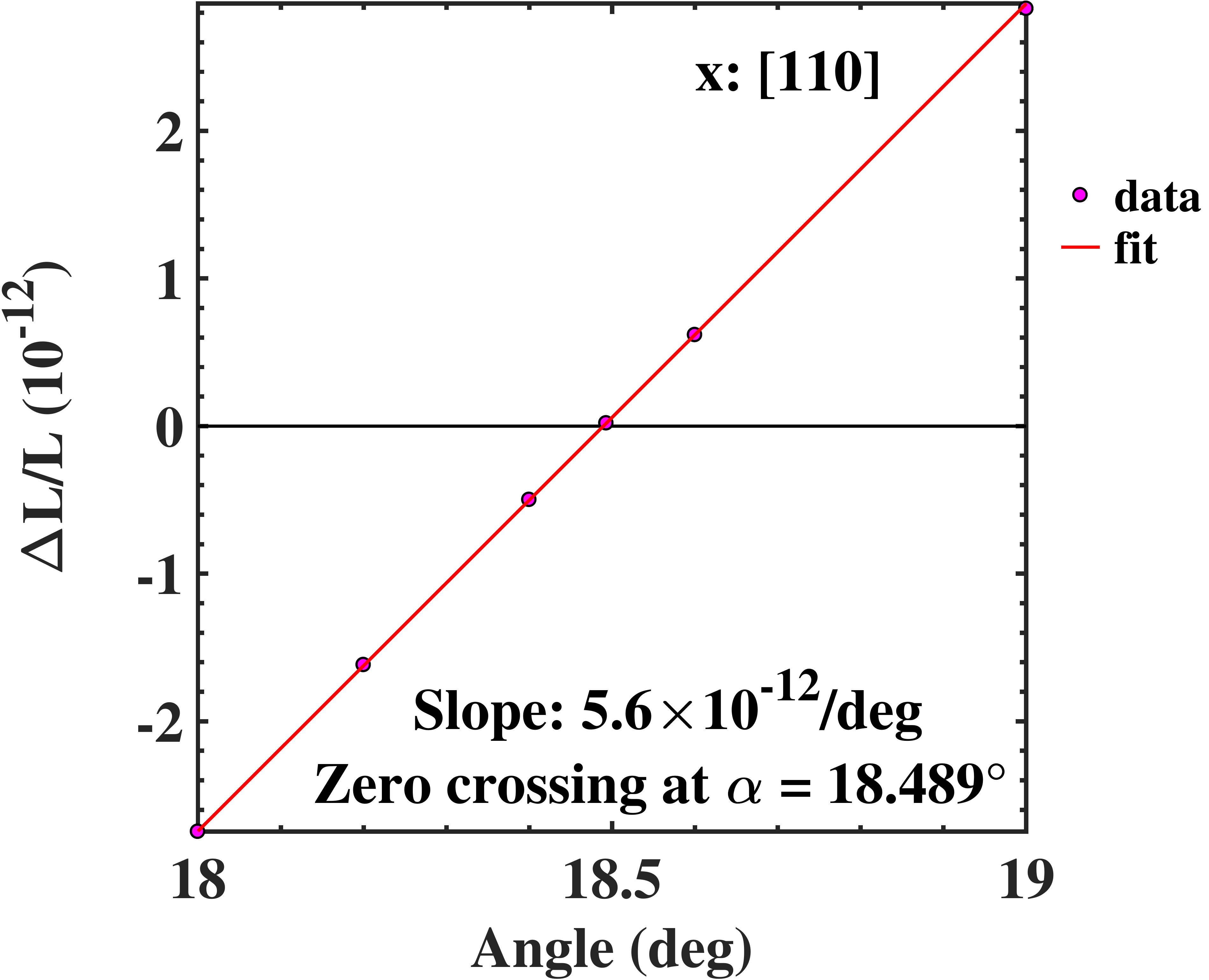}
\par\end{center}%
\end{minipage}
\par\end{center}%
\end{centering}
\caption{\label{fig:Si100 and Si110 Optimal Angle}Determination of an optimal
angle and cut depth for the zero fractional length change of the $x$-cavity
of the resonator and {[}100{]} silicon orientation (top) and $z$-cavity (or $y$-cavity with an angle $\alpha$ shifted by $+90$
degrees) and {[}110{]} orientation (bottom). A force of 1~N is applied
at each support. }
\end{figure}

This difference should be reflected in the dependence of the fractional
length change on the variation in cut depth and in angle of rotation
around their optimal values. To obtain these numbers we performed
a series of simulations displayed in Fig.~\ref{fig:Si100 and Si110 Optimal Angle}.
We find that for the $x:$~{[}100{]} orientation (top row) both $\Delta L_{x}/L$
and its sensitivity to cut depth variation are zero at a cut depth
of 7.54~mm and at angle $\alpha=33.23$~degree. This case is denoted
by Si-II in the following. The fractional length change varies by
$12\times10^{-12}$/deg around the optimum angle and by $3.5\times10^{-12}$/mm
around the optimum cut depth.  

The bottom row of panels shows that the cube with $x:$~{[}110{]}
orientation has a zero sensitivity  of the $z$-cavity at an angle
of 18.49~degree and at a cut depth of 6.88~mm (geometry denoted
by Si-III). The $y$-cavity displays a zero crossing at the same cut
depth but at an angle which is shifted by $90$ degrees from that
of the $z$-cavity (see Fig.~\ref{fig:Resonator rotation}, bottom
left panel)  The fractional sensitivity variations are $5.6\times10^{-12}$/deg
and $2.8\times10^{-12}$/mm for the variations in angle and cut depth,
respectively.  

\section{Effect of imperfections\label{sec:Effect-of-imperfections}}

We evaluated the effect of additional imperfections on the sensitivity
to the supporting force and on the acceleration insensitivity, besides
the already considered cut depth deviation and orientation deviation
(for anisotropic materials). For silicon we consider only the geometries
Si-I, Si-II, Si-III introduced above, and for ULE and $\beta$-SiC
the shapes of Sec.~\ref{sec:Cubic-cavities-conventional-optical-materials}.
The results are presented in Table~\ref{tab:Sensitivity to Error in Manufacturing-1}.
Item 1 in the table reports results already discussed above. 

The cut depth of the individual resonator vertices may vary due to
the accuracy of manufacture. We analyze the case that only one vertex
(also serving as support) deviates in cut depth from the other seven.
 Tab.~\ref{tab:Sensitivity to Error in Manufacturing-1}, item 2,
shows that for a cut depth precision of 0.1~mm, the cavity length
deformation effects are at the level of $3\times10^{-12}$ per N support
force or smaller. In the gravity field this imperfection introduces
an acceleration sensitivity Max$(k_{xx},\,k_{xz})$ on the order of $6\times10^{-12}/{\rm g}$
or smaller.

Misplacement of the mirrors with respect to the symmetry axes of the
cube can occur during assembly. As a result, the light propagation
occurs along an axis shifted with respect to the symmetry axis. This
breaks the symmetry assumed in the concept of the cubic block. Different
deformation of the opposing mirrors at the intersection of the axis
and the mirror surface introduces an additional length change and
degrades the acceleration sensitivities $k_{ij}$. The calculation
of the degradation was performed for a $\epsilon=1$~mm shift of
the optical axis in the direction having the largest deformation.
The value 1~mm is larger than errors in fabrication and only serves
as an example. Our result for ULE, given in the table (item 3), appears
consistent with the FEA value reported by \cite{Webster2011}. For
silicon, we find acceleration sensitivities up to $3\times10^{-11}/{\rm g}$.

The sensitivity to orientation of the Si crystal with respect to the
resonator is reported under item 4; we find the acceleration sensitivities
to be rather small if an error of 1~degree is assumed. 

Asymmetrical mounting of the resonator in the frame with the supports
displaced from their optimal position is another source of error (items
5,~6 in the table). We see that the effects are not negligible. For
an offset of $\epsilon=0.1$~mm, in the case of ULE the cavity length
changes fractionally by $\simeq1\times10^{-11}$ for a 1~N support
force, and a sensitivity to acceleration perpendicular to the cavity
$k_{xz}=7\times10^{-12}/{\rm g}$ arises. For silicon the numbers
are similar.

The above results make it clear that great care should be taken in
mounting the resonator in the supporting frame. Together with the
offsets of mirrors from the respective symmetry axes, mounting errors
appear to be a major potential cause of the degradation of sensitivity
compared to the ideal. Comparing silicon with ULE we find that silicon
is less sensitive to errors, but in several respects only by a factor
approximately 2. Comparing the three silicon resonator geometries,
Si-I, Si-II, and Si-III, we find Si-II to be more advantageous, in
particular with respect to one critical error, the offset from the
optical axis. 

\begin{table}[tb]
\raggedright{}{\tiny{}}%
\begin{tabular}{|l|l|c|c|c|c|c|}
\hline 
\multirow{2}{*}{{\footnotesize{} \makecell{Type of geometry \\ change $\epsilon$} }} & \multirow{2}{*}{{\footnotesize{}Quantity$(10^{-11})$}} & \multicolumn{5}{c|}{{\footnotesize{}Material and orientation}}\tabularnewline
\cline{3-7} 
 &  & {\footnotesize{}ULE} & {\footnotesize{}$\beta$-SiC} & {\footnotesize{}Si-I} & {\footnotesize{}Si-II} & {\footnotesize{}Si-III}\tabularnewline
\hline \hline
\multirow{2}{*}{{\footnotesize{}\makecell{1. Cut depth,\\ all vertices} }} & {\footnotesize{}$S_{\epsilon}\,({\rm mm}^{-1})$} & {\footnotesize{}$3.7$} & {\footnotesize{}$4.1$} & {\footnotesize{}$11$} & {\footnotesize{}$0.35$} & {\footnotesize{}$0.28$}\tabularnewline
\cline{2-7} 
 & {\footnotesize{}$k\,({\rm g}^{-1})$} & {\footnotesize{}0} & {\footnotesize{}0} & {\footnotesize{}0} & {\footnotesize{}0} & {\footnotesize{}0}\tabularnewline
\hline 
\multirow{2}{*}{{\footnotesize{}\makecell{2. Cut depth,\\ one vertex} }} & {\footnotesize{}$S_{\epsilon}\,({\rm mm}^{-1})$} & {\footnotesize{}$0.8$} & {\footnotesize{}$1.1$} & {\footnotesize{}$3.3$} & {\footnotesize{}$0.01$} & {\footnotesize{}$0.41$}\tabularnewline
\cline{2-7} 
 & {\footnotesize{}$k_{xx},\,k_{xz}\,({\rm g}{}^{-1})$ ~} & {\footnotesize{}$6.4$, $3.6$} & {\footnotesize{}$1.5$, $0.13$} & {\footnotesize{}$1.01$, $0.2$} & {\footnotesize{}$2.2$, $0.61$} & {\footnotesize{}$4.2$, $2.2$}\tabularnewline
\hline 
\multirow{2}{*}{{\footnotesize{}\makecell{3. Offset of the\\optical axis} }} & {\footnotesize{}$S_{\epsilon}\,({\rm mm}^{-1})$} & {\footnotesize{}$2.7$} & {\footnotesize{}$0.07$} & {\footnotesize{}$0.6$} & {\footnotesize{}$0.7$} & {\footnotesize{}$3.5$}\tabularnewline
\cline{2-7} 
 & {\footnotesize{}$k_{xx},\,k_{xz}\,({\rm g}^{-1})$ ~} & {\footnotesize{}$0$, $5$} & {\footnotesize{}$0$, $0.4$} & {\footnotesize{}$0.23$, $3.2$} & {\footnotesize{}$0$, $1$} & {\footnotesize{}$0$, $0.8$}\tabularnewline
\hline 
\multirow{2}{*}{{\footnotesize{}\makecell{4. Orientation\\ of material} }} & {\footnotesize{}$S_{\epsilon}\,({\rm deg}^{-1})$} & {\footnotesize{}-} & {\footnotesize{}-} & {\footnotesize{}$0.3$} & {\footnotesize{} $1.2$} & {\footnotesize{}$0.56$}\tabularnewline
\cline{2-7} 
 & {\footnotesize{}$k_{xx},\,k_{xz}\,({\rm g}{}^{-1})$} & {\footnotesize{}-} & {\footnotesize{}-} & {\footnotesize{}$0$, $0.002$} & {\footnotesize{}$0.05$, $0.002$} & {\footnotesize{}$0$, $0.15$}\tabularnewline
\hline 
\multirow{2}{*}{{\footnotesize{}\makecell{5. Horizontal offset\\ of one support} }} & {\footnotesize{}$S_{\epsilon}\,({\rm mm}^{-1})$} & {\footnotesize{}$13$} & {\footnotesize{}$2.9$} & {\footnotesize{}$14.4$} & {\footnotesize{}$7.7$} & {\footnotesize{}$5.7$}\tabularnewline
\cline{2-7} 
 & {\footnotesize{}$k_{xx},\,k_{xz}\,({\rm g}{}^{-1})$} & {\footnotesize{}$5.4$, $7.0$} & {\footnotesize{}$1.4$, $0.12$} & {\footnotesize{}$4.7$, $0.13$} & {\footnotesize{}$3.7$, $3.8$} & {\footnotesize{}$5.0$, $2.9$}\tabularnewline
\hline 
\multirow{2}{*}{{\footnotesize{}\makecell{6. Vertical offset\\ of one support} }} & {\footnotesize{}$S_{\epsilon}\,({\rm mm}^{-1})$} & {\footnotesize{}$7.5$} & {\footnotesize{}$2.0$} & {\footnotesize{}$8.41$} & {\footnotesize{}$3.6$} & {\footnotesize{}$3.8$}\tabularnewline
\cline{2-7} 
 & {\footnotesize{}$k_{xx},\,k_{xz}\,({\rm g}{}^{-1})$} & {\footnotesize{}$3.2$, $3.2$} & {\footnotesize{}$1.2$, $0.84$} & {\footnotesize{}$3$, $1.43$} & {\footnotesize{}$1.4$, $1.8$} & {\footnotesize{}$1.9$, $2.1$}\tabularnewline
\hline 
\end{tabular}\caption{\label{tab:Sensitivity to Error in Manufacturing-1}Sensitivity $S_{\epsilon}=\partial(\Delta L(F_{c})/L)/\partial\epsilon$
of the length change caused by a $F_{c}=1$~N force to possible manufacturing
errors $\epsilon$, for different cube geometries. The geometries
are specified in the text. Also listed are values of acceleration sensitivity
$k$ in the presence of an error $\epsilon$ of 1~mm or 1 degree. Sensitivity
to cut depth of one vertex as well as to the offset of one support
were evaluated using the vertex and the support at the location defined
by the vector $\vec{v}=(1,-1,1)$. For the offset of the optical axis
we considered as symmetry axis the $x$-cavity axis in the case of
ULE, polycrystalline $\beta$-SiC and Si {[}100{]} and the $y$-cavity
axis in the case of Si {[}110{]}. The direction of the offset was assumed
to be along the direction with the largest mirror deformation.}
\end{table}

\section{Error evaluation}
\label{sec:Error_Evaluation}
To determine the influence of finite mesh size on the optimum cut depth we performed simulations of a ULE block applying different mesh density. Assuming that simulations with infinitely small mesh size adequately represent the reality, we extrapolated our results toward decreasing mesh size and obtained an error of less then $0.08$~mm for the optimum cut depth.

We also evaluated the scatter of data points in different simulation results by fitting them with a polynomial of high degree and plotting the distribution of the residuals. This evaluation indicates an error in sensitivity $\Delta L/L$ of $\pm2\times10^{-12}$ for the simulations where no acceleration is involved and an error of $\pm1\times10^{-11}$ whenever an acceleration is applied. This error was found to have approximately cubic dependence on mesh size.  

Another way to validate our simulation procedures is to compare with published results. In \cite{Matei2016}, a vertically oriented, biconical silicon resonator was simulated and the results experimentally validated. Our simulations are in good agreement.

\section{Summary and conclusion}
\label{sec:Conclusion}

We analyzed the sensitivity to support forces of the three-cavity
cubic block made of different materials. For isotropic materials,
we identified a ``magic'' range for Poisson's ratio, $0.13\le\nu\le0.23$,
for which the three cavity lengths become insensitive to the strength
of the support force. Because of this particular range, apart from
ULE, only fused silica and $\beta$-SiC are suitable materials among
the common isotropic materials used in the optics industry. Silicon,
as anisotropic material, offers multiple suitable orientations for
providing zero sensitivity. Based on FEA simulations, we identified
two orientations, {[}100{]} and {[}110{]}, to be particularly suitable.
Compared to ULE, they provide one cavity with more robustness to the
errors in manufacturing: the acceleration sensitivity is reduced by
a factor of approximately $2$ or more compared to ULE, depending on
the error.

We thus showed that silicon spacers with octahedral symmetry can provide
a favorable option for cryogenic, support-force-insensitive and vibration-insensitive
cavities. Particularly attractive is the fact that there exists one
geometry, with {[}100{]} orientation of the crystal, which provides
simultaneously three nominally insensitive cavities in the same spacer.
This geometry could be useful for certain applications, e.g. tests
of Lorentz Invariance. Nevertheless, even with only 0.1~mm imprecision
in manufacturing and mounting, a residual sensitivity to support force
at the level of $14\times10^{-12}/{\rm N}$ level can occur. The corresponding
residual vibrational sensitivity can be as high as $5\times10^{-12}/{\rm g}$.
Achieving a suitable design and production of the frame that provides
stable support forces will be an important additional aspect of the
overall system.



\begin{acknowledgements}
We thank T. Legero (PTB) for providing us with the design of the biconical Si resonator allowing us to test our simulations, A. Nevsky for stimulating discussions, and D. Sutyrin for his help with the simulations.
This work was performed in the framework of project SCHI 431/21-1 of the Deutsche Forschungsgemeinschaft.
\end{acknowledgements}




\end{document}